\documentclass[onecolumn]{emulateapj} 
 
\usepackage{verbatim} 
\usepackage{natbib} 
\usepackage{amsmath} 
\usepackage{amsbsy} 
\usepackage{lscape} 
\usepackage{epstopdf} 
 
\slugcomment{Submitted to ApJ} 
\shorttitle{UV Background Spectrum} 
\shortauthors{Faucher-Gigu\`ere et al.} 
 
\newcommand{\Lya}{\mbox{Ly$\alpha$}}

\begin{document} 
 
\title{A New Calculation of the Ionizing Background Spectrum and the Effects of HeII Reionization} 
 
\author{Claude-Andr\'e Faucher-Gigu\`ere\altaffilmark{1}, Adam Lidz\altaffilmark{1}, Matias Zaldarriaga\altaffilmark{1,2}, Lars Hernquist\altaffilmark{1},} 
\altaffiltext{1}{Department of Astronomy, Harvard University, Cambridge, MA, 02138, USA; cgiguere@cfa.harvard.edu.} 
\altaffiltext{2}{Jefferson Physical Laboratory, Harvard University, Cambridge, MA, 02138, USA.} 
 
\begin{abstract}
The ionizing background determines the ionization balance and the thermodynamics of the cosmic gas.
It is therefore a fundamental ingredient to theoretical and empirical studies of both the IGM and galaxy formation.
We present here a new calculation of its spectrum that satisfies the empirical constraints we recently obtained by combining state-of-the-art luminosity functions and intergalactic opacity measurements.

In our preferred model, star-forming galaxies and quasars each contribute substantially to the HI ionizing field at $z<3$, with galaxies rapidly overtaking quasars at higher redshifts as quasars become rarer.
In addition to our fiducial model, we explore the physical dependences of the calculated background and clarify how recombination emission contributes to the ionization rates.
We find that recombinations do not simply boost the ionization rates by the number of reemitted ionizing photons as many of these rapidly redshift below the ionization edges and have a distribution of energies.
A simple analytic model that captures the main effects seen in our numerical radiative transfer calculations is given.

Finally, we discuss the effects of HeII reionization by quasars on both the spectrum of the ionizing background and on the thermal history of the IGM.
In regions that have yet to be reionized, the spectrum is expected to be almost completely suppressed immediately above 54.4 eV while a background of higher-energy ($\gtrsim0.5$ keV) photons permeates the entire universe owing to the frequency-dependence of the photoionization cross section.
We provide an analytical model of the heat input during HeII reionization and its effects on the temperature-density relation.
\end{abstract}

\section{INTRODUCTION}
\label{introduction}
The cosmic baryons give the ultraviolet (UV) background a particularly important standing among radiation backgrounds.
In fact, the ionization potentials of both hydrogen and helium\footnote{13.6, 24.6, and 54.6 eV for HI, HeI, and HeII, respectively.}, which together account for 99\% of the baryonic mass density \citep[e.g.,][]{2001ApJ...552L...1B}, correspond to electromagnetic wavelengths in the UV regime.
The UV background therefore governs the ionization state of intergalactic gas and furthermore plays a key role in its thermal evolution through photoheating.
As such, it is an essential input to cosmological hydrodynamic simulations \citep[e.g.,][]{1996ApJ...457L..51H, 1996ApJ...457L..57K, 1999ApJ...511..521D, 2003MNRAS.339..312S} as well as to observational studies of the intergalactic medium (IGM).\\ \\
The ionizing background can for example suppress the abundance of dwarf galaxies and the amount of cool gas in low-mass galaxies that do form both by modifying the cooling function through the ionization balance and by heating the gas before it collapses \citep[][]{1992MNRAS.256P..43E, 1996MNRAS.278L..49Q,1996ApJ...465..608T, 1997ApJ...477....8W}.
It is also crucially important for any simulation of the \Lya~forest, since the absence of a \cite{1965ApJ...142.1633G} trough in the spectra of quasars up to $z\sim6$ \citep[e.g.,][]{2002AJ....123.1247F, 2006AJ....132..117F, 2007ApJ...662...72B} indicates that the IGM is highly ionized up to at least that redshift.
Since the optical depth of the \Lya~forest is directly tied to the hydrogen photoionization rate \citep[e.g.,][]{1997ApJ...489....7R, 2001ApJ...549L..11M, 2003MNRAS.342.1205M, 2004ApJ...617....1T, 2005MNRAS.360.1373K, 2005MNRAS.357.1178B, 2008ApJ...682L...9F, 2008ApJ...688...85F}, it is important to know the latter accurately.
In addition, the UV background determines the photoionization rate of helium, which is of particular relevance given the growing interest in studying HeII reionization, which may occur at redshifts $z\sim3-4$ for which a wealth of observational data are already available and upcoming (\S \ref{reionization}).
The full spectrum of the UV background is perhaps most important in the study of metal ions, such as SIV and CIV, where relating the ionic abundances to elemental abundances or cosmic metal mass density requires ionizing corrections \citep[e.g.,][]{1995AJ....109.1522C, 1996AJ....112..335S, 2001ApJ...561L.153S, 2003ApJ...596..768S, 2003astro.ph..7557B, 2004ApJ...602...38A, 2004ApJ...606...92S, 2007arXiv0712.1239A, 2008arXiv0807.4638F}.
Finally, the spectrum of the UV background obviously depends on its sources and its study can therefore teach much about the sources responsible for keeping the IGM ionized, as well as reionizing it \citep[e.g.,][]{2003ApJ...597...66M, 2007MNRAS.382..325B, 2008ApJ...682L...9F, 2008ApJ...688...85F}.\\ \\
Following early work \citep[][]{1990ApJ...350....1M, 1994ApJ...427...25S, 1996ApJS..102..191G}, \cite{1996ApJ...461...20H} \citep[see also][]{1998AJ....115.2206F} pioneered calculations of the UV background spectrum in their study of radiative transfer in a clumpy universe.
Their model and some variants \citep[e.g.,][]{2001cghr.confE..64H} have since been extensively used in several hundreds of studies in the literature.
Over a decade after their original calculation, the empirical constraints on the UV backgrounds and its sources have however improved dramatically.
Larger and deeper surveys at all wavelengths have constrained the quasar luminosity function to both fainter magnitudes and higher redshifts \citep[e.g.,][]{2000MNRAS.317.1014B, 2000A&A...353...25M, 2001AJ....121...54F, 2003ApJ...598..886U, 2004AJ....128..515F, 2004MNRAS.349.1397C, 2005MNRAS.360..839R, 2005AJ....129..578B, 2005A&A...441..417H, 2006AJ....131.2766R, 2006ApJ...638...88B, 2007ApJ...654..731H, 2007A&A...461...39F, 2008ApJ...675...49S}.
At the same time, our understanding of the population of high-redshift star-forming galaxies has tremendously expanded thanks to the application of the Lyman break selection technique to ever more ambitious surveys \citep[e.g.,][]{1999ApJ...519....1S, 2006ApJ...642..653S, 2006ApJ...653..988Y, 2007ApJ...670..928B, 2008ApJS..175...48R}.
Detailed studies of the absorption properties of the IGM, particularly by HI and HeII, have also provided particularly valuable constraints on the UV background.
These constraints are especially relevant for the UV background as the IGM is sensitive to the integral of the UV photons emitted by all sources, regardless of whether these are directly detected.
Moreover, the IGM constraints probe the density of ionizing photons and thus circumvent the need to assume an escape fraction relating the luminosity of quasars and galaxies measured redward of the Lyman limit to their net output of ionizing radiation.\\ \\
In a series of previous papers, we have measured the intergalactic \Lya~opacity \citep[][]{2008ApJ...681..831F} and derived empirical constraints on the UV background and its sources incorporating also information on the reionization of HI and HeII, as well as $N_{\rm HeII}/N_{\rm HI}$ column density ratios \citep[][]{2008ApJ...682L...9F, 2008ApJ...688...85F}.
Specifically, we found that the HI photoionization rate is remarkably constant over the redshift interval $z=2-4$.
Since the quasar luminosity function peaks strongly around $z=2$, star-forming galaxies most likely dominate the ionizing background beyond $z\approx3$.
The column density ratios however indicate that quasars likely do contribute a large fraction of the ionizing background at their peak.
In this paper, we use these constraints as a basis for a new calculation of the full spectrum of the UV background.
In addition to the improved empirical input, we reexamine many of the assumptions entering the original \cite{1996ApJ...461...20H} calculation.
As we will show, we find that the original calculation likely overestimated the contribution of recombination emission to the ionizing background by a factor of a few.
Of perhaps greatest interest, the original calculation completely neglected the effects of HeII reionization, so that simulators have usually resorted to artificial prescriptions to complement the \cite{1996ApJ...461...20H} spectrum.
Here, we explicitly discuss the effects of HeII reionization on the UV background spectrum as well as on the thermal history of the IGM and provide a physical framework to implement them.\\ \\
We review the basic equations of cosmological radiative transfer and the column density distribution of HI absorbers in \S \ref{cosmological radiative transfer}.
In \S \ref{individual absorbers}, we study the ionization structure of individual absorbers and derive approximations to be used in the cosmological solution.
\S \ref{recombinations} is devoted to the calculation of the contribution recombinations to the cosmological emissivity.
Empirically calibrated calculations of the UV background spectrum, derived quantities, and their dependences on input parameters are presented in \S \ref{calculations}.
In \S \ref{dependences}, we investigate how the calculated spectra (including recombination emission) and the corresponding ionization rates depend on input parameters. 
The effects of HeII reionization are investigated in \S \ref{reionization}.
We finally compare our results with previous work in \S \ref{comparison} and conclude in \S \ref{discussion}.\\ \\
A series of appendices supplement the main text with technical details.
In Appendix \ref{photoionization appendix}, we describe our photoionization code.
Appendix \ref{recombination emission appendix} contains technical aspects of our treatment of recombination emission, while Appendix \ref{recombinations analytical model} presents an analytic model of how this recombination emission boosts the photoionization rates.
Appendix \ref{spectral filtering} analytically discusses spectral filtering in different regimes to aid in interpreting our results.
Appendix \ref{atomic physics values} finally references atomic physics quantities used in our calculations.\\ \\
Throughout, we assume a cosmology with $(\Omega_{\rm m},~\Omega_{\rm b},~\Omega_{\Lambda},~h,~\sigma_{8})=(0.28,~0.046,~0.72,~0.70,~0.82)$, as inferred from the \emph{Wilkinson Microwave Anisotropy Probe} (WMAP) five-year data in combination with baryon acoustic oscillations and supernovae \citep[][]{2008arXiv0803.0547K}.
Unless otherwise stated, all error bars are $1\sigma$.
Table \ref{definitions} defines many symbols used here.
 
\begin{deluxetable}{lccl}
\tablewidth{0pc}
\tablecaption{Symbols used in this work\label{definitions}}
\tabletypesize{\footnotesize}
\tablehead{\colhead{Symbol} & \colhead{Definition}}
\startdata
$n_{i}$                   & number density of species $i$ \\
$N_{i}$                   & column density of species $i$ \\
$\tau_{\nu}$              & optical depth at frequency $\nu$ \\
$\bar{\tau}$   & effective optical depth \\
$I_{\nu}$                 & specific intensity along a ray \\
$J_{\nu}$                 & angle-averaged specific intensity \\
$\alpha_{\nu}$\tablenotemark{a}            & absorption coefficient \\
$j_{\nu}$                 & emission coefficient \\
$T$                       & gas temperature \\
$\sigma_{i}$         & photoionization cross section of species $i$ \\
$\Gamma_{i}$              & photoionization rate of species $i$ \\
$\alpha_{i}^{A,B}$        & case A or B recombination coefficient to species $i$ \\
$\alpha_{i,n=j}$          & recombination coefficient directly to level $n=j$ of species $i$ \\
$\phi(\nu)$               & line profile \\
$X$                       & mass fraction of hydrogen \\
$Y$                       & mass fraction of helium \\
$x_{\rm II}$              & fraction of hydrogen in HII \\
$y_{\rm II},~y_{\rm III}$ & fractions of helium in HeII, HeIII \\
\enddata
\tablenotetext{a}{The symbol $\alpha$ is also sometimes used as a recombination coefficient or a spectral index.
The meaning should be transparent from the context.}
\end{deluxetable} 
 
\section{COSMOLOGICAL RADIATIVE TRANSFER}
\label{cosmological radiative transfer}
\subsection{Radiative Transfer Equations}
\label{radiative transfer equations}
In this work, we are first concerned with the specific intensity of the diffuse cosmological UV background averaged over both space and angle, which we denote by $J_{\nu}$.
The basic equations of cosmological radiative transfer were particularly well summarized by \cite{1996ApJ...461...20H}, on which we base our treatment below. 
The specific intensity satisfies the radiative transfer equation,
\begin{equation}
\label{transfer equation}
\left(
\frac{\partial}{\partial t} - \nu H \frac{\partial}{\partial \nu}
\right)
J_{\nu}
=
-3 H J_{\nu}
-c \alpha_{\nu} J_{\nu}
+ \frac{c}{4\pi}\epsilon_{\nu},
\end{equation}
where $H(t)$ is the Hubble parameter, $c$ is the speed of light, $\alpha_{\nu}$ is the proper absorption coefficient, and $\epsilon_{\nu}$ is the proper emissivity.
Integrating equation (\ref{transfer equation}) and expressing the result in terms of redshift gives
\begin{equation}
\label{transfer equation solution}
J_{\nu_{0}}(z_{0})
=\frac{1}{4 \pi}
\int_{z_{0}}^{\infty}
dz
\frac{dl}{dz}
\frac{(1+z_{0})^{3}}{(1+z)^{3}} \epsilon_{\nu}(z) \exp[-\bar{\tau}(\nu_0, z_{0}, z)],
\end{equation}
where $\nu=\nu_{0}(1+z)/(1+z_{0})$, the proper line element $dl/dz=c/[(1+z)H(z)]$, and the ``effective optical depth'' $\bar{\tau}$\footnote{This quantity is often denoted by $\tau_{\rm eff}$. We use a different notation here to distinguish it from the effective optical depth $\tau_{\rm eff}(z)=-\ln{\langle F \rangle(z)}$ owing to \Lya~absorption measured from quasar spectra \citep[e.g.,][]{2008ApJ...681..831F}.} quantifies the attenuation of photons of frequency $\nu_{0}$ at redshift $z_{0}$ that were emitted at redshift $z$ by the relation $e^{\bar{\tau}}=\langle e^{-\tau} \rangle$, where the average is over all lines of sights from $z_{0}$ to $z$.
For Poisson-distributed absorbers, each of column density $N_{\rm HI}$,
\begin{equation}
\label{taueff poisson expression}
\bar{\tau}(\nu_{0}, z_{0}, z) =
\int_{z_{0}}^{z} dz'
\int_{0}^{\infty}
dN_{\rm HI}
\frac{\partial^{2}N}{\partial N_{\rm HI} \partial z'}
(1 - e^{-\tau_{\nu}}),
\end{equation}
where $\partial^{2}N/\partial N_{\rm HI} \partial z'$ is the column density distribution versus redshift \citep{1980ApJ...240..387P}.\\ \\
Note that these expressions neglect the clustering of sources and sinks of radiation in both the Poisson distribution assumption and in assuming that the spatial average $\langle \epsilon_{\nu}(z) \exp[-\tau(\nu_0, z_{0}, z)] \rangle$ separates into $\langle \epsilon_{\nu}(z) \rangle \langle \exp[-\tau(\nu_0, z_{0}, z)] \rangle$ in the integrand of equation \ref{transfer equation solution}.
The Poisson distribution assumption should be very good since the mean free path of ionizing photons of  hundreds of comoving Mpc at most redshifts of interest (see \S \ref{reionization}) far exceeds the correlation length $\lesssim5$ comoving Mpc of the \Lya~forest absorbers \citep[e.g.,][]{2000ApJ...543....1M, 2008ApJ...673...39F}.
While absorbers likely do cluster around sources, this effect can be viewed as being incorporated into the definition of the escape fraction.
The above formalism will however break down in certain regimes where sources are rare and in particular during HeII reionization.
We discuss these cases in \S \ref{reionization}.
As in equation \ref{transfer equation solution}, we henceforth drop the explicit averaging brackets around the emissivity $\epsilon_{\nu}$.\\ \\ 
The optical depth $\tau_{\nu}$ shortward of the Lyman limit will be dominated by the photoelectric opacity of hydrogen and helium,
\begin{equation}
\tau_{\nu} = N_{\rm HI}\sigma_{\rm HI}(\nu) + N_{\rm HeI}\sigma_{\rm HeI}(\nu) + N_{\rm HeII}\sigma_{\rm HeII}(\nu),
\end{equation}
where the $N_{i}$ and $\sigma_{i}$ are the column densities and photoionization cross sections of ion $i$.
Only the distribution of $N_{\rm HI}$ is reasonably well determined over a large redshift interval.
We will therefore make use of relations between $N_{\rm HI}$ and the column densities of helium established in \S \ref{individual absorbers}.
In our calculations, we will prescribe $\partial^{2}N/\partial N_{\rm HI} \partial z$ as well as the specific emissivity of the ionizing sources, $\epsilon_{\nu}^{\rm src}$, based on our previous empirical studies \citep{2008ApJ...681..831F, 2008ApJ...682L...9F, 2008ApJ...688...85F}. 

\subsection{HI Column Density Distribution}
\label{column density distribution}
Following previous work and consistent with empirical constraints, we parameterize the column density distribution with power laws in $N_{\rm HI}$ and $z$:
\begin{equation}
\label{column density distribution eq}
\frac{\partial^{2}N}{\partial z \partial N_{\rm HI}}=
\left\{
\begin{array}{ll}
N_{0,{\rm low}}N_{\rm HI}^{-\beta}(1+z)^{\gamma_{\rm low}} & z\leq z_{\rm low} \\
N_{0}N_{\rm HI}^{-\beta}(1+z)^{\gamma} & z>z_{\rm low}
\end{array}
\right.
.
\end{equation}
The transition at $z_{\rm low}$ accounts for the flattening of the redshift evolution observed at $z\lesssim1.5$ \citep[e.g.,][]{1998ApJ...506....1W, 2001A&A...373..757K, 2002MNRAS.335..555K} theoretically understood to arise from the drop in intensity of the ionizing background at low redshifts \citep[][]{1998MNRAS.297L..49T, 1999ApJ...511..521D, 2001A&A...376....1B, 2002ApJ...571..665S}.
We fix $\gamma_{\rm low}=0.2$.\\ \\ 
For a steep column density distribution with $\beta<2$, most of the contribution to $\bar{\tau}$ at the Lyman limit arises from systems of optical depth near unity ($N_{\rm HI}\approx10^{17.2}$ cm$^{-2}$).
We thus focus on the values of the power-law indices $\beta$ and $\gamma$ that are most appropriate in this neighborhood.
\cite{1995ApJ...444...64S} find that $dN/dz=C(1+z)^{\gamma}$ with $C=0.25$ and $\gamma=1.5$ provides a good fit at least up to $z=4.1$ for systems with $N_{\rm HI}\geq10^{17.2}$ cm$^{-2}$, whereas the column density power law is well-fitted by $\beta=1.4$
\citep[][]{2007AJ....134.1634M}.
The constant $N_{0}$ in equation \ref{column density distribution eq} is related to $C$ by $N_{0}=(\beta-1)C N_{\rm HI,min}^{\beta-1}$, where $N_{\rm HI,min}=\sigma_{\rm HI}^{-1}=10^{17.2}$ cm$^{-2}$.
We use these values, with $z_{\rm low}=1.5$, in fiducial calculations but explore varying these parameters in \S \ref{dependences}.\\ \\
Before proceeding, we note that column density distribution is largely unconstrained above $z=4.1$, the highest redshift at which the Lyman limit system abundance has been measured.
We therefore simply extrapolate from lower redshifts and caution that our calculation of the ionizing background may become inaccurate in this regime.
In particular, we expect the extrapolation to become unreliable at $z\approx5.7$, where the evolution of the effective optical depth measured from the spectra of $z\geq6$ quasars diverges rapidly from the power law fitting the data below this redshift \citep[][though see Becker et al. (2007)\nocite{2007ApJ...662...72B} for an opposing point of view]{2006AJ....132..117F}, perhaps owing to HI reionization.

\section{THE IONIZATION STRUCTURE OF INDIVIDUAL ABSORBERS}
\label{individual absorbers}

\subsection{Overview}
\label{photoionization overview}
In this section, we study the photoionization equilibrium structure of individual cosmic absorbers composed of hydrogen and helium (with mass fraction 75\% and 25\%, respectively) as a function of the illuminating radiation background.
This serves two purposes: close the set of equations of cosmological radiative transfer (\S \ref{cosmological radiative transfer}) and allow us to more realistically calculate the contribution of recombination lines to the ionizing background spectrum (\S \ref{recombinations}).
To this end, we have developed a code that self-consistently solves the photoionization equilibrium balance, including the influence of recombination radiation.
This code provides more accurate solutions than previous approximations with semi-infinite geometry and an escape probability formalism \citep[][]{1996ApJ...461...20H} or gray cross sections \citep[][]{1998AJ....115.2206F}.
To alleviate the text, the details of our photoionization calculations are provided in Appendix \ref{photoionization appendix}.
We assume our absorbers to be slabs of thickness equal to the Jeans scale of the gas, which is a function of the assumed temperature $T=2\times10^{4}$ K \cite[][]{2001ApJ...559..507S}.
This temperature is consistent with the line-fitting analysis of \cite{2001ApJ...562...52M} and with the \Lya~forest power spectrum analysis of \cite{2001ApJ...557..519Z} for the $z\sim2-4$ IGM at mean density.

\subsection{$N_{\rm HeII}$ and $N_{\rm HeI}$ from $N_{\rm HI}$}
\label{column density relations}
\begin{figure}[ht] 
\begin{center} 
\includegraphics[width=1.0\textwidth]{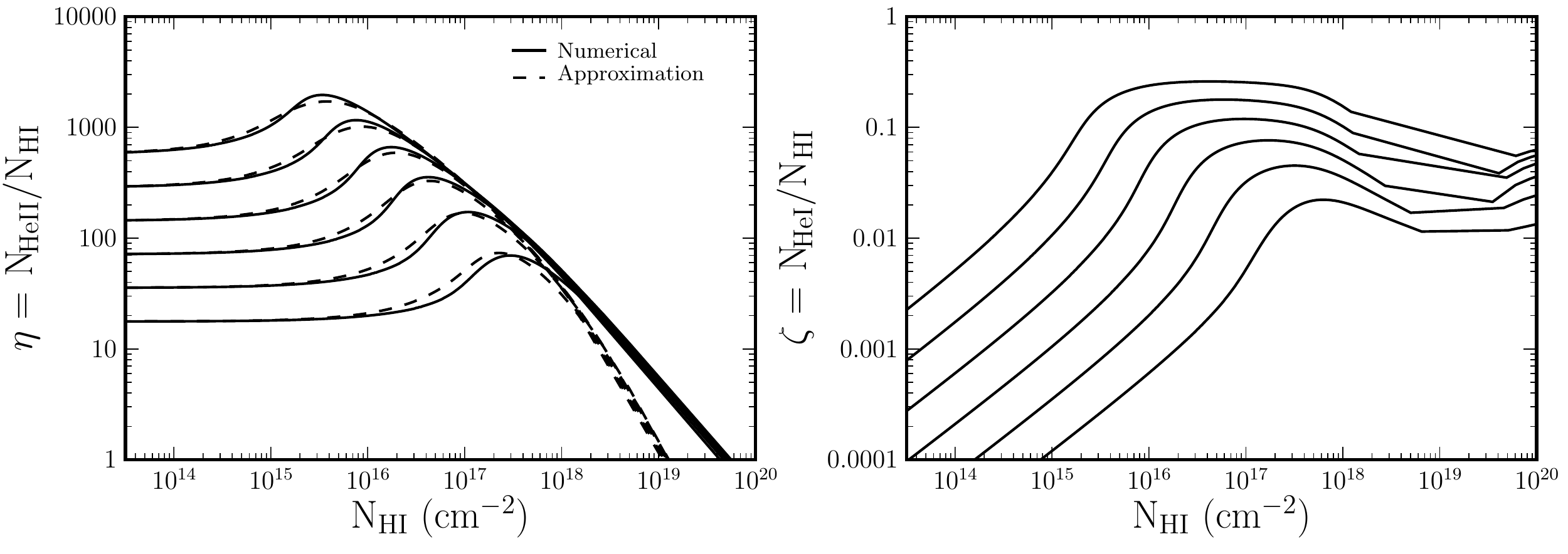} 
\end{center} 
\caption{Ratio $\eta=N_{\rm HeII}/N_{\rm HI}$ (left) and $\zeta=N_{\rm HeI}/N_{\rm HI}$ (right) as a function of HI column density.
The solid curves show full numerical photoionization calculations and the dashed ones show analytical approximations based on equation \ref{fardal fitting formula} for $\eta$.
From the bottom up, power-law external spectra $J_{\nu}=10^{-21} {\rm~erg~s^{-1}~cm^{-2}~Hz^{-1}~sr^{-1}} (\nu/\nu_{\rm HI})^{-\alpha}$ with $\alpha=0.0,~0.5,~1.0,~1.5,~2.0,{\rm~and~}2.5$, suppressed by a factor of 10 above the HeII ionization edge, are assumed.} 
\label{eta vs nhi comp} 
\end{figure}

As only the column density distribution of HI is reasonably well constrained, the first application of our photoionization calculations is to obtain relations giving $N_{\rm HeII}$ and $N_{\rm HeI}$ in terms of $N_{\rm HI}$.
In Figure \ref{eta vs nhi comp}, we show the numerical results for external spectra $J_{\nu}=10^{-21} {\rm~erg~s^{-1}~cm^{-2}~Hz^{-1}~sr^{-1}} (\nu/\nu_{\rm HI})^{-\alpha}$ with $\alpha=0.0,~0.5,~1.0,~1.5,~2.0,{\rm~and~}2.5$ from bottom up.
To ensure that these test spectra are representative of the UV background, we suppress the power laws by a factor of 10 above the HeII ionization edge.\\ \\
Since these relations enter within three nested integrals (eqs. \ref{transfer equation} and \ref{transfer equation solution}), it is necessary to develop analytical approximations that are fast to evaluate.
It would be impractical to use self-consistent numerical photoionization calculations at each redshift and for each column density in the cosmological solution.
Defining $\eta \equiv N_{\rm HeII}/N_{\rm HI}$, when both HI and HeII are optically thin and in the limit of nearly complete ionization we have
\begin{equation}
\label{eta optically thin}
\eta_{\rm thin} = 
\frac{\Gamma_{\rm HI}}{\Gamma_{\rm HeII}}
\frac{\alpha_{\rm HeII}^{\rm A}}{\alpha_{\rm HI}^{\rm A}}
\frac{Y}{4X}.
\end{equation}
For fixed external background and increasing $N_{\rm HI}$, an absorber first becomes optically thick in HeII, at which point $\eta$ increases rapidly with $N_{\rm HI}$.
The absorber then becomes optically thick in HI as well and, owing to the greater abundance of hydrogen, $N_{\rm HI}$ finally rapidly overtakes $N_{\rm HeII}$.
This leads to the plateau, increase, and then decrease of $\eta$ with respect to $N_{\rm HI}$ seen in the numerical calculations.
Similar behavior is found in three-dimensional radiative transfer simulations of the IGM \citep[][]{2005MNRAS.364.1429M}.\\ \\
\cite{1998AJ....115.2206F} give a fitting formula derived under the assumptions of negligible $N_{\rm HeI}$ and $n_{\rm HI}/n_{\rm H}\ll 1$:
\begin{equation}
\label{fardal fitting formula}
\frac{Y}{16X}
\frac{\tau_{\rm HI}}{1+A\tau_{\rm HI}}
I_{\rm HI}
=
\tau_{\rm HeII}
+
\frac{\tau_{\rm HeII}}{1+B\tau_{\rm HeII}}
I_{\rm HeII},
\end{equation}
where $I_{\rm i}\equiv \Gamma_{\rm i}^{\rm ext}/n_{e} \alpha_{\rm i}^{\rm A}$, $\tau_{\rm HI}=\sigma_{\rm HI}N_{\rm HI}$ and $\tau_{\rm HeII}=\sigma_{\rm HeII}N_{\rm HeII}$, and we have generalized their result to allow for arbitrary coefficients $A$ and $B$.
These fitting coefficients depends on, in particular, the relation between $N_{\rm HI}$ and $n_{e}$ and we will not be using the same model as these authors.
Although our numerical calculations do not a priori assume a relation between $N_{\rm HI}$ and $n_{e}$, we must assume one in order to make use of the analytic approximation in equation \ref{fardal fitting formula}.
The approximation curves in Figure \ref{eta vs nhi comp} can be reproduced by taking $n_{e}=1.4\times10^{-3}$ 
cm$^{-3}$ $(N_{\rm HI}/10^{17.2}~$ cm$^{-2})^{2/3} (\Gamma_{\rm HI}/10^{-12}~$s$^{-1})^{2/3}$.
This relation is approximately derived under the assumptions of Jeans length thickness of the absorbers and optically thin photoionization equilibrium at $T=2\times10^{4}$ K.
Figure \ref{eta vs nhi comp} shows that $A=0.15$ and $B=0.2$ give a good fit to our numerical results for a wide range of external illuminating spectra.
The fitting formula has the exact optically thin limit; the asymptotic divergence from the numerical results as $N_{\rm HI}\rightarrow\infty$ is unimportant as most of the HeII opacity arises in systems with $\tau_{\rm HI}\lesssim1$.
Although we have assumed specific (but varying) spectral shapes in determining the fitting parameters $A$ and $B$, the \cite{1998AJ....115.2206F} derivation of the functional form in equation \ref{fardal fitting formula} illustrates how the relation between $N_{\rm HeII}$ and $N_{\rm HI}$ depends principally on the photoionization rates $\Gamma_{\rm HI}$ and $\Gamma_{\rm HeII}$.
The relation should therefore hold well in general.
\\ \\
Obtaining a physically-motivated analytic approximation to $\zeta \equiv N_{\rm HeI}/N_{\rm HI}$ is more difficult since $y_{\rm HeII}$ is not readily known (for $\eta$, we know that $x_{\rm HII}\approx y_{\rm III}\approx1$ in the almost-completely ionized case, so that the ionized fractions do not appear explicitly in eqs \ref{eta optically thin} and \ref{fardal fitting formula}).
Because the ionization potential of HeI is relatively close to that of HI, their ionization states are similar and since helium is less abundant by a factor of 12 by number, HeI should contribute relatively little to the ionizing opacity.
This intuition is supported by the right panel of Figure \ref{eta vs nhi comp}, which shows that $\zeta$ versus $N_{\rm HI}$ is $\ll1$ for illuminating spectra considered.
After hardening by IGM filtering, both star-forming galaxies and quasars are expected to produce roughly flat spectra between the HI and HeI ionization edges (\S \ref{calculations}), so that a representative case is $\alpha\approx0$, yielding $\zeta\lesssim10^{-3}$. 
We therefore approximate $\zeta=0$ in our cosmological calculations and verify using toy cases of constant $\zeta$ that this is justified in \S \ref{dependences}.\\ \\

\section{RECOMBINATION EMISSION}
\label{recombinations}

\subsection{Cosmological Emissivity}
The cosmic absorbers not only act as sinks but also as sources of ionizing radiation as a certain fraction of ionizations are followed by the reemission of other ionizing photons via recombinations \citep[][]{1993ApJ...415..524F, 1996ApJ...461...20H, 1998AJ....115.2206F}.
This recombination emission must be taken into account because it may boost the photoionization rates and also since line reemission can imprint significant narrow features in the ionizing background spectrum.
Our approach to include this recombination contribution is based on the self-consistent numerical calculations of recombination emission from individual absorbers using the code outlined in the previous section and detailed in Appendix \ref{photoionization appendix}.
This again differs from \cite{1996ApJ...461...20H}, who used an analytical escape probability formalism and made the assumption of a constant source function within the absorbers (which breaks down in the very optically thick systems), and from the treatment of \cite{1998AJ....115.2206F} and thus provides a check of these results.\\ \\
For each recombination process of interest, we calculate the emergent specific intensity $I_{\nu}^{\rm rec}(N_{\rm HI})$ owing to the process given the external illuminating spectrum numerically using our photoionization code.
The cosmological recombination emissivity for this process is then an average over the column density distribution:
\begin{equation}
\label{epsilon nu rec}
\epsilon_{\nu}^{\rm rec} = 
4\pi
\frac{dz}{dl}
\int_{0}^{\infty}
dN_{\rm HI}
\frac{\partial^{2}N}{\partial z \partial N_{\rm HI}}
I_{\nu}^{\rm rec}(N_{\rm HI}).
\end{equation}
Since in general $I_{\nu}^{\rm rec}(N_{\rm HI})$ depends on the spectrum of the ionizing background, which is not known \emph{a priori} and evolves at each step in the redshift integration, it is again necessary to obtain an analytical approximation for this function that scales appropriately with the external background, as it is not practical to perform self-consistent numerical calculations in the cosmological solution.
We develop these analytical approximations in Appendix \ref{recombination emission appendix} for each recombination process of interest.

\begin{figure}[ht] 
\begin{center} 
\includegraphics[width=1.0\textwidth]{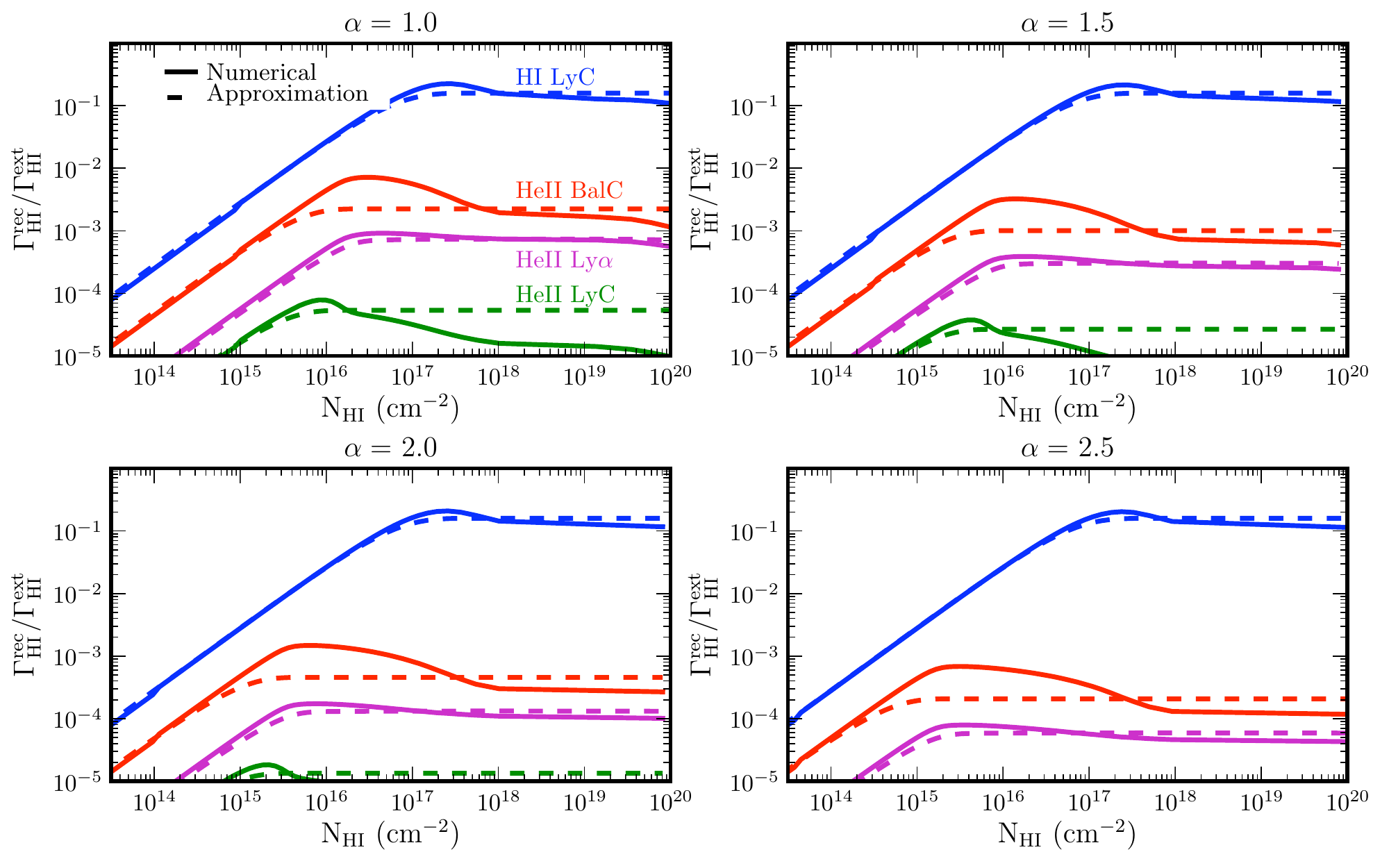} 
\end{center} 
\caption{Ratio of the HI photoionization rate outside an individual absorber that is contributed by different recombination processes (HI LyC in blue, HeII BalC in red, HeII Ly$\alpha$ in magenta, HeII LyC in green) to the external photoionization rate.
The solid curves show full numerical integrations over photo-ionized slabs (eq. \ref{recombination transfer solution}) and the dashed ones show analytical approximations based on the optically thin limit (eq. \ref{irec optically thin}) and saturation in the optically thick regime described in \S \ref{recombinations analytic approximations}.
External power-law spectra $J_{\nu}^{\infty}=10^{-21} {\rm~erg~s^{-1}~cm^{-2}~Hz^{-1}~sr^{-1}} (\nu/\nu_{\rm HI})^{-\alpha}$ with $\alpha=1.0,~1.5,~2.0,{\rm~and~}2.5$, suppressed by a factor of 10 above the HeII ionization edge, are assumed in the different panels.} 
\label{gammarec vs nhi} 
\end{figure}

\begin{figure}[ht] 
\begin{center} 
\includegraphics[width=0.5\textwidth]{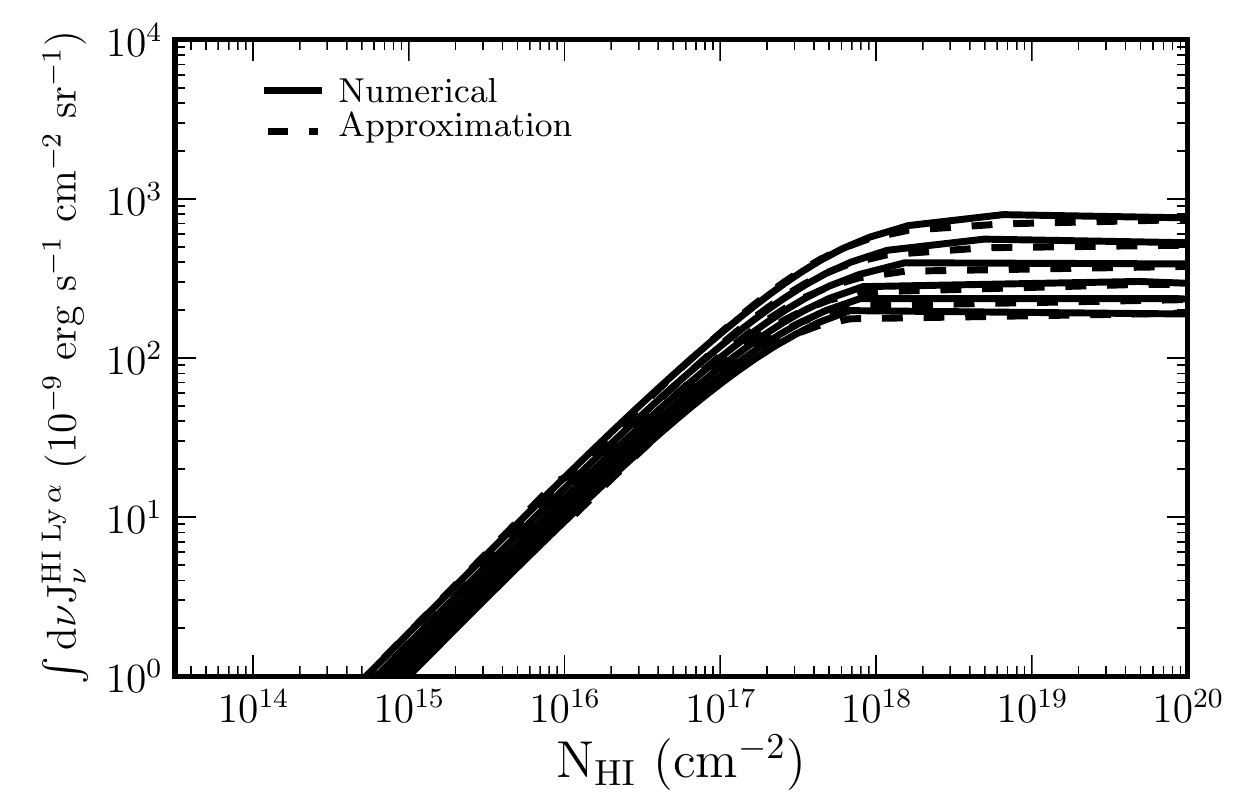} 
\end{center} 
\caption{Frequency integral of HI Ly$\alpha$ reemission as function of HI column density.
The solid curves show full numerical photoionization calculations and the dashed ones show analytical approximations based on equation \ref{Lya reemission approximation}.
From the bottom up, power-law external spectra $J_{\nu}^{\infty}=10^{-21} {\rm~erg~s^{-1}~cm^{-2}~Hz^{-1}~sr^{-1}} (\nu/\nu_{\rm HI})^{-\alpha}$ with $\alpha=0.0,~0.5,~1.0,~1.5,~2.0,{\rm~and~}2.5$, suppressed by a factor of 10 above the HeII ionization edge, are assumed.} 
\label{HI Lya reem} 
\end{figure}

\subsection{Recombination Processes}
\label{recombination processes}
For hydrogen, the only ionizing process is direct recombination to the ground state, which produces a 1 Ryd HI LyC photon.
For helium, both recombinations to HeII and to HeI can in principle produce ionizing photons.
In cosmological conditions, HeI plays a negligible role \citep[\S \ref{column density relations} and \ref{dependences};][]{1996ApJ...461...20H} and we will ignore it in our reemission calculations.
Three permitted HeII recombination channels lead to the reemission of ionizing photons: HeII LyC recombinations directly to the ground state, indirect recombinations leading to HeII Ly$\alpha$ emission, and recombinations to the $n=2$ excited level resulting in Balmer continuum (BalC) photons.
These respectively give photons of energy 4, 3, and 1 Ryd.
Higher HeII Lyman-series photons could also produce ionizing photons, but we assume case B conditions in which these are ultimately degraded into lower-energy photons, the only ones of which that can ionize hydrogen being HeII \Lya.
We do not include forbidden two-photon recombination processes as these are energetically subdominant and do not result in distinctive emission features.\\ \\
In calculating the contribution of reemission to the photoionization rates, it is important to model the finite width of the recombination lines.
If HI LyC reemission is incorrectly modeled as a $\delta-$function, the reemission photons are immediately redshifted below the HI ionization edge and are lost as contributors to the ionizing background.
For continuum recombinations, the line profile is well approximated by
\begin{equation}
\label{recombination line profile text}
\phi_{\rm rec}(\nu) = \frac{(\nu/\nu_{\rm rec})^{-1}\exp{(-h\nu/kT})}{\Gamma(0,~h\nu_{\rm rec}/kT)} \frac{\theta(\nu-\nu_{\rm rec})}{\nu_{\rm rec}},
\end{equation}
where $T$ is the temperature of the gas and $\theta(\Delta \nu)$ is the Heaviside function which is 1 for $\Delta \nu \geq 0$ and 0 otherwise (Appendix \ref{atomic physics values}).
Electrons in gas at higher temperature tend to have large kinetic energy and so give rise to higher-energy recombination photons that take longer to redshift below the ionization edge.
In Appendix \ref{atomic physics values}, we show that broadening owing to thermal and peculiar motion is negligible relative the width of the profile in equation \ref{recombination line profile text}.
For Ly$\alpha$ emission, either by HI or HeII, a $\delta-$function profile $\phi_{\rm HI/HeII~Ly\alpha}(\nu)=\delta(\nu-\nu_{\rm HI/HeII~Ly\alpha})$ is however appropriate because of the narrow intrinsic line width (much smaller than the mean free path by which photons are redshifted before being reabsorbed) and its distance from the ionization edges.
While resonant scattering radiative transfer effects can broaden \Lya~emission lines by $\sim10-1000$ km s$^{-1}$ \citep[e.g.,][]{1990ApJ...350..216N, 2002ApJ...578...33Z, 2006ApJ...649...14D, 2006A&A...460..397V}, this width is negligible in comparison to the cosmological redshift broadening.\\ \\
The analytical approximations for the recombination emission from individual absorbers are compared to the full numerical calculations in Figure \ref{gammarec vs nhi} for the ionizing processes.
In all cases, $\Gamma_{\rm HI}^{\rm rec}/\Gamma_{\rm HI}^{\rm ext}$ (where we define $\Gamma_{\rm HI}^{\rm rec}(N_{\rm HI})\equiv 4 \pi \int_{\nu_{\rm HI}}^{\infty} d\nu/(h \nu) I_{\nu}^{\rm rec}(N_{\rm HI}) \sigma_{\rm HI}(\nu)$) is maximum for HI LyC reemission, as expected since hydrogen recombinations are more frequent owing to its greater abundance and these recombination photons have the largest photoionization cross section, and is equal to about 10\%.
The helium recombination processes all contribute at the $10^{-3}$ level or less.
Note, however, that HeII LyC reemission will contribute more significantly to the HeII ionizing background and that processes which contribute negligibly to the photoionization rates can still imprint important narrow features in the background spectrum that can be important for metal line studies.
The agreement between the numerical calculations and analytical approximations is generally good and the approximations scale well for different spectral indices.
Discrepancies of a factor of a few exist over some column density intervals, particularly for the HeII BalC and HeII LyC processes.
These processes are complex in their details that depend on the non-monotonic relative ionization of hydrogen and helium (Fig. \ref{eta vs nhi comp}) but their contributions are nevertheless reasonably well captured and fortunately subdominant to the photoionization rates.
In contrast, the dominant contribution of reemission to the hydrogen ionizing background, HI LyC emission, involves only hydrogen and is accurately and robustly approximated.\\ \\
Figure \ref{HI Lya reem} compares the analytical approximation for HI \Lya~approximation to the full numerical solutions.
In this case,  both the optically thin and optically thick limits are accurately captured, resulting in an excellent approximation at all column densities that scales correctly with the external illuminating spectrum.\\ \\

\section{EMPIRICALLY CALIBRATED SPECTRA}
\label{calculations}

\subsection{Quasar and Stellar Emissivities}
\label{quasar and stellar emissivities}
Having established efficient approximations for the radiative transfer within individual absorbers (\S \ref{individual absorbers} and \ref{recombinations}), we proceed to include these in the solution of the cosmological radiative transfer problem (\S \ref{cosmological radiative transfer}).
Our prescriptions for the sources of ionizing radiation are based on the empirical constraints obtained in \cite{2008ApJ...681..831F, 2008ApJ...682L...9F, 2008ApJ...688...85F}.
Note, however, that these prescriptions can easily be modified to accommodate further constraints: 
our numerical code can compute the ionizing background for arbitrary input emissivities.
We explore variations about these fiducial parameters in \S \ref{dependences}.
Here, we consider two dominant known sources of ionizing radiation: quasars and star-forming galaxies.\\ \\
For the quasar emissivity, we use the quasar luminosity function of \cite{2007ApJ...654..731H} based on a large set of observed quasar luminosity functions in the infrared, optical, soft and hard X-rays, as well as emission line measurements.
Denoting by $\epsilon_{B}$ the emissivity at 4400~\AA~and assuming $L_{B}\equiv \nu L_{\nu}|_{4400~{\rm \AA}}$,
\begin{equation}
\label{quasar B emissivity}
\epsilon_{B}^{\rm QSO,com} =
\int_{0}^{\infty} dL_{B} 
\frac{d\phi}{dL_{B}}
\frac{L_{B}}{\nu|_{\rm 4400~\AA}},
\end{equation}
where $d\phi/dL_{B}$ is the $B$-band luminosity function in comoving units.
The emissivity shortward of 4400~\AA~is calculated assuming that quasars have a spectral index $\alpha=0.3$ at 2500-4400~\AA, 0.8 at 1050-2500~\AA~\citep[][]{1999ApJ...514..648M}, and $\alpha_{\rm QSO}$ shortward of 1050~\AA.
In order to match the total HI photoionization rate measured from the \Lya~forest and to account for uncertainties in converting from the emissivity at 4400~\AA~to the photoionization rate, we allow this emissivity to be normalized by a constant factor (see \S \ref{spectra results}).
In our fiducial model, $\alpha_{\rm QSO}=1.6$ \citep[][]{2002ApJ...565..773T} but note that other studies have found both softer and harder spectra, with a significant variance about the mean \citep[e.g.,][]{1997ApJ...475..469Z, 2004ApJ...615..135S}.\\ \\
For the stellar emissivity, we assume that the emissivity is proportional to the star formation rate density,
\begin{equation}
\epsilon^{\star {\rm,com}}_{\nu}  =
K \dot{\rho}_{\star}^{\rm com},
\end{equation}
with the observationally-calibrated proportionality constant accounting for the efficiency of conversion of mass into ionizing photons.
We use the theoretical star formation history of \cite{2003MNRAS.341.1253H} developed from a combination of hydrodynamical simulations \citep[][]{2003MNRAS.339..312S} and simple analytical arguments. 
In \cite{2008ApJ...688...85F}, we found that this model provides a better fit at high redshifts to the opacity of the \Lya~forest over $z=2-4.2$, is easier to reconcile with hydrogen reionization completing by $z=6$, and is in better agreement with the rate of long gamma-ray bursts observed by \emph{Swift} than many of the existing measurements based on galaxy surveys, among which there is still a wide dispersion.
We use equation 45 of \cite{2003MNRAS.341.1253H} to scale their fiducial model to the $WMAP5$ cosmology assumed in this work.
We assume, motivated by the theoretical starburst calculations of \cite{2001ApJ...556..121K}, that star-forming galaxies have a spectral index $\alpha_{\star}=1$ between 1 and 4 Ryd.
This model is applicable for the stellar populations calculated with the PEGASE code using the \cite{1987MNRAS.228..759C} atmosphere models for Wolf-Rayet stars.
While different theoretical assumptions lead to significant variance in the 1$-$4 Ryd spectrum, this model  provides the best observational match to the hard starburst spectra inferred by optical line diagnostics by \cite{2001ApJ...556..121K}.
We assume that they effectively emit no harder photons, the theoretical calculations showing a break of several orders of magnitude at the HeII ionization edge.
In the UV spectrum redward of 1 Ryd, we take $\alpha_{\star}=0$, consistent with the LBGs observed by \cite{2003ApJ...588...65S}.
Finally, we assume that the stellar emissivity has a discontinuity of a factor of 4 at the Lyman limit.
While this factor is neither well constrained empirically or observationally, it only affects our predicted spectra (normalized to the measured ionizing background) at energies less than 1 Ryd, which we do not attempt to accurately model in this work.\\ \\
The emissivities are converted to proper units before being inserted in the solution to the cosmological radiative transfer solution in equation \ref{transfer equation solution} and the total emissivity is then $\epsilon_{\nu}(z) = \epsilon_{\nu}^{\rm QSO}(z) + \epsilon_{\nu}^{\star}(z) + \epsilon_{\nu}^{\rm rec}(z)$.

\begin{figure}[ht] 
\begin{center} 
\includegraphics[width=1.0\textwidth]{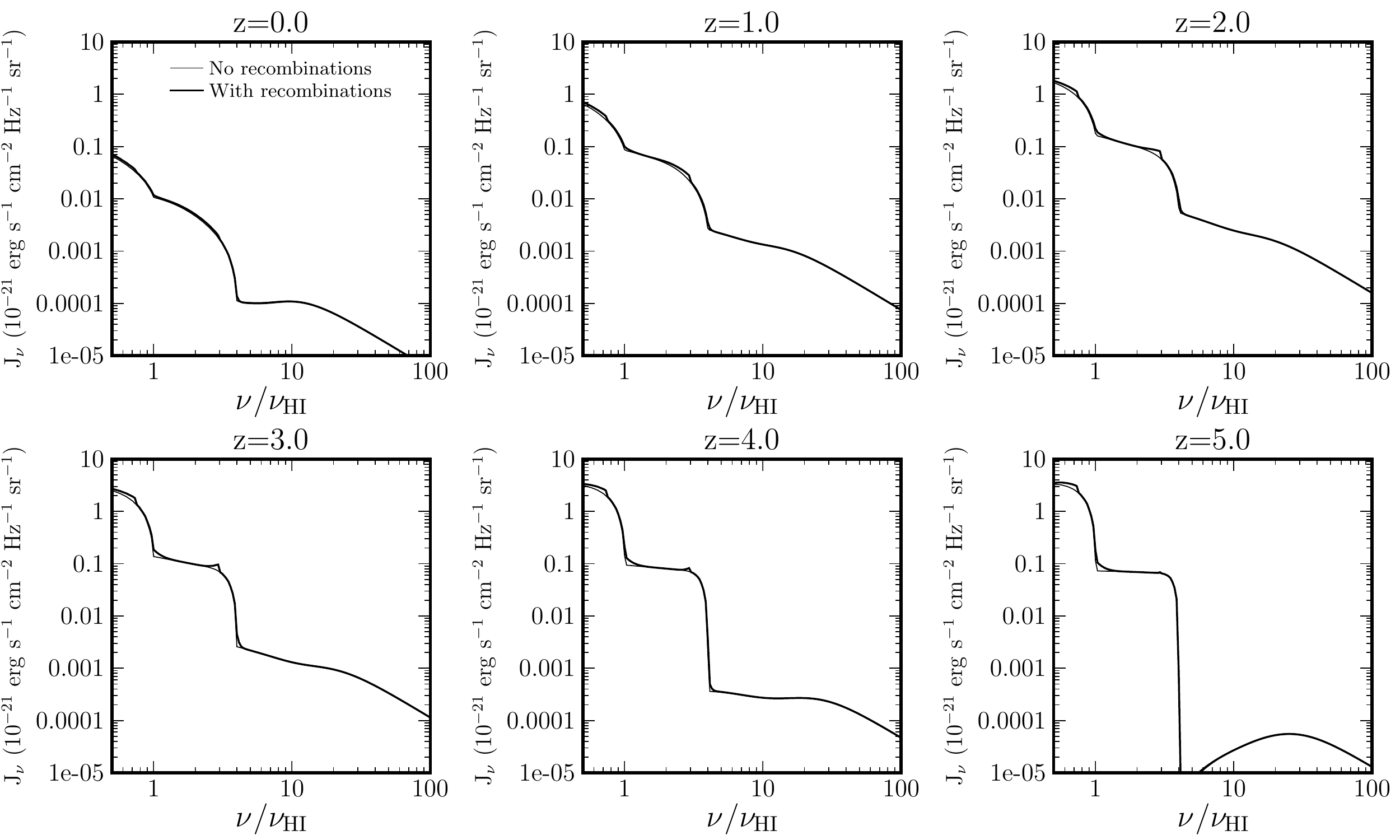} 
\end{center} 
\caption{Spectra of the UV background obtained by solving the cosmological radiative transfer equation (eq. \ref{transfer equation}) at different redshifts for our empirically-calibrated fiducial model with star-forming galaxies and quasars described in \S \ref{calculations}.
The thin curves ignore recombination emission by intergalactic absorbers and the thicker curves include this contribution.
Star-forming galaxies dominate the HI photoionization rate at $z\gtrsim3$, with the quasar contribution becoming more important as the $z\sim2$ peak of the quasar luminosity function is approached.
Only quasars are assumed to produce HeII ionizing photons and only they contribute to HeII recombination processes.
The integrated photoionization rates are given in Figure \ref{gammas vs z}.
} 
\label{spectra} 
\end{figure}

\begin{figure}[ht] 
\begin{center} 
\includegraphics[width=1.0\textwidth]{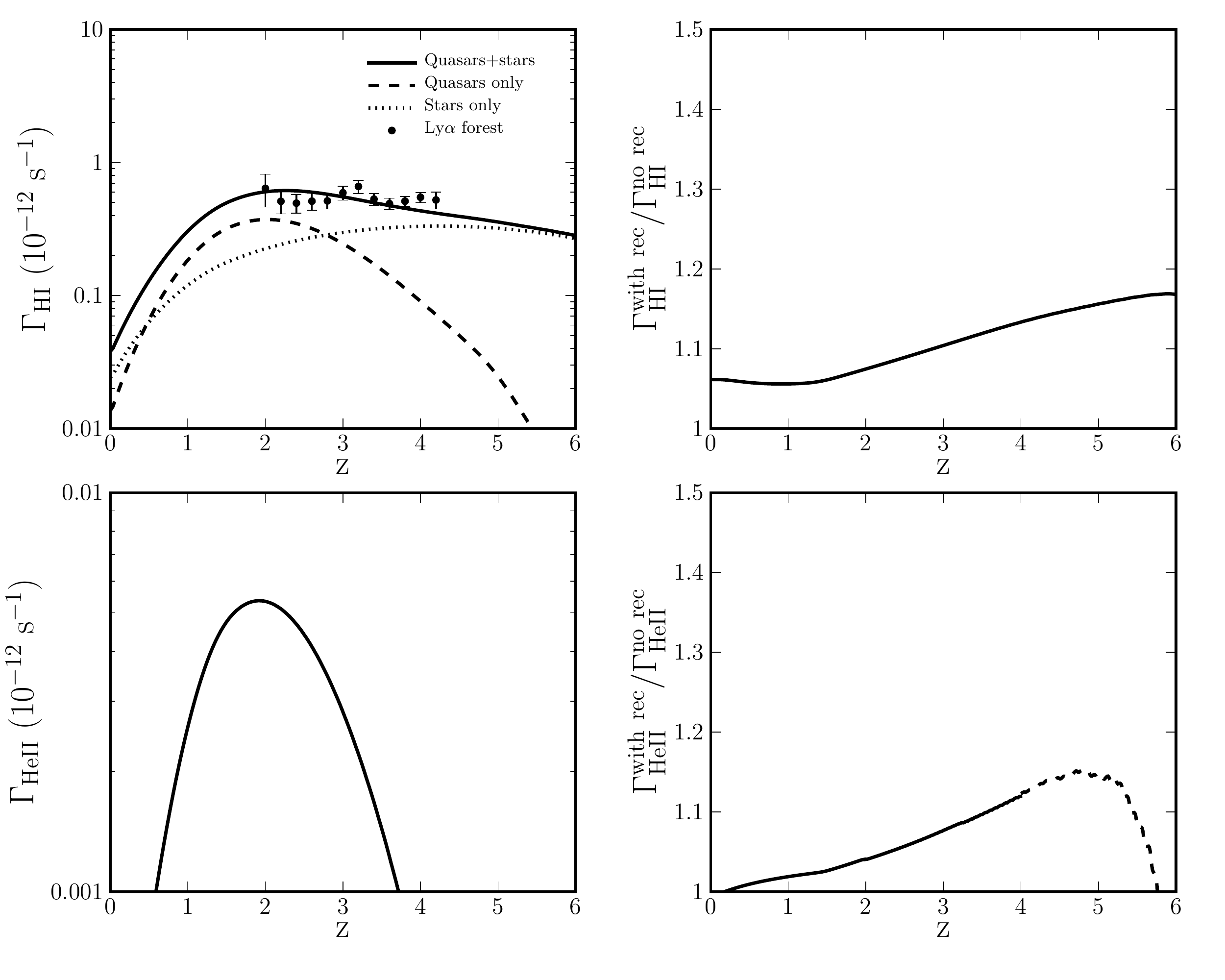} 
\end{center} 
\caption{Photoionization rates for the UV background shown in Figure \ref{spectra}.
\emph{Top left:} HI photoionization rates including the contribution from recombinations.
The total (quasars+stars) photoionization rate is compared to the value inferred from the \Lya~forest forest flux decrement \citep[][]{2008ApJ...682L...9F, 2008ApJ...688...85F}.
\emph{Top right:} ratio of the total HI photoionization rate to the value value obtained neglecting recombination emission.
The bottom panels are analogous but for HeII ionizing radiation.
In this case, quasars are the only contributors.
The dashed portion of the $\Gamma_{\rm HeII}^{\rm with rec}/\Gamma_{\rm HeII}^{\rm no rec}$ curve at $z>4$ indicates a regime of poor numerical convergence (see \S \ref{spectra results}) and the turnover at $z\sim5$ is likely an artifact.
} 
\label{gammas vs z} 
\end{figure}

\subsection{Results}
\label{spectra results}
In Figure \ref{spectra}, we show the calculated cosmological UV background spectra at $z=0-5$ for the fiducial model above, with and without the recombination processes included.
Since only quasars are assumed to contribute photons above the 4 Ryd HeII ionization edge, only them contribute to the HeII recombination lines and photoionization rate.\\ \\
In Figure \ref{gammas vs z}, we show the integrated photoionization rates of HI and HeII, as well as the fractional contribution of recombination lines with respect to the total background including both stars and quasars.
The quasar contribution to the HI ionizing background increases toward $z\sim2$ as the peak of the quasar luminosity function is approached; the $z\gtrsim3$ photoionization rate is dominated by stellar emission.
The fractional recombination contribution to the HI photoionization rate ranges from 5\% to 17\% over the interval $z=0-6$, significantly smaller than the $\alpha_{\rm HI,n=1}(T)/\alpha_{\rm HI}^{\rm A}(T)|_{T=20,000~{\rm K}}\approx48$\% fraction of HI recombinations that are directly to the ground state.
The relatively small contribution of recombinations to the ionizing background owes to a combination of the saturation of reemission in optically thick systems (Fig. \ref{gammarec vs nhi}), leakage of the reemitted photons at the ionizing edge, and the frequency dependence of the photoionization cross section (\S \ref{recombination contribution} and Appendix \ref{recombinations analytical model}).\\ \\
The fractional contribution of HeII recombinations to the HeII photoionization rate is also relatively small for the same reason, but is more difficult to calculate accurately at redshifts $z\gtrsim4$ in our model.
In order to obtain an accurate value, in addition to the HeII LyC reemission line to be well resolved on the computational frequency grid, the mean free path of HeII ionizing photons must also be well resolved by the redshift grid.
In our calculation, quasars produce a negligible and rapidly dropping HeII photoionization rate at $z\gtrsim4$ while star-forming galaxies maintain a roughly constant HI photoionization rate.
In these conditions, the ratio $\eta=N_{\rm HeII}/N_{\rm HI}$ tends to infinity and the HeII mean free path to zero, making it exceedingly difficult to resolve it.
Fortunately, the total HeII photoionization rate in this regime is so small that its fractional enhancement from recombinations is of little practical importance.
Moreover, in this regime HeII reionization may well be still underway and the HeII ionizing background consequently modified, as elaborated on in \S \ref{reionization}.
In Figure \ref{gammas vs z}, we indicate the portion of poor convergence by a dashed curve segment; the turnover of $\Gamma_{\rm HeII}^{\rm with~rec}/\Gamma_{\rm HeII}^{\rm no~rec}$ around $z\sim5$ is likely an artifact and we in fact expect it to continue to increase slightly toward higher redshifts owing to the reduced leakage (\S \ref{recombination contribution}).
\\ \\
The total HI photoionization rate matches the value $\Gamma_{\rm HI}=(0.5\pm0.1)\times10^{-12}$ s$^{-1}$ derived from the \Lya~forest at $z=2-4.2$, subject to the constraint that quasars must contribute a large fraction near their peak \citep[][]{2008ApJ...682L...9F, 2008ApJ...688...85F}.
This was done by normalizing the nominal quasar contribution (\S \ref{quasar and stellar emissivities}) by a factor of 0.36 and normalizing the stellar contribution so as to provide the rest of the ionizing photons.
The renormalization of the quasar contribution can be justified by uncertainties in the mean free path of HI ionizing photons (a direct product of the prescribed HI column density distribution), in their escape fraction, and in the quasar spectral template \citep[see discussion in][]{2008ApJ...688...85F}.
These uncertain factors enter in the conversion from the quasar luminosity to the photoionization rate.
Since we wish to reproduce the more robustly constrained photoionization rate measured from the \Lya~forest, we adjust the normalization accordingly.\\ \\
Although our calculations are normalized to match the hydrogen photoionization rate measured from the \Lya~forest, it is important to emphasize that this measurement and hence the normalization of the spectra calculated here are somewhat uncertain.
The measurement was obtained using the flux decrement method \citep[e.g.,][]{1997ApJ...489....7R}, in which we solve for the value of $\Gamma_{\rm HI}$ needed to produce the measured mean transmission of the \Lya~forest.
Two important sources of systematic uncertainty are the assumed IGM temperature (since the flux decrement constrains only the combination $\Gamma_{\rm HI}/\alpha_{\rm HI}^{A}(T)$) and the gas density distribution (whose details depend on the cosmological parameters and thermal history).
Another potential worry is that the measured \Lya~forest mean transmission may be increasingly biased high toward high redshifts (inducing a redshift-dependent error) as the continuum level is increasingly absorbed and difficult to estimate directly. 
We have however quantified and corrected for this effect in our measurement \citep[][]{2008ApJ...681..831F} and so it should not affect our results.
In the end, we expect the measured $\Gamma_{\rm HI}$ to be accurate within a factor $\sim2$, with the possible errors mostly systematic and weakly dependent on redshift.
For a more exhaustive discussion of the uncertainties of the measured $\Gamma_{\rm  HI}$, see \S 3 of \cite{2008ApJ...688...85F}.\\ \\
Finally, we must also note that for precise work with hydrodynamical simulations, the simulated \Lya~forest mean transmission should always be compared with the measured value and the photoionization rates renormalized if necessary.
In fact, even if the correct ionizing background (with the correct normalization) is prescribed, the simulated mean transmission may be slightly off if, for example, the temperature of the IGM is incorrect reproduced.
This is particularly likely to occur if the effects of HI and HeII reionization (see \S \ref{reionization}) are not explicitly modeled.

\begin{figure}[ht]
\begin{center} 
\includegraphics[width=0.75\textwidth]{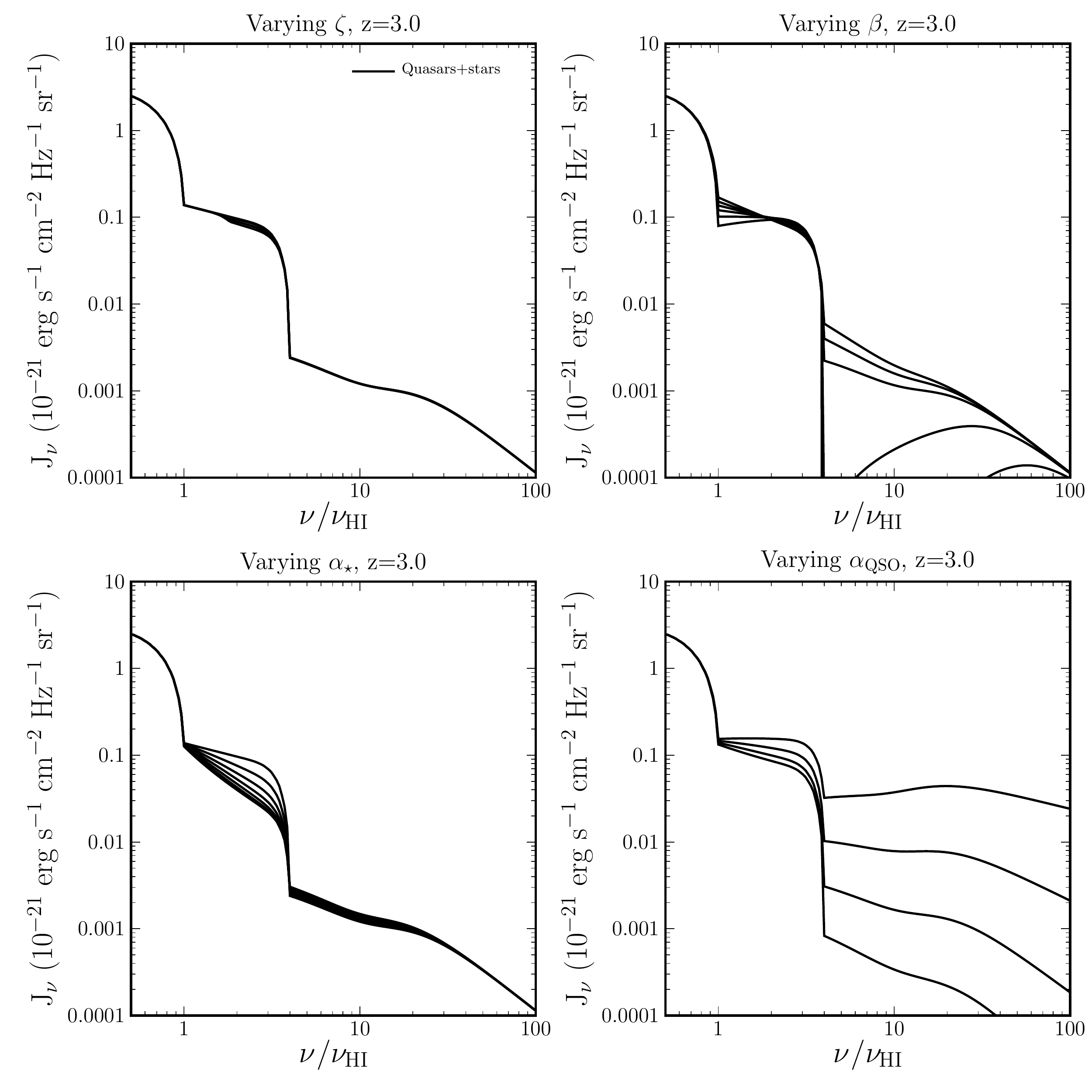} 
\end{center} 
\caption{Effects of parameters on the calculated spectrum.
\emph{Top left:} varying the constant ratio $\zeta=N_{\rm HeI}/N_{\rm HI}=0,~0.001,~0.005,~0.01,~0.05,~{\rm and}~0.1$.
\emph{Top right:} varying the HI column density distribution power-law index $\beta=1.2,~1.3,~1.4,~1.5,~1.6~{\rm and}~1.7$.
\emph{Bottom left:} varying the stellar spectral index $\alpha_{\star}=1.0,~1.5,~2.0,~2.5,~3.0,~3.5,~{\rm and}~4.0$.
\emph{Bottom right:} varying the quasar spectral index $\alpha_{\rm QSO}=0.5,~1.0,~1.5,~{\rm and}~2.0$.
In each panel, the curves correspond to these values from the top to the bottom.
For realistic values of $\zeta$ (\S \ref{column density relations} and Fig. \ref{eta vs nhi comp}) for a background spectrum arising from stars and quasars, the effect of HeI is small.
The column density distribution power law $\beta$ determines the spectral hardening just shortward of the ionization edges.
The stellar and quasar spectral indices determine the spectral slope of the background.
For fixed emissivity at the Lyman limit, the stellar spectral index has only a modest effect on the amplitude of the spectrum because it is truncated at 4 Ryd.
The quasar spectral index, assumed to extend to infinity, has a more drastic overall impact toward high energies.
Recombination emission has been omitted for clarity of presentation.
}
\label{Jnu dependences} 
\end{figure}

\begin{figure}[ht]
\begin{center} 
\includegraphics[width=1.0\textwidth]{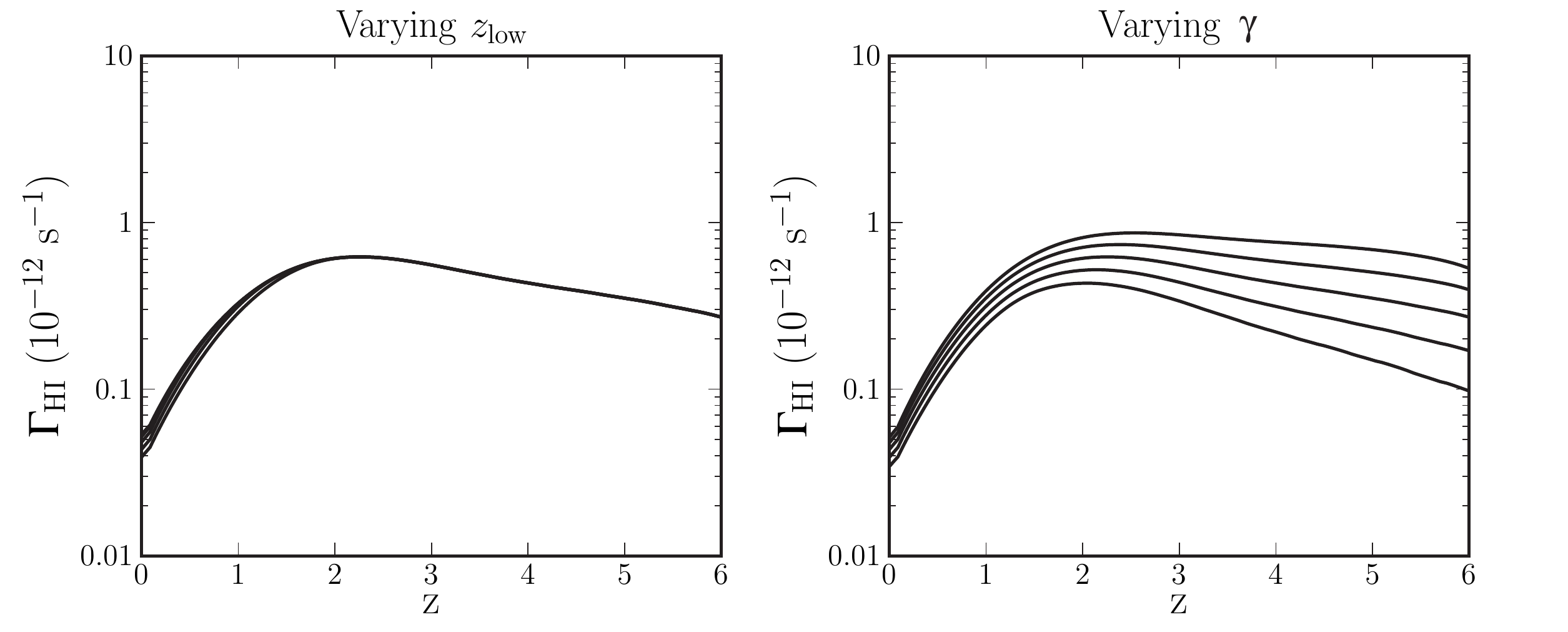} 
\end{center} 
\caption{Effects of the redshift evolution of the HI column density distribution on the hydrogen photoionization rate.
Left: varying the redshift at which the column density distribution transitions to relatively flat evolution $z_{\rm low}=0,~0.5,~1.0,~1.5,~2.0$ (bottom up).
Right: varying the power-law index of redshift evolution $\gamma=1.0,~1.25,~1.5,~1.75,~{\rm and}~2.0$ (top-bottom) of the column density distribution at $z>z_{\rm low}$.
Varying $z_{\rm low}$ at $z\leq2$ has little impact on $\Gamma_{\rm HI}$ since at low redshift the HI ionizing mean free path is sufficiently large that the spectral intensity is limited by the cosmological horizon.
The high-redshift $\Gamma_{\rm HI}$ declines more rapidly with more a rapid increase in the abundance of absorbers with redshift, or large $\gamma$, translating into a more rapidly diminishing mean free path.
}
\label{Gamma vs zlow gamma} 
\end{figure}

\section{DEPENDENCES ON INPUT PARAMETERS}
\label{dependences}
Even after fixing the stellar and quasar emissivities for our fiducial model, the spectrum calculations depend on a number of parameters.
It is useful to investigate how the calculated spectrum and its integrals depend on these as their values are only known to limited precision.
This also provides a physical understanding of the shape of the calculated spectra.
We begin by considering the dependences of the overall ionizing background spectrum in \S \ref{overall spectrum} and focus on the recombination contribution in \S \ref{recombination contribution}.\\ \\

\subsection{Overall Spectrum}
\label{overall spectrum}
In Figure \ref{Jnu dependences}, we show how the spectrum changes when the constant ratio $\zeta=N_{\rm HeI}/N_{\rm HI}$, the HI column density distribution power-law index $\beta$,  and the stellar and quasar spectral indices $\alpha_{\star}$ and $\alpha_{\rm QSO}$ are individually varied.
In each case, all other parameters are fixed to the fiducial model of the previous section.
Even for a constant ratio $\zeta=0.1$, a factor more than one hundred times that expected in our fiducial calculation (\S \ref{column density relations}), HeI absorbs only a very small fraction of the spectrum shortward of its ionization edge.
It is therefore a good approximation to neglect it in our cosmological calculations.
The HI column density distribution power-law index $\beta$ determines the spectral hardening just above the ionization edges following $\alpha \to \alpha - 3(\beta-1)$ (Appendix \ref{spectral filtering}) as well as the depth of the absorption edges.
Note that the depth of the HeII absorption edge is more sensitive to $\beta$; this arises because the column density distribution is normalized to the abundance of HI Lyman limit systems (\S \ref{column density distribution}) so that it is fixed in these calculations while the abundance of the HeII Lyman limit systems varies.
The stellar and quasar spectral indices simply determine the spectral slopes of the background prior to hardening.
For fixed emissivity at the Lyman limit, the stellar spectral index has only a modest effect on the amplitude of the spectrum because it is truncated at 4 Ryd.
The quasar spectral index, assumed to extend to infinity, has a more drastic overall impact toward high energies: as $\nu \to \infty$ and spectral hardening becomes negligible, different spectral indices result in a $J_{\nu}$ ratio of $(\nu/\nu_{\rm HI})^{\alpha_{\rm QSO,1}-\alpha_{\rm QSO,2}}$.
At 10 keV, for example, this ratio is 735 for $\alpha_{\rm QSO,1}=1.5$ and $\alpha_{\rm QSO,2}=0.5$; redshifted from $z=2$ to $z=0$, this falls in the bandpass of x-ray observatories such as \emph{Chandra} and \emph{XMM-Newton}.
As most ($\gtrsim80$\%) of the soft x-ray background has already been resolved into AGNs \citep[e.g.,][]{2005A&A...441..417H}, the x-ray background is a powerful probe of the high-energy quasar spectral energy distribution, although a proper analysis requires the inclusion of obscured quasars, which we do not explicitly consider in this work \citep[e.g.,][]{2007A&A...463...79G}.\\ \\
In Figure \ref{Gamma vs zlow gamma}, we explore how the hydrogen photoionization rate is affected by the redshift evolution of the column density distribution.
In the left panel, we vary the redshift at which the redshift evolution of the column density distribution flattens (\S \ref{column density distribution}) from $z_{\rm low}=0$ to $z_{\rm low}=2$.
Interestingly, this has a minimal impact on the redshift evolution of the photoionization rate even if it does significantly change the mean free path of HI ionizing photons at these redshifts.
This is easily understood as a consequence of the fact that the universe effectively becomes transparent at a ``breakthrough'' redshift $z_{\rm bt}\sim2$ \citep[][]{1999ApJ...514..648M}, below which the mean free path becomes so large the local ionizing background is not limited by the latter but by the cosmological horizon.
As shown in the right panel, the high-redshift $\Gamma_{\rm HI}$ declines more rapidly with more a rapid increase in the abundance of absorbers with redshift, or large $\gamma$, translating into a more rapidly diminishing mean free path.
At present, although more than a decade old, the best constraints on the abundance of the Lyman limit systems \citep[][]{1994ApJ...427L..13S, 1996MNRAS.282.1330S, 1995ApJ...444...64S} are relatively loose and mostly nonexistent beyond $z=4$.
As future measurements refine these and push toward higher redshifts, it is possible that these will give more credence to one of the alternative values of $\gamma$ plotted here.\\ \\

\begin{figure}[ht]
\begin{center} 
\includegraphics[width=1.0\textwidth]{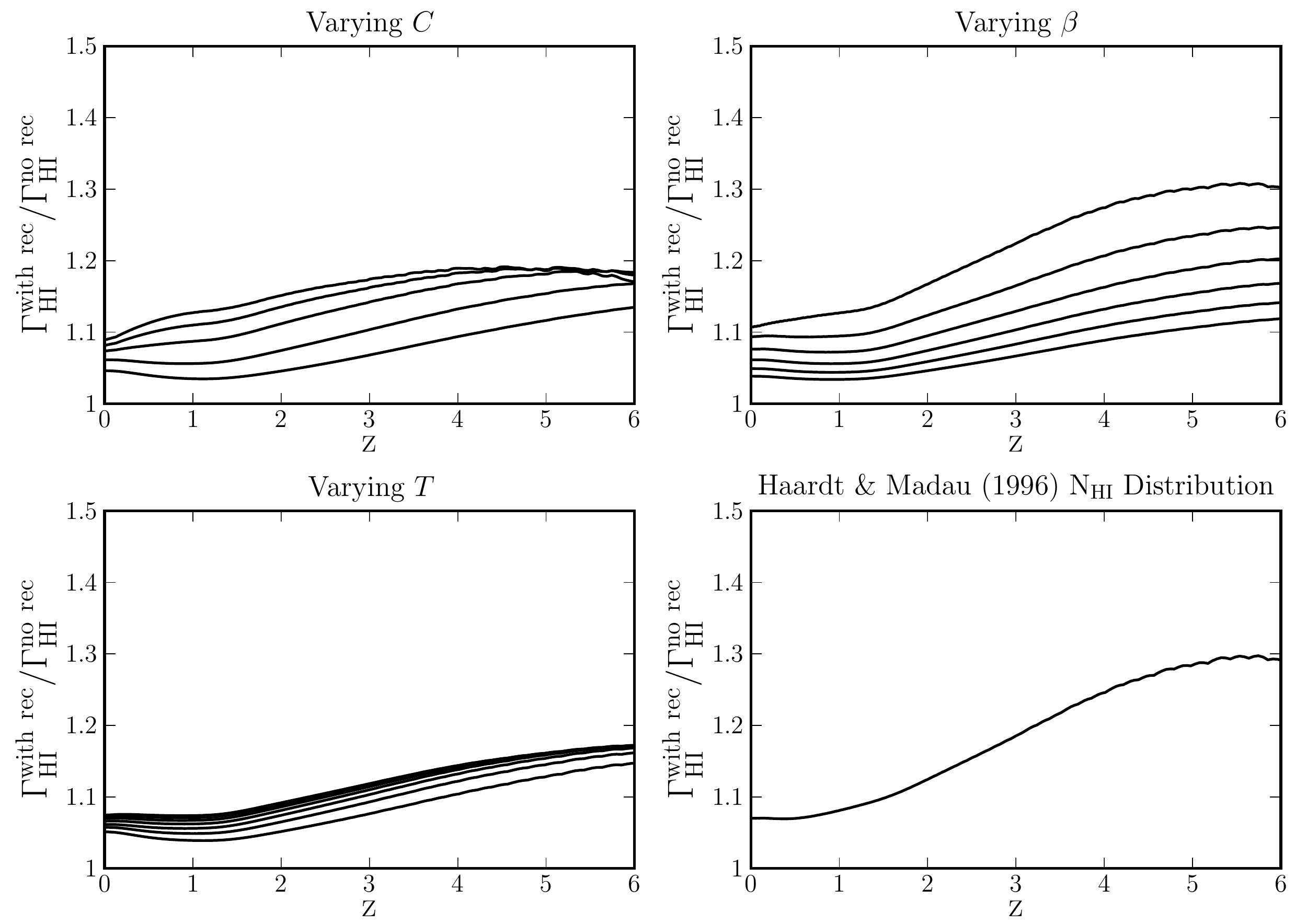} 
\end{center} 
\caption{Dependences of the recombination contribution on HI photoionization rate.
Shown are ratios of the total HI photoionization rate, including recombination emission, to the same calculation ignoring recombination emission.
In all cases, the sources of the ionizing background are fixed to the fiducial model of \S \ref{calculations}, but we vary the parameters of the column density distribution and the temperature of the absorbers.
\emph{Top left:} $C=0.125,~0.25,~0.5,~0.75,~\textrm{and}~1$ from bottom up.
\emph{Top right:} $\beta=1.2,~1.3,~1.4,~1.5,~1.6~\textrm{and}~1.7$ from bottom up.
\emph{Bottom left:} $T=1.0,~1.5,~2.0,~2.5,~3.0,~3.5,~\textrm{and}~4.0\times10^{4}$ K from bottom up.
\emph{Bottom right:} \cite{1996ApJ...461...20H} column density distribution with different redshift evolutions in the optically thin and optically thick regimes.}
\label{Gammarec dependences} 
\end{figure}

\subsection{Recombination Contribution}
\label{recombination contribution}
The contribution of recombinations to the photoionization rates, $\Gamma^{\rm with~rec}/\Gamma^{\rm no~rec}$, is a subtle question as it depends on several factors.
It not only depends on the number of reemitted ionizing photons integrated over the distribution of absorbers (\S \ref{recombinations}) but also crucially on the energy at which these photons are reemitted as well as on their redshifted energy at the point of evaluation of the photoionization rate.\\ \\
The LyC recombination line processes, most important for the boosting the ionization rates, reemit ionizing photons just above the ionization edges of HI or HeII.
Since the ionizing background at a given point is sourced along its past light cone, its photons have generally redshifted slightly from their emission energy.
As a result, many recombination photons with initial energy just above their corresponding ionization edges quickly redshift below these edges and are lost as contributors to the ionization rates.
The fraction of ionizing recombination photons lost in this way depends on two factors: 1) the recombination line profile which determines how far above the ionization edge an ionizing photon is reemitted and 2) the mean free path of ionizing photons which determines how long the photons have to redshift before they are absorbed.
In the limit of a mean free path of zero length, the recombination photons cannot redshift before they are reabsorbed and no photons are lost; as the mean free path increases, more photons leak out of the ionizing range.
Similarly, a narrow line profile concentrates the recombination photons just above the ionization edges, leading to a high probability of leakage, while a wider one allows them to remain longer in the ionizing range.
Since the mean free path of ionizing photons is determined by the column density distribution and the recombination line width is determined by the temperature of the absorbers, these are important parameters for the recombination contribution.
Finally, as recombination photons are reemitted at energies above the ionization edges and subsequently redshift, the frequency dependence $\sigma_{i}(\nu)\propto \nu^{-3}$ of the photoionization cross section changes the weight they receive in the photoionization rates.\\ \\
All these effects are self-consistently treated when solving the radiative transfer equation \ref{transfer equation}.
Figure \ref{Gammarec dependences} shows how much the HI photoionization rate is increased by recombination emission as a function of the normalization $C$ of the column density distribution, its power-law slope $\beta$ (see eq. \ref{column density distribution eq}), and the temperature $T$ of the absorbers.
We also show the case of the column density distribution assumed by \cite{1996ApJ...461...20H}, in which optically thin and optically thick absorbers have different redshift evolutions, leading to a redshift-dependent effective slope of the distribution (steeper at high redshifts).\\ \\
Because the mean free path decreases with increasing abundance of Lyman limit systems (the normalization $C$), fewer recombination photons leak out of the ionizing range and so the ratio of the photoionization rate with and without recombinations, $\Gamma_{\rm HI}^{\rm with~rec}/\Gamma_{\rm HI}^{\rm no~rec}$, increases.
The ratio also increases with the steepness of the column density distribution since more recombinations occur in optically thin systems, from which practically all the recombination photons escape into into the IGM, as opposed to in optically thick systems that trap a large fraction.
Higher gas temperatures result in wider recombination lines so that fewer photons are lost owing to redshifting as well as a higher fraction of recombinations directly to the ground state ($\alpha_{\rm HI,n=1}(T)/\alpha_{\rm HI,n=1}^{\rm A}(T)$; Appendix \ref{recombinations analytical model}).
A competing effect is that the recombination photons of high-temperature absorbers tend to have higher energies and receive less weight in the photoionization rate.
The net effect is however relatively weak on $\Gamma_{\rm HI}^{\rm with~rec}/\Gamma_{\rm HI}^{\rm no~rec}$ for the relevant temperature $T\sim2.0\times10^{4}$ K.
Note that for any given set of parameters the ratio tends to increase toward higher redshifts since the mean free path is lower at the higher cosmological densities.
This behavior is however not seen below the breakthrough redshift $z_{\rm bt}\lesssim2$, where the photoionization rate is limited by the cosmological horizon rather than by the mean free path.\\ \\
In Appendix \ref{recombinations analytical model}, we develop a quantitative analytic model that captures and clarifies these effects and agrees well with the full numerical calculations presented here.

\section{REIONIZATION EVENTS}
\label{reionization}
The previous calculations have implicitly assumed that the universe is reionized in both HI and HeII.
This assumption is most evident in the case of HI, for which we have used a column density distribution measured from the $z\lesssim4$ \Lya~forest.
The assumption creeps in for HeII reionization during which there are large HeII patches the inside which the HeII photoionization rate is very low in comparison to within ionized bubbles.
In each region, the mapping between $N_{\rm HI}$ and $N_{\rm HeII}$ depends on the local spectrum and will in general be very inhomogeneous.
The IGM opacity to HeII ionizing photons will therefore be poorly approximated by using a globally-averaged spectrum to map from $N_{\rm HI}$ to $N_{\rm HeII}$.
In this work, we do not attempt to model HI reionization (and the likely simultaneous reionization of HeI), which occurs at the limit of the present observational reach at $z>6$ \citep[e.g.,][]{2006ARA&A..44..415F, 2008arXiv0803.0586D}.\\ \\
The reionization of HeII was however likely delayed until the rise of the quasar luminosity function, at redshifts that are immediately accessible to observations and some understanding of its effects can be obtained by studying the ratio $\eta=N_{\rm HeII}/N_{\rm HI}$.
In fact, while stellar spectra are theoretically expected to have a strong break at the HeII ionization edge and therefore have little impact on the HeII ionization state, quasars have power-law far-UV spectra that extend well into the x-rays (\S \ref{quasar and stellar emissivities}).
Theoretical calculations based on the quasar luminosity function in fact indicate that quasars can reionize HeII by $z\sim3-4$ \citep[e.g.,][]{2002MNRAS.332..601S, 2003ApJ...586..693W, 2008ApJ...681....1F, 2008ApJ...688...85F}.
A number of lines of evidence, based HI and HeII \Lya~forests as well as on the evolution of metal line ratios, also suggest that the IGM is undergoing changes that could be associated with HeII reionization at these redshifts \citep[for a review of these lines of evidence, see][]{2008ApJ...681..831F}.
While alternative candidate sources of HeII reionization exist -- such as possible HeII ionizing emission from galaxies \citep[e.g.,][]{2008ApJ...681....1F}, high-redshift x-rays \citep[e.g.,][]{2001ApJ...553..499O, 2004MNRAS.352..547R}, or thermal emission from shock heated gas \citep[][]{2004MNRAS.348..964M} -- quasars are the best established and most likely.
Large fluctuations observed in the HeII ionizing background toward $z=3$, which can be explained by the small number density of bright objects, also lend support to the quasar hypothesis \citep[][]{2004ApJ...605..631Z, 2004ApJ...600..570S, 2006MNRAS.366.1378B} and we will therefore concentrate on this scenario.

\begin{figure}[ht]
\begin{center} 
\includegraphics[width=1.0\textwidth]{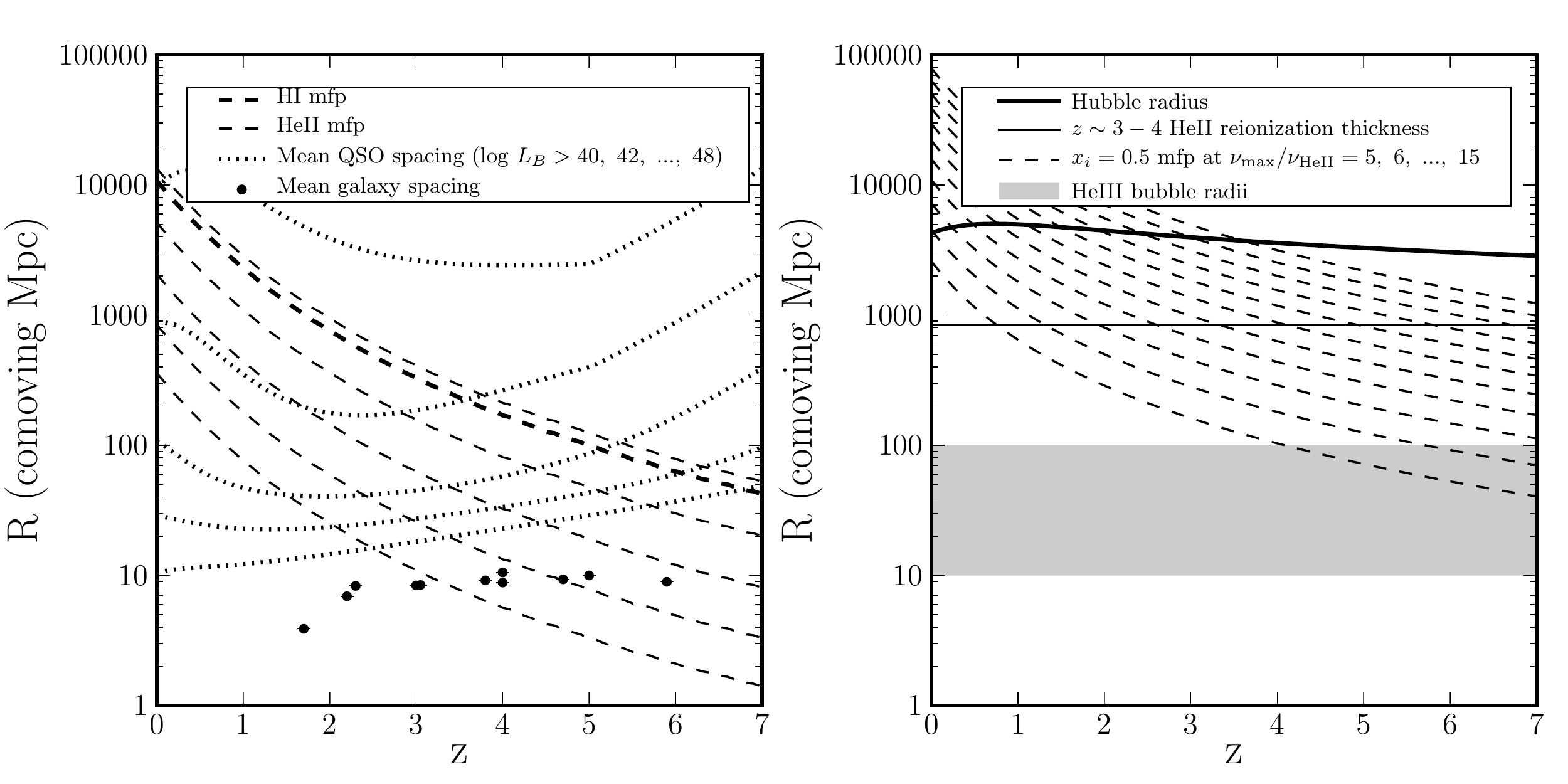} 
\end{center} 
\caption{Important physical scales for cosmological radiative transfer.
The left panel compares the mean free paths of 1 Ryd (HI ionizing; thick dashed) and 4 Ryd (HeII ionizing; thin dashed) photons to the mean separation between the sources of the ionizing background.
The different HeII ionizing mean free path curves correspond to different $\Gamma_{\rm HeII}$ assumed in the calculation and ignore HeII reionization.
From top down, $\Gamma_{\rm HeII}=10^{-13},~10^{-14},~10^{-15},~10^{-16},{\rm~and}~10^{-17}$ s$^{-1}$, assuming a constant $\Gamma_{\rm HI}=0.5\times10^{-12}$ s$^{-1}$.
The mean separation between $L^{\star}$ galaxies versus redshift is shown by the black points and calculated from measured galaxy UV luminosity functions \citep[][]{1999ApJ...519....1S, 2006ApJ...642..653S, 2006ApJ...653..988Y, 2007ApJ...670..928B, 2008ApJS..175...48R}.
For the mean spacing between quasars, the dotted curves correspond to different lower $B$-band luminosity cuts and are calculated using the \cite{2007ApJ...654..731H} luminosity function.
From bottom up, $L_{B}\geq10^{40},~10^{42},~10^{44},~10^{46},~{\rm and }~10^{48}$ erg s$^{-1}$.
The right panel shows scales relevant to understanding the possible effects of HeII reionization by quasars at $z\sim3-4$.
The comoving Hubble radius $c(1+z)/H(z)$ is indicated by the thick solid curve and the comoving distance between $z=3$ and $z=4$ spatial surfaces, labeled the ``thickness of HeII reionization'', is shown by the thinner solid curve.
The dashed curves show the mean free path of high-energy HeII ionizing photons versus redshift assuming a constant ionized fraction $x_{i}=0.5$ and that the HeII is uniform distributed, $\Delta l_{\rm mfp}(z)\equiv[\sigma_{\rm HeII}(\nu)n_{\rm HeII}(z)]^{-1}$.
The curves, from bottom up, correspond to individual frequencies $\nu_{\rm max}=(5,~6,~{\rm, ...},~15)\nu_{\rm HeII}$ (see \ref{heat input calculation} for the significance of $\nu_{\rm max}$).
The gray shaded area indicate typical HeIII ionized bubble radii during HeII reionization.}
\label{scales} 
\end{figure}

\subsection{The Ionizing Background During HeII Reionization}
\label{spectrum during heii reion}
Recently, \cite{2008arXiv0807.2799M} performed detailed radiative transfer simulations of HeII reionization in large boxes up to 430 comoving Mpc on a side \citep[for previous simpler treatments, analytic and numerical, see][]{2002MNRAS.332..601S, 2004MNRAS.348L..43B, 2005MNRAS.361.1399G, 2007arXiv0711.1904P, 2008ApJ...681....1F, 2008arXiv0807.2447B}.
These simulations used realistic models for the quasar sources based on the luminosity function of \cite{2007ApJ...654..731H} and with physically and empirically motivated prescriptions for the triggering of quasars in massive halos as well as of quasar light curves
\citep[see, e.g.,][]{2005ApJ...630..705H, 2005ApJ...625L..71H, 2006ApJS..163....1H, 2008ApJS..175..356H}.
A striking result of this work is the remarkable complexity of HeII reionization, in particular of the HeII ionizing radiation field, likely rendering the detailed resulting structure beyond analytic tractability.
Nevertheless, some intuition on the spectrum and magnitude of the ionizing background during HeII reionization can be gained by considering idealized cases.
We consider two such cases: 1) a single quasar at the center of an isolated ionized bubble and 2) a point in a large HeII patch that has yet to be reionized.\\ \\
Key insight into the ionizing background is gained by considering relevant physical scales.
The left panel of Figure \ref{scales} compares the mean free paths (calculated as in Appendix \ref{spectral filtering}) of 1 Ryd (HI ionizing) and 4 Ryd (HeII ionizing) photons to the mean separation between the sources of the ionizing background, while the right panel shows scales relevant to understanding the possible effects of HeII reionization by quasars at $z\sim3-4$.
The HeII ionizing mean free paths are calculated by converting the HI column densities to HeII assuming a constant $\Gamma_{\rm HI}=0.5\times10^{-12}$ s$^{-1}$ and varying $\Gamma_{\rm HeII}$.
Since emission from star-forming galaxies provides most of the hydrogen photoionization rate at $z\gtrsim3$ (\S \ref{calculations}) and the mean separation between $L^{\star}$ galaxies is much smaller than the HI ionizing mean free path at all redshifts $z\lesssim6$ considered,\footnote{At redshifts $z\gtrsim4$, the estimated mean free path relies on an extrapolation of the measured column density distribution and so the conclusion should accordingly be treated with caution.
In particular, the conclusion is likely to break down if HI reionization ends at $z\approx6$ \citep[e.g.,][]{2002AJ....123.1247F, 2006AJ....132..117F}.}
it is a good approximation to treat the stellar emissivity as a uniform volume average as in equation \ref{transfer equation solution}.
It is also similarly the case for HI ionizing quasar emissivity at redshifts $z\lesssim3$, where quasars are relatively abundant and the HI ionizing mean free path large, though with larger fluctuations expected from the smaller number of quasars within each mean free path \citep[for more detailed studies of UV background fluctuations, see][]{1992MNRAS.258...45Z, 1992MNRAS.258...36Z, 1993ApJ...415..524F, 1999ApJ...520....1C, 2002ApJ...581...20C, 2002MNRAS.334..107G, 2003MNRAS.342.1205M, 2004MNRAS.350.1107M, 2004ApJ...610..642C}.
Thus, it is a reasonable approximation at all redshifts to calculate the ionizing background between 1 and 4 Ryd using a volume average emissivity.\\ \\
The situation is however quite different beyond 4 Ryd, where continuum opacity owing to HeII dominates.
In fact, the mean free path of HeII ionizing photons at these energies, which depends on the local HeII photoionization rate, is generally smaller than the mean free path of 1 Ryd HI ionizing photons since even quasars produce relatively few photons above 4 Ryd.
For example, near the peak of the quasar luminosity function at $z=2.1$, $S=\Gamma_{\rm HI}/\Gamma_{\rm HeII}=140$ \citep[][]{2006MNRAS.366.1378B}.
The relative rarity of quasars and shortness of the HeII ionizing mean free path combine to create a situation in which often a single bright quasar contributes to the local HeII ionizing flux.
This is the case even after HeII reionization has completed and results in substantial fluctuations in the ionizing background above 4 Ryd.
These fluctuations could be important for metal absorption line studies and will be addressed in future work \citep[for recent studies of the HeII ionizing background fluctuations, see][]{2006MNRAS.366.1378B, 2008arXiv0812.3411F, 2009arXiv0901.2584F}.
Prior to the complete reionization of HeII, ionized bubble walls will further limit the exposure of a given point to the HeII ionizing fields of distant quasars.
The radii of HeIII bubbles depend, at least until they percolate, on the ionizing luminosity of the central quasars and the duration for which these have been shining.
This gives rise to a wide range of scales, depending on the specific quasar model, but by the middle of HeII reionization (determined by an ionized fraction $x_{i}\sim0.5$) bubble radius $R_{b}\sim10-100$ comoving Mpc (corresponding to a few to 20-25 proper Mpc at $z=3-4$) are representative \citep[e.g.,][]{2008ApJ...681....1F, 2008arXiv0807.2799M}.
A volume average uniform emissivity is then clearly inappropriate.

\begin{figure}[ht]
\begin{center} 
\includegraphics[width=1.0\textwidth]{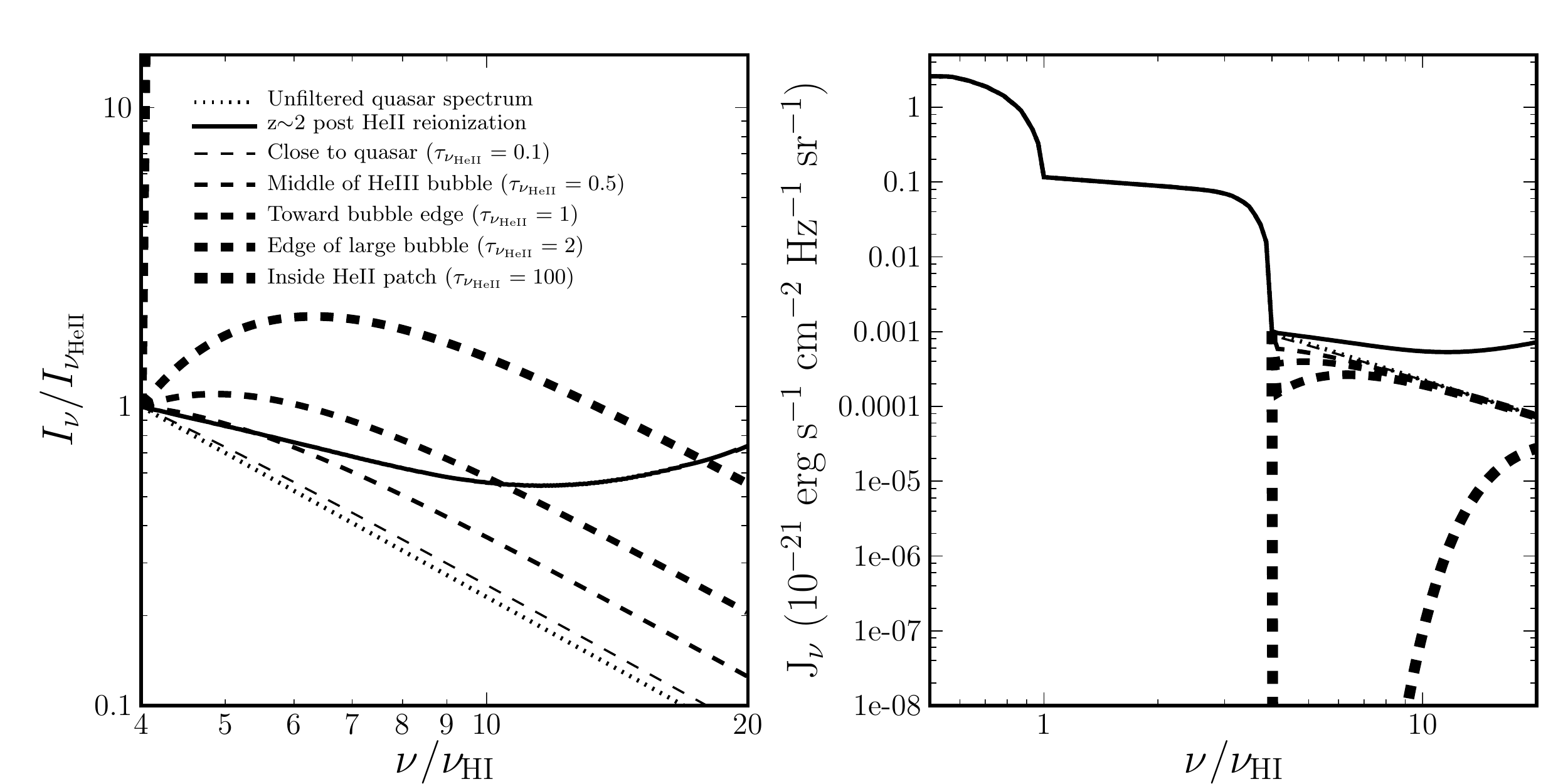} 
\end{center} 
\caption{Left: Normalized spectral hardening above the HeII ionization edge.
The unhardened quasar spectrum (dotted curve) is assumed to be a power law $I_{\nu}\propto \nu^{-1.6}$ \citep[e.g.,][]{2002ApJ...565..773T}.
The solid $z\sim2$ post HeII reionization curve shows the limit in which the mean free path is sufficiently large to contain several quasar sources, with the HI column density distribution and softness parameter $S=\Gamma_{\rm HI}/\Gamma_{\rm HeII}=140$ measured at $z\approx2$ \citep[][]{2006MNRAS.366.1378B}.
This limit is representative of the hardening in the calculations of \S \ref{calculations}.
The dashed curves show the hardened spectrum at different optical depths from a central quasar at the center of an isolated HeIII bubble.
From the thinnest to the thickest curve, $\tau_{\nu_{\rm HeII}}=0.1,~0.5,~1,~2,~\textrm{and}~100$.
The curves are pictorially labeled assuming a HeII ionizing mean free path comparable to the bubble size, from $close$ to the quasar to $toward~the~edge$ of a large bubble and $inside~a~HeII~patch$.
Right: Illustration of the absolute effect of HeII attenuation on the full background spectrum.
In addition to the spectral hardening, the spectrum is exponentially suppressed above the HeII ionization edge.
Recombination reemission is omitted here but discussed in \S \ref{recombinations during heii reion}.}
\label{heii spectral hardening} 
\end{figure}

\subsubsection{Quasar Within an Isolated Ionized Bubble}
\label{quasar within ionized bubble}
Consider first a point within an isolated HeIII bubble occupied by a single quasar at the center, $r=0$.
Locally neglecting cosmological effects, the specific intensity of a radial ray is given by
\begin{equation}
\label{quasar intensity}
I_{\nu}= I_{\nu}(r=0) e^{-\tau_{\nu}(r)}.
\end{equation}
In addition to the intensity being exponentially suppressed, the spectral shape is altered by the frequency dependence of the optical depth:
\begin{equation}
\label{quasar spectral hardening}
\frac{I_{\nu}}{I_{\nu_{\rm HeII}}} =
\frac{e^{-\tau_{\nu}}}
{e^{-\tau_{\nu_{\rm HeII}}}}
= e^{\tau_{\nu_{\rm HeII}}}
e^{-\tau_{\nu_{\rm HeII}}[\sigma_{\rm HeII}(\nu)/\sigma_{\rm HeII}(\nu_{\rm HeII})]}
\propto
\left( e^{-\tau_{\nu_{\rm HeII}}} \right)
^{\sigma_{\rm HeII}(\nu)/\sigma_{\rm HeII}(\nu_{\rm HeII})}
\approx
\left( e^{-\tau_{\nu_{\rm HeII}}} \right)
^{(\nu/\nu_{\rm HeII})^{-3}},
\end{equation}
where the last equality holds approximately just above the HeII ionization edge and we have neglected the fractionally small HI continuum opacity.
It follows that the magnitude of the specific intensity is set by the optical depth at the HeII ionization edge to the source, with the spectral shape entirely determined by the frequency dependence of the photoionization cross section at a given optical depth, in addition to the intrinsic spectrum of the source.\\ \\
In the left panel of Figure \ref{heii spectral hardening}, we show a quasar spectrum $I_{\nu}(r=0)\propto \nu^{-1.6}$ is hardened as a function of $\tau_{\nu_{\rm HeII}}$, the optical depth at the HeII ionization edge from the source.
The curves are pictorially labeled assuming a HeII ionizing mean free path comparable to the HeIII bubble size, so that $\tau_{\nu_{\rm HeII}}\sim1$ is near the edge of an isolated bubble centered on the quasar.
We also show a $z\sim2$ post HeII reionization case in which the mean free path is sufficiently large to contain several quasar sources, with the HI column density distribution and softness parameter $S=\Gamma_{\rm HI}/\Gamma_{\rm HeII}=140$ measured at $z\approx2$ \citep[][]{2006MNRAS.366.1378B}.
This limit is representative of the hardening in the calculations of \S \ref{calculations}.
Note that as $\tau_{\nu_{\rm HeII}}\to \infty$, the spectral shape can be arbitrarily hardened just above the HeII ionization edge.
As $\nu \to \infty$ and $\sigma_{\rm HeII}(\nu)\to0$, the spectrum returns to the unfiltered case.\\ \\
The rarity of quasars implies that around an individual object the specific intensity obeys equation \ref{quasar intensity}, in which a single source is attenuated with distance, rather than a solution involving a volume average emissivity as in equation \ref{transfer equation solution}.
Why, though, does the ordinary optical depth $\tau_{\nu}$ enter in equation \ref{quasar intensity} instead of the effective optical depth $\bar{\tau}$ as in equation \ref{transfer equation solution}?
Any given light ray is always attenuated according to the intervening ordinary optical depth $\tau_{\nu}$.
However, the optical depth between two points separated by a fixed distance (at fixed frequency and redshifts) fluctuates depending on their particular spatial positions because of the stochastic nature of the intervening absorbers.
The effective optical depth captures the average attenuation through $e^{-\bar{\tau}}=\langle e^{-\tau} \rangle$.
It is an appropriate quantity for the ionizing background between 1 and 4 Ryd, where the local intensity is an average over the light received from sources in all directions within one mean free path.
The radiation above 4 Ryd at a given point in the vicinity of a quasar prior to and during HeII reionization will often be dominated by the local quasar and therefore be uniquely attenuated as in equation \ref{quasar intensity}.

\subsubsection{Point in a Large HeII Patch}
A point within a HeII patch that has not yet been reionized\footnote{In reality, the HeII ionization fronts are quite smooth and extended since they are driven by a hard spectrum \citep[e.g.,][]{2008arXiv0807.2799M}. Except at the very beginning, few points have been truly untouched by HeII reionization, but the discussion holds wherever the ionized fraction has not exceeded, say, $\sim1/2$.}
 will see a similarly hardened spectrum, but with a much stronger suppression at the HeII ionization edge owing to the large intervening optical depth.
The optical depth at the HeII ionization edge, as a function of redshift and path length $L$, in a medium in which all the helium is assumed to be in the form of HeII is given by
\begin{equation}
\tau_{\nu_{\rm HeII}}^{\rm neutral} = 
\sigma_{\rm HeII}(\nu_{\rm HeII}) n_{\rm HeII}(z) L
=
318
\left( \frac{1+z}{4.5} \right)^{4}
\left( \frac{L}{{\rm10~comoving~Mpc}} \right).
\end{equation}
In HeII patch, the intensity of the background at the ionization edge is therefore expected be almost entirely suppressed.
As $\nu \to \infty$ and $\sigma_{\rm HeII}(\nu)\to0$, however, the optical depth drops quickly and the intensity of the background recovers.
The corresponding increase of the mean free path with energy leads to the presence of a spatially smooth high-energy radiation background permeating most of the cosmic volume, as seen for example in the numerical simulations of \cite{2008arXiv0807.2799M}.
The right panel of Figure \ref{heii spectral hardening} shows how the fiducial spectrum of the ionizing background at $z=3.5$, as calculated in \S \ref{calculations}, is altered shortward of the HeII ionization edge as a function of the intervening optical depth.
Note, in particular, the tremendous HeII edge suppression even in the moderate case of $\tau_{\nu_{\rm HeII}}=100$. 

\subsection{Recombination Lines during HeII Reionization}
\label{recombinations during heii reion}
The photoionization rate and ionization state of hydrogen are unaffected by the presence of HeII before and during HeII reionization apart from a small contribution by photons above 4 Ryd.
Consequently, only the HeII recombination processes are significantly modified.
Of these, the most important is HeII~\Lya~which imprints a distinctive line feature at 3 Ryd (Fig. \ref{spectra}); HeII~LyC and BalC only slightly smooth the spectrum at the HeII and HI ionization edges, respectively, and contribute only marginally to the photoionization rates (Figs. \ref{gammarec vs nhi} and \ref{gammas vs z}).\\ \\
Equations \ref{Nabs eq} and \ref{Lya reemission approximation} compactly capture the behavior to HeII~\Lya~reemission.
As explained in the previous section, before HeII reionization begins the background spectrum is almost completely suppressed above 4 Ryd by the large optical depth at these energies.
Since HeII~\Lya~reemission scales with the HeII ionizing spectrum (with saturation in the optically thick limit), it will be absent before the start of HeII reionization.
Similarly, no HeII~\Lya~should arise within HeII patches during HeII reionization.
However, HeII~\Lya~will be reemitted within ionized bubbles illuminated by the local quasars.
As HeII reionization proceeds, the distance between neighboring bubbles should quickly become smaller than the mean free path for the LyC absorption of 3 Ryd photons by HI (at 1 Ryd, Fig. \ref{scales} shows the mean free path to be about 200 comoving Mpc at $z=3.5$; at 3 Ryd, eq. \ref{mean free path analytical equation} predicts the mean free path to be longer by a factor of $3^{3(\beta-1)}\approx5$) that governs the attenuation of HeII~\Lya~radiation.
In the regime in which this mean free path contains several ionized bubbles, the volume fraction of reionized HeII can be viewed as the fraction of the IGM reemitting in HeII~\Lya~and the HeII~\Lya~reemission line of the background spectrum can be expected to be about this fraction times the fully reionized value.

\subsection{Heat Input During HeII Reionization}
The photons that ionize HeII atoms in general carry more energy than the $h \nu_{\rm HeII}$ required.
The residual energy is converted into kinetic energy of the resulting free electron and HeIII nucleus, with the frequent Coulomb collisions leading to rapid thermalization.
This process of photoheating is at work at all times and for all species present.
Its effect is however much more important during reionization, when atoms are being ionized at a much greater rate.
The effects of HeII reionization on the thermal state of the gas in cosmological simulations has so far generally be modeled by artificially boosting the photoheating rate calculated from a prescribed spatially homogeneous background instantaneously\citep[e.g.,][]{2000ApJ...534...57B, 2002ApJ...574L.111T, 2005MNRAS.361...70J}, or ignoring it altogether.
This approach, of limited physical basis, is a serious limitation of these simulations given the growing body of evidence that HeII reionization occurs at observable redshifts $z\sim3-4$ and is certain to manifest itself to some extent.\\ \\
While cosmological radiative transfer simulations are beginning to self-consistently treat gas thermodynamics during HeII reionization \citep[e.g.,][]{2007arXiv0711.1904P, 2008arXiv0807.2799M}, it is likely that the vast majority of simulations performed in the near to moderate future will not explicitly incorporate radiative transfer, either due to the computational cost or to the unavailability of an appropriate code.
It therefore remains important to develop ways of approximately treating the effects of HeII reionization in those simulations.
We examine this problem in this section.
Specifically, we consider the questions: How much does HeII reionization heat the IGM? Over what timescale? And how can we approximately model its effects in standard cosmological N-body and hydrodynamical codes such as GADGET \citep[][]{2001NewA....6...79S, 2005MNRAS.364.1105S}, Hydra \citep[][]{1997NewA....2..411P}, or Enzo \citep[][]{2004astro.ph..3044O}?\\ \\
The simple analytic models that follow are motivated by and owe much to the physical picture of HeII reionization suggested by the radiative transfer calculations of \cite{2008arXiv0807.2799M}.
We refer to that work for many original insights.

\begin{figure}[ht]
\begin{center} 
\includegraphics[width=0.5\textwidth]{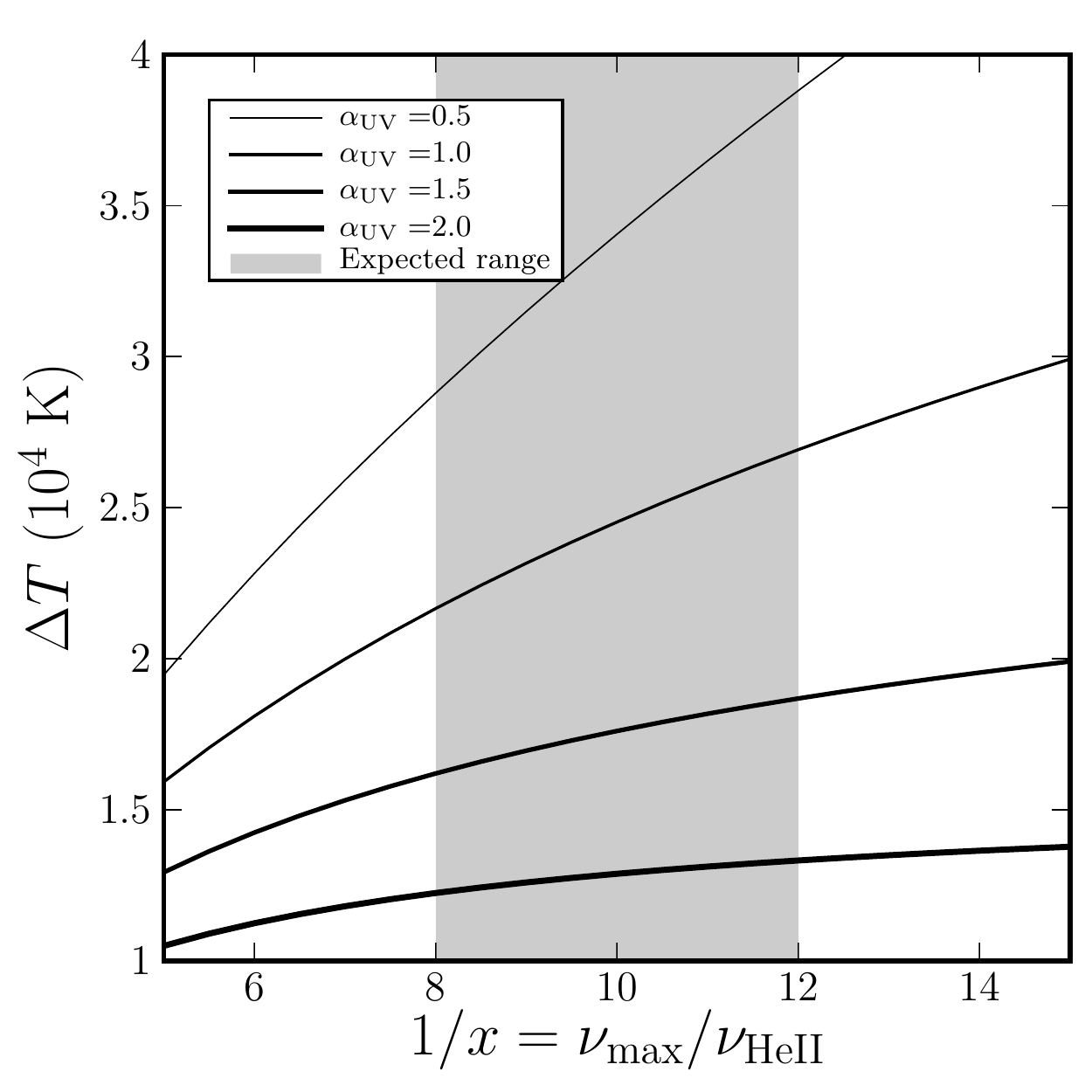} 
\end{center} 
\caption{Total temperature increase owing to HeII reionization photoheating as a function of the maximum absorbed frequency for different spectral indices $\alpha_{\rm UV}$ of the reionizing sources.
The ``expected range'' corresponds to the bulk of HeII reionization occurring between $z=3$ and $z=4$ and is shown only for suggestive purposes.
Figure \ref{heii thermal history} presents more rigorous calculations that avoid fixing a hard cutoff frequency based on the HeII reionization history calculated from the quasar luminosity function.
}
\label{heii heating} 
\end{figure}

\subsubsection{Heat Input Calculation}
\label{heat input calculation}
\begin{figure}[ht]
\begin{center} 
\includegraphics[width=1.0\textwidth]{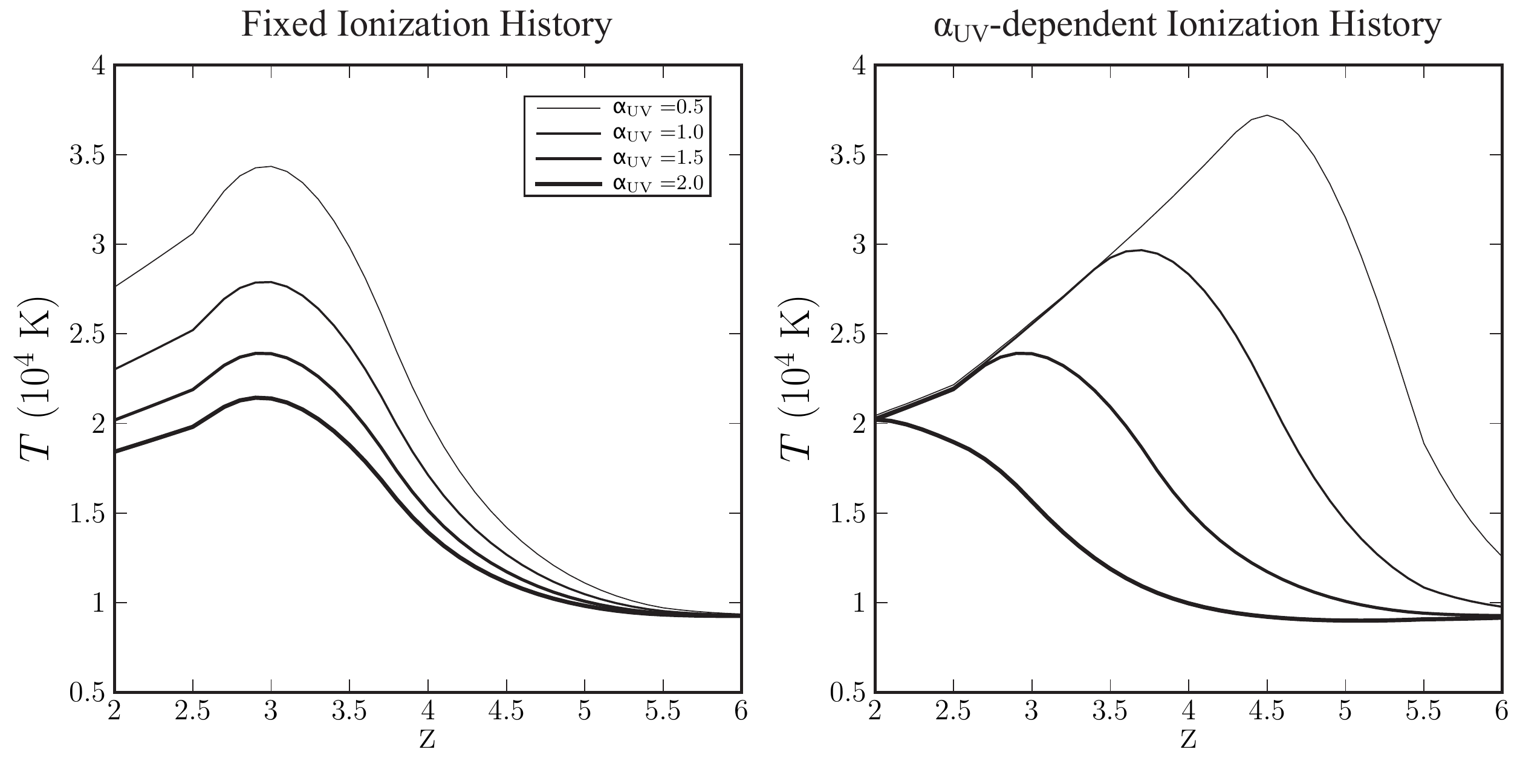} 
\end{center} 
\caption{Evolution of the IGM temperature for different quasar spectrum indices $\alpha_{\rm UV}$ in the early HI reionization limit in which the temperature would follow the thermal asymptote (eq. \ref{thermal asymptote power law}) with $\alpha_{\rm bg}=0$ in the absence of HeII reionization.
HeII reionization is taken to occur through the action of quasars, with the $B-$band luminosity function of \cite{2007ApJ...654..731H}.
In the left panel, the evolution of the HeIII ionized fraction $y_{\rm III}(z)$ is artificially fixed to the value calculated for a quasar spectral index $\alpha_{\rm UV}=1.5$.
In the right panel, the ionization history is calculated consistent with the spectral index of the sources, with harder spectra resulting in higher ionizing photon output rates and thus earlier reionization.}
\label{heii thermal history} 
\end{figure}

In order to gain physical intuition, we begin with a simplified model.
Suppose that all the photons up to frequency $\nu_{\rm max}$ emitted by a population of sources with intrinsic spectral index $\alpha_{\rm UV}$ are absorbed by HeII atoms.
Then the mean energy injected into the IGM per ionization is given by
\begin{equation}
\label{Delta E cut off}
\langle E_{i} \rangle =
\frac{\int_{\nu_{\rm HeII}}^{\nu_{\rm max}}d\nu/(h\nu)(h\nu-h\nu_{\rm HeII})\nu^{-\alpha_{\rm UV}}}
{\int_{\nu_{\rm HeII}}^{\nu_{\rm max}}d\nu/(h\nu) \nu^{-\alpha_{\rm UV}}} =
h\nu_{\rm HeII}
\left[
\frac{\alpha_{\rm UV}}{\alpha_{\rm UV}-1}
\frac{(1-x^{\alpha_{\rm UV}-1})}
{(1-x^{\alpha_{\rm UV}})}
-1
\right] \approx
\frac{h\nu_{\rm HeII}}{\alpha_{\rm UV}-1}
(1 - \alpha_{\rm UV}x^{\alpha_{\rm UV}-1}),
\end{equation}
where $x\equiv \nu_{\rm HeII}/\nu_{\rm max}$ and the last equality holds approximation for $x\ll1$ and $\alpha_{\rm UV}>0$.
This equation neglects redshifting of the photons before absorption, which is a reasonable approximation if HeII reionization lasts $\Delta z\approx1$ at $z\sim3-4$ \citep[e.g.,][]{2008ApJ...681....1F, 2008ApJ...688...85F, 2008arXiv0807.2799M}.
Here, the effects of spectral filtering \citep[e.g.,][]{1999ApJ...520L..13A, 2004MNRAS.348L..43B, 2007MNRAS.380.1369T, 2008arXiv0807.2447B} are incorporated in the prescribed frequency cutoff $\nu_{\rm max}$.\\ \\
If all the helium is initially in the form of HeII and hydrogen is fully ionized, the temperature increase is obtained by distributing the injected energy over all particles.
After thermal equilibrium has been reached,
\begin{equation}
\Delta T _{\rm HeII}= \frac{2}{3k}
\frac{n_{\rm He}}{n_{\rm tot}}
\langle E_{i} \rangle
= 15550{\rm~K}
\left[
\frac{\alpha_{\rm UV}}{\alpha_{\rm UV}-1}
\frac{(1-x^{\alpha_{\rm UV}-1})}
{(1-x^{\alpha_{\rm UV}})}
-1
\right]
\approx
31100~{\rm K}
\left(
\frac{0.5}{\alpha_{\rm UV}-1}
\right)
(1 - \alpha_{\rm UV}x^{\alpha_{\rm UV}-1}).
\end{equation}
Here, $n_{\rm tot}=2n_{\rm H}+3n_{\rm He}$ is the total number density of particles including free electrons.
Note that the total number of particles is slightly less before HeII reionization owing to the smaller number of free electrons.
The fractional change of $1/16$ is however negligible.
Although the use of a sharp frequency cutoff $\nu_{\rm max}$ is a simplification of the radiative transfer, \cite{2008arXiv0807.2799M} show that a simple argument like this one gives a good estimate of the heat input determined from detailed radiative transfer simulations.\\ \\
What is the relevant value of $\nu_{\rm max}$?
A reasonable guess is the value such that the mean free path of photons of this frequency equals the ``thickness'' of HeII reionization.
Photons of higher frequency (and therefore longer mean free path) will typically not be absorbed before HeII reionization is complete.
The right panel of Figure \ref{scales} shows where the mean free path intersects the thickness of HeII reionization, assuming that the bulk of the latter takes place between $z=3$ and $z=4$, for different values of $\nu_{\rm max}$.
For this purpose, we calculate the mean free path $R_{\rm mfp}(\nu_{\rm max})=[n_{\rm HeII}\sigma_{\rm HeII}(\nu_{\rm max})]^{-1}$ assuming homogeneously distributed 50\% ionized HeII at $z=3.5$.
Under these conditions (accounting for some uncertainty on the thickness of HeII reionization) we expect $x^{-1}\sim8-12$. 
Figure \ref{heii heating} shows the corresponding heat input owing to HeII reionization for different value of the spectral index $\alpha_{\rm UV}$.
For spectral indexes $\alpha_{\rm UV}\sim1.5$ (\S \ref{quasar and stellar emissivities}), the heat input depends only weakly on our rough estimate of $x^{-1}$.\\ \\
Having obtained simple estimates for the total heat input during HeII reionization, we proceed to make the derivation more rigorous, which also allows us to trace the time evolution of the heat injection.
Specifically, we replace the sharp frequency cutoff by a calculation taking into account the fraction of photons emitted at each frequency at any given redshift that is absorbed during HeII reionization:
\begin{equation}
\label{Delta E rigorous}
\langle E_{i} \rangle(z)
=\frac{
\int_{z}^{\infty}dt
\int_{\nu'_{\rm HeII}}^{\infty}
d\nu'/(h\nu')(h\nu'-h\nu'_{\rm HeII})
\epsilon_{\nu'}^{\rm QSO,com}(z')
[1-e^{-\tau(\nu',~z,~z'(t))}]
}
{
\int_{z}^{\infty}dt
\int_{\nu'_{\rm HeII}}^{\infty}
d\nu'/(h\nu')
\epsilon_{\nu'}^{\rm QSO,com}(z')
[1-e^{-\tau(\nu',~z,~z'(t))}]
},
\end{equation}
where $dt=(dz'/c)(dl/dz')$ and
\begin{equation}
\tau(\nu',~z,~z')=
\int_{z}^{z'}dz''
\frac{dl}{dz''}
n_{\rm He}(z'')[1-y_{\rm III}(z'';~\alpha_{\rm UV})]
\sigma_{\rm HeII}
\left(
\nu''= \nu \frac{(1+z'')}{(1+z')}
\right)
\end{equation}
is the optical depth encountered by a photon of frequency $\nu'$ emitted at redshift $z'$ before reaching redshift $z$.
Here, $n_{\rm He}$ is the proper number density of helium atoms and a fraction $1-y_{\rm III}$ given by the reionization state is assumed to be homogeneously distributed in the form of HeII, taken to be the dominant source of opacity.
Equation \ref{Delta E rigorous} is similar to equation \ref{Delta E cut off}, but with the high-frequency cutoff replaced by the smoothly varying fraction of photons absorbed $1-e^{-\tau}$ for each frequency.
In addition, the mean energy injected per ionization is calculated as a function of redshift, which allows us to trace the heat input over time.
While the homogeneous IGM approximation is obviously a simplification, it is a reasonable assumption for this heuristic calculation.
In fact, the potential error introduced by neglecting the inhomogeneities is most important for $\tau \sim 1$.
However, the strong frequency dependence of the HeII photoionization cross section implies that the range of photon energy for which $\tau \sim 1$ is narrow.
Morever, for a quasar spectral index $\alpha_{\rm UV}\approx1.5$, the cruder estimate of Fig. \ref{heii heating} indicates that the heat input is only weakly sensitive to the exact maximum energy of the absorbed photons.
These effects combine to make the uniform IGM approximation relatively robust for this particular calculation.
Ultimately, though, the calculation is motivated by the fact that it reproduces the results of the full radiative transfer calculations of \cite{2008arXiv0807.2799M} well.\\ \\
Since at a given redshift $z$, only a fraction $y_{\rm III}$ of the HeII has been reionized, the temperature increase contributed by HeII reionization at that redshift, neglecting cooling, is given by
\begin{equation}
\label{Delta T HeII}
\Delta T_{\rm HeII}(z)= \frac{2}{3k}
\frac{n_{\rm He}}{n_{\rm tot}}
y_{\rm III}(z)\langle E_{i} \rangle(z).
\end{equation}
The temperature of a cosmic gas parcel is in general determined by all the processes by which it gains heat and cools as it evolves, including adiabatic heating and cooling, shock heating, photoheating, Compton cooling off microwave background photons, and recombination cooling \citep[e.g.,][]{1997MNRAS.292...27H}.
Instructive intuition can however be gained from idealized solutions.\\ \\
In the limit of early HI reionization (with the reionization of HeI assumed to proceed simultaneously), the temperature at mean density $T_{0}$ reaches a ``thermal asymptote'' determined by the competition between adiabatic cooling and photoheating and whose value depends on the HeII ionization state.
For a power-law background spectrum $J_{\nu}\propto \nu^{-\alpha_{\rm bg}}$ just above the ionization edges, a good approximation to the thermal asymptote is given by
\begin{equation}
\label{thermal asymptote power law}
T_{0}^{\rm asymp}(z)=2.49\times10^{4}{\rm~K} (0.464 + 0.536 y_{\rm III}) (2 + \alpha_{\rm bg})^{-1/1.7}
\left(
\frac{1+z}{4.9}
\right)^{0.53}
\end{equation}
\citep[][]{2003ApJ...596....9H}.
To first order and ignoring inhomogeneities, the effect of HeII reionization is to inject additional heat to each gas parcel.
As the universe expands, the extra heat is diluted by adiabatic cooling, $T(z)=T(z')[(1+z)/(1+z')]^{2}$, so that an estimate of the overall temperature evolution in the early HI reionization limit accounting for HeII reionization heat input is given by
\begin{equation}
\label{thermal history approx}
T(z) \approx T_{0}^{\rm assymp}(z) + 
\int_{\infty}^{z}
dz'
\frac{d\Delta T_{\rm HeII}(z')}{dz'}
\left(\frac{1+z}{1+z'}\right)^{2}.
\end{equation}\\ \\
Figure \ref{heii thermal history} shows thermal histories calculated using this equation assuming a background spectral index $\alpha_{\rm bg}=0$ and the \cite{2007ApJ...654..731H} quasar luminosity function in the $B-$band for different spectral indices of the HeII ionizing sources $\alpha_{\rm UV}$.
Harder spectral indices are seen to result in greater heat injections, which simply owes to the larger fraction of ionizations caused by high-energy photons.
Moreover, the magnitude of the total heat input as a function of spectral index is consistent with the simpler estimates using a sharp frequency cutoff shown in Figure \ref{heii heating}.
At fixed $B-$band luminosity, harder spectral indices result in higher ionizing photon output rates and thus earlier HeII reionization.\\ \\
The HeIII fraction $y_{\rm III}$ in the above equations is obtained by counting the number of HeII ionizing photons emitted by quasars as in \cite{2008ApJ...688...85F} and we have assumed a gas clumping factor $C=5$.

\subsubsection{Scatter in the Temperature-Density Relation}
\label{temperature density relation scatter}
\begin{figure}[ht]
\begin{center} 
\includegraphics[width=1.0\textwidth]{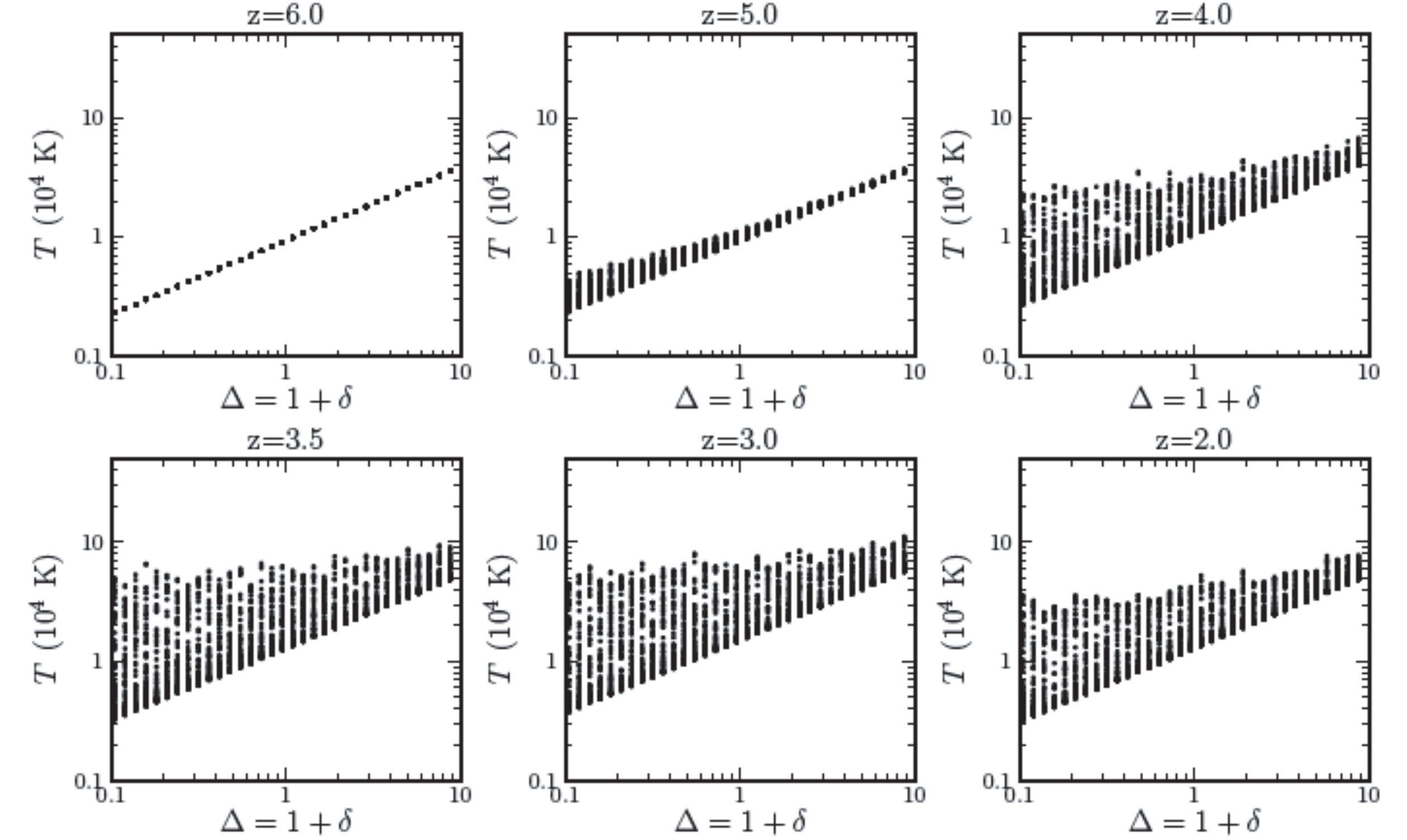} 
\end{center} 
\caption{Evolution of the temperature-density relation in the model of \S \ref{temperature density relation scatter} for the case of a quasar spectral index $\alpha_{\rm UV}=1.5$, with $\alpha_{\rm bg}=0$.
The mean temperature at mean density traces the corresponding curve in Figure \ref{heii thermal history} by construction.
The slope and scatter of the temperature-density relation are however manifest.
The underlying slope is set by the early HI reionization limit, $T=T_{0}\Delta^{0.62}$, but the HeII reionization heat input somewhat flattens the mean slope.
The flattening arises because at low densities that are optically thin to the high-energy photons the heat deposition is density-independent in the sense that gas parcels of different densities are heated by the same amount.
Since the lower-density elements are initially cooler, the fractional temperature increase is larger for these.
HeII reionization does not produce an isothermal temperature-density relation in this model because the initial heat in gas parcels above mean density is significant compared to the HeII reionization heat injection and so the imprint of this initial heat is not erased.
}
\label{t delta relation} 
\end{figure}

The thermal history calculations of the previous section implicitly assumed that the universe is homogeneous at a mean density and that HeII reionization happens simultaneously throughout.
In reality, the IGM is characterized by density fluctuations and the quasars that putatively drive HeII reionization turn on at different times at different locations owing to cosmic variance.
These inhomogeneities imply that the IGM temperature is not fully described by a single redshift dependent number $T(z)$ but in reality exhibits a temperature-density relation $T(z;~\Delta)$ with some scatter about the mean at each redshift.
\\ \\
In the absence of HeII reionization, \cite{1997MNRAS.292...27H} showed that the temperature-density relation at $z=2-4$ is well approximated by a power law $T(z;~\Delta)=T_{0}\Delta^{\beta}$.
In the limit of early HI reionization, $\beta \to 0.62$ as a result of the competition between photoheating and adiabatic cooling.
We are interested in how this result is modified by HeII reionization.
Equation \ref{thermal history approx} can be generalized to
\begin{equation}
T(z) \approx T_{0}^{\rm assymp}(z) \Delta^{\beta} + 
\kappa \int_{\infty}^{z}
dz' \frac{d\Delta T_{\rm HeII}(z')}{dz'}
\left(\frac{1+z}{1+z'}\right)^{2},
\end{equation}
where $\beta$ is set to the value that would be obtained without HeII reionization and $\kappa$ is a stochastic factor that accounts for the fact that different regions are heated at different times by HeII reionization photoheating.
Our task is then reduced to determining the distribution function of $\kappa$ to estimate the scatter in the temperature-density relation.\\ \\
One of the results highlighted by the radiative transfer simulations of \cite{2008arXiv0807.2799M} is that much of the heating during HeII reionization by quasars results from ionizations by the diffuse background of high-energy photons with large mean free paths that penetrate into HeII patches before these are actually reionized by softer photons \citep[for a different picture, see][]{2008arXiv0807.2447B}.
In this picture, the longer a given region is exposed to the high-energy background before it is reionized, the more heat it receives; regions that are reionized last tend to be hotter.
As an ansatz, again motivated by the work of \cite{2008arXiv0807.2799M}, we may thus posit that $\kappa \propto t_{\rm exp,eff}$, where $t_{\rm exp,eff}$ is an effective exposure time to the high-energy background.
Note, though, that this will not be correct at the very beginning of HeII reionization before the background has had time to diffuse.
We denote by $z_{\rm HeII}$ the redshift at which a given gas parcel is reionized in HeII and set
\begin{equation}
\label{effective exposure time ansatz}
t_{\rm exp,eff}(z_{\rm HeII}) \equiv \int_{\infty}^{z_{\rm HeII}}dt y_{\rm III}(z)(1 - y_{\rm III}(z)).
\end{equation}
The effective exposure time is thus the age of the universe at reionization of the gas parcel, weighted by the time-dependent ionized fraction, and saturating as the latter reaches order unity.
The motivation for the weighting is that the heat injection is not only proportional to the raw exposure time, but also to the intensity of the high-energy background.
The ionized fraction $y_{\rm III}$ counts the number of ionizing photons emitted and is therefore a tracer of this high-energy background.
The $1-y_{\rm III}$ saturation factor approximates the fact the rate of heat input also scales with $n_{\rm HeII}$ and is thus suppressed toward the end of reionization.\\ \\
The PDF of reionization redshifts is also straightforwardly approximated from the ionized fraction evolution since the probability of reionization during a redshift interval scales as the rate at which ionizations occur at that time:
\begin{equation}
P(z_{\rm HeII};~z) = 
\left\{
\begin{array}{cl}
y_{\rm III}(z_{\rm HeII})^{-1} \frac{dy_{\rm III}}{dz}(z_{\rm HeII}) & z_{\rm HeII} \geq z \\
0 & z_{\rm HeII}<z
\end{array}
\right.
.
\end{equation}
Energy conservation requires $\langle \kappa \rangle=1$, so we set $\kappa = t_{\rm exp,eff}/\langle t_{\rm exp,eff} \rangle$ and in this model the scatter in the temperature-density relation ultimately is calculable from the quasar luminosity function.
We wish to emphasize that the effective exposure time ansatz in equation \ref{effective exposure time ansatz} was obtained heuristically and certainly does not capture the full complexity of HeII reionization, though it does agree reasonably well with the results of the \cite{2008arXiv0807.2799M} simulations and provides a simple way to understand them.
\\ \\
Figure \ref{t delta relation} shows the evolution of the temperature-density relation in this model for the case of a quasar spectral index $\alpha_{\rm UV}=1.5$, with $\alpha_{\rm bg}=0$.
The mean temperature at mean density, $\langle T(z;~\Delta=1) \rangle$, then traces the corresponding curve in Figure \ref{heii thermal history} by construction.
The slope and scatter of the temperature-density relation are however manifest.
The underlying slope is set by the early HI reionization limit $T=T_{0}^{0.62}$, but the HeII reionization heat input somewhat flattens the mean slope.
The flattening arises because at low densities that are optically thin to the high-energy photons the heat deposition is density-independent in the sense that gas parcels of different densities receive the same temperature increment.
Since the lower-density elements are initially cooler, the fractional temperature increase is larger for these.
In agreement with the simulations of \cite{2008arXiv0807.2799M}, HeII reionization does not produce an isothermal ($\beta=0$) temperature-density relation.
This is, similarly, simply because the initial heat in gas parcels above mean density ($\Delta>1$) is significant compared to the HeII reionization heat injection and so the imprint of this initial heat is not erased.
Finally, it is worth reemphasizing that this model predicts a large scatter in the temperature-density relation, which may have important consequence for interpreting \Lya~forest data.
In this model, though, the scatter arises from the scatter in the HeII reionization times for different gas parcels.
Locally, neighboring points will be reionized at similar times and we therefore expect the scatter in the temperature-density relation to be significantly reduced.

\begin{deluxetable}{ccccccccc}
\tablewidth{0pc}
\tablecaption{Photoionization and Photoheating Rates for our Fiducial Model\label{rates table}}
\tabletypesize{\footnotesize}
\tablehead{\colhead{$z$} & \colhead{$\Gamma_{\rm HI}$\tablenotemark{a}} & \colhead{$\Gamma_{\rm HeI}$\tablenotemark{a}} & \colhead{$\Gamma_{\rm HeII}$\tablenotemark{a}} & \colhead{$\dot{q}_{\rm HI}$\tablenotemark{a}} & \colhead{$\dot{q}_{\rm HeI}$\tablenotemark{a}} & \colhead{$\dot{q}_{\rm HeII}$\tablenotemark{a}} & \colhead{$\Delta T_{\rm HeII}$\tablenotemark{b} } \\ 
\colhead{} & \colhead{10$^{-12}$ s$^-1$} & \colhead{10$^{-12}$ s$^-1$} & \colhead{10$^{-12}$ s$^-1$} & \colhead{10$^{-12}$ eV s$^-1$} & \colhead{10$^{-12}$ eV s$^-1$} & \colhead{10$^{-12}$ eV s$^-1$} & \colhead{$T$ (K)}}
\startdata

0.0  & 0.0384  & 0.0213  & $1.231\times10^{-4}$  & 0.158  & 0.141  & 0.0032                & 14269 \\
0.25 & 0.0728  & 0.0443  & $2.956\times10^{-4}$  & 0.311  & 0.299  & 0.0073                & 14269 \\
0.5  & 0.1295  & 0.0860  & $6.845\times10^{-4}$  & 0.569  & 0.600  & 0.0156                & 14269 \\
0.75 & 0.2082  & 0.1471  & $1.361\times10^{-3}$  & 0.929  & 1.064  & 0.0292                & 14269 \\
1.0  & 0.3048  & 0.2241  & $2.317\times10^{-3}$  & 1.371  & 1.677  & 0.0476                & 14269 \\
1.25 & 0.4074  & 0.3076  & $3.401\times10^{-3}$  & 1.841  & 2.371  & 0.0676                & 14269 \\
1.5  & 0.4975  & 0.3843  & $4.288\times10^{-3}$  & 2.260  & 3.037  & 0.0837                & 14269 \\
1.75 & 0.5630  & 0.4446  & $4.744\times10^{-3}$  & 2.574  & 3.579  & 0.0920                & 14269 \\
2.0  & 0.6013  & 0.4856  & $4.811\times10^{-3}$  & 2.768  & 3.974  & 0.0926                & 14269 \\
2.25 & 0.6142  & 0.5076  & $4.511\times10^{-3}$  & 2.852  & 4.196  & 0.0867                & 14269 \\
2.5  & 0.6053  & 0.5132  & $3.939\times10^{-3}$  & 2.839  & 4.274  & 0.0757                & 14269 \\
2.75 & 0.5823  & 0.5074  & $3.223\times10^{-3}$  & 2.762  & 4.246  & 0.0622                & 14269 \\
3.0  & 0.5503  & 0.4942  & $2.479\times10^{-3}$  & 2.642  & 4.161  & 0.0480                & 14269 \\
3.25 & 0.5168  & 0.4781  & $1.812\times10^{-3}$  & 2.511  & 4.047  & 0.0351                & 11392 \\
3.5  & 0.4849  & 0.4617  & $1.245\times10^{-3}$  & 2.384  & 3.933  & 0.0243                & 8007 \\
3.75 & 0.4560  & 0.4469  & $7.907\times10^{-4}$  & 2.272  & 3.830  & 0.0159                & 5519 \\
4.0  & 0.4320  & 0.4329  & $4.818\times10^{-4}$  & 2.171  & 3.737  & $9.862\times10^{-3}$  & 3802 \\
4.25 & 0.4105  & 0.4203  & $2.618\times10^{-4}$  & 2.083  & 3.657  & $5.639\times10^{-3}$  & 2619 \\
4.5  & 0.3917  & 0.4080  & $1.351\times10^{-4}$  & 2.002  & 3.580  & $3.133\times10^{-3}$  & 1770 \\
4.75 & 0.3743  & 0.3948  & $7.271\times10^{-5}$  & 1.921  & 3.493  & $1.624\times10^{-3}$  & 1170 \\
5.0  & 0.3555  & 0.3800  & $3.724\times10^{-5}$  & 1.833  & 3.393  & $8.087\times10^{-4}$  & 713 \\
5.25 & 0.3362  & 0.3647  & $2.060\times10^{-5}$  & 1.745  & 3.284  & $3.654\times10^{-4}$  & 394 \\
5.5  & 0.3169  & 0.3494  & $8.924\times10^{-6}$  & 1.661  & 3.169  & $1.448\times10^{-4}$  & 210 \\
5.75 & 0.3001  & 0.3327  & $5.992\times10^{-6}$  & 1.573  & 3.042  & $6.795\times10^{-5}$  & 112 \\
6.0  & 0.2824  & 0.3160  & $4.033\times10^{-6}$  & 1.487  & 2.906  & $3.461\times10^{-5}$  & 58 \\
6.25 & 0.2633  & 0.2987  & $1.657\times10^{-6}$  & 1.399  & 2.762  & $1.381\times10^{-5}$  & 30 \\
6.5  & 0.2447  & 0.2801  & $9.978\times10^{-7}$  & 1.305  & 2.606  & $6.409\times10^{-6}$  & 15 \\
6.75 & 0.2271  & 0.2620  & $5.898\times10^{-7}$  & 1.216  & 2.450  & $3.022\times10^{-6}$  & 8 \\
7.0  & 0.2099  & 0.2439  & $3.430\times10^{-7}$  & 1.127  & 2.292  & $1.418\times10^{-6}$  & 4 \\

\enddata
\tablenotetext{a}{Optically thin rates.}
\tablenotetext{b}{Cummulative temperature increase owing to HeII reionization, which should be incrementally added to gas elements and let cooling to model the heat input (\S \ref{hydro implementation}).}
\end{deluxetable}

\subsubsection{Implementation in Hydrodynamical Codes}
\label{hydro implementation}
Cosmological hydrodynamical simulations usually incorporate the effects of a prescribed UV background on the thermal history of the gas under the assumption of an optically thin plasma \citep[for the relevant equations, see][]{1996ApJS..105...19K}.
This assumption manifestly breaks down during HeII reionization and has led simulators to artificially increase photoheating rates as a rough approximation of the radiative transfer effects.
A more physically motivated approach is to increase the temperature of each gas element by an amount $(d\Delta T_{\rm HeII}(z)/dz) \Delta z$ (which is subsequently let cooling) at each time step $\Delta z$ in the simulation, where $\Delta T_{\rm HeII}(z)$ is pre-computed given the desired HeII reionization history as in equation \ref{Delta T HeII}.
This approach misses the scatter and inhomogeneity of the temperature-density relation discussed in the previous section but has the advantage of only requiring an additional term in the temperature equation and adding negligible computational overhead while capturing the time scale and magnitude of the heat input more realistically.
It could conceivably be extended to account for spatial inhomogeneities and the scatter in reionization times similarly to semi-analytic models being applied to efficiently model HI reionization \citep[e.g.,][]{2007ApJ...654...12Z}.
An alternative approach would be to replace the optically thin photoionization and heating rates by ``effective'' values that can be substituted into the usual optically thin equations to yield the desired result.\\ \\
Table \ref{rates table} tabulates both the optically thin photoionization and photoheating rates, and the $\Delta T_{\rm HeII}$ values to use to model the effects of HeII reionization under our prescription for the quasar model with $\alpha_{\rm QSO}=1.6$ employed in the fiducial background spectrum calculations of \S \ref{calculations}.
In general, the photoheating rate for species $i$ is given by
\begin{equation}
\dot{q}_{i} = 
\frac{1}{4\pi}
\int_{\nu_{i}}^{\infty}
\frac{d\nu}{h \nu}
J_{\nu} \sigma_{i}(\nu) (h\nu - h\nu_{i}).
\end{equation}
The corresponding ionized fraction versus redshift is shown by the $C=5$ curve in Fig. 7 of \cite{2008ApJ...688...85F}.
In this model, HeII reionization completes by $z=3$, with $\sim80$\% of the ionization occurring between $z=3$ and $z=4$.

\begin{figure}[ht]
\begin{center} 
\includegraphics[width=0.65\textwidth]{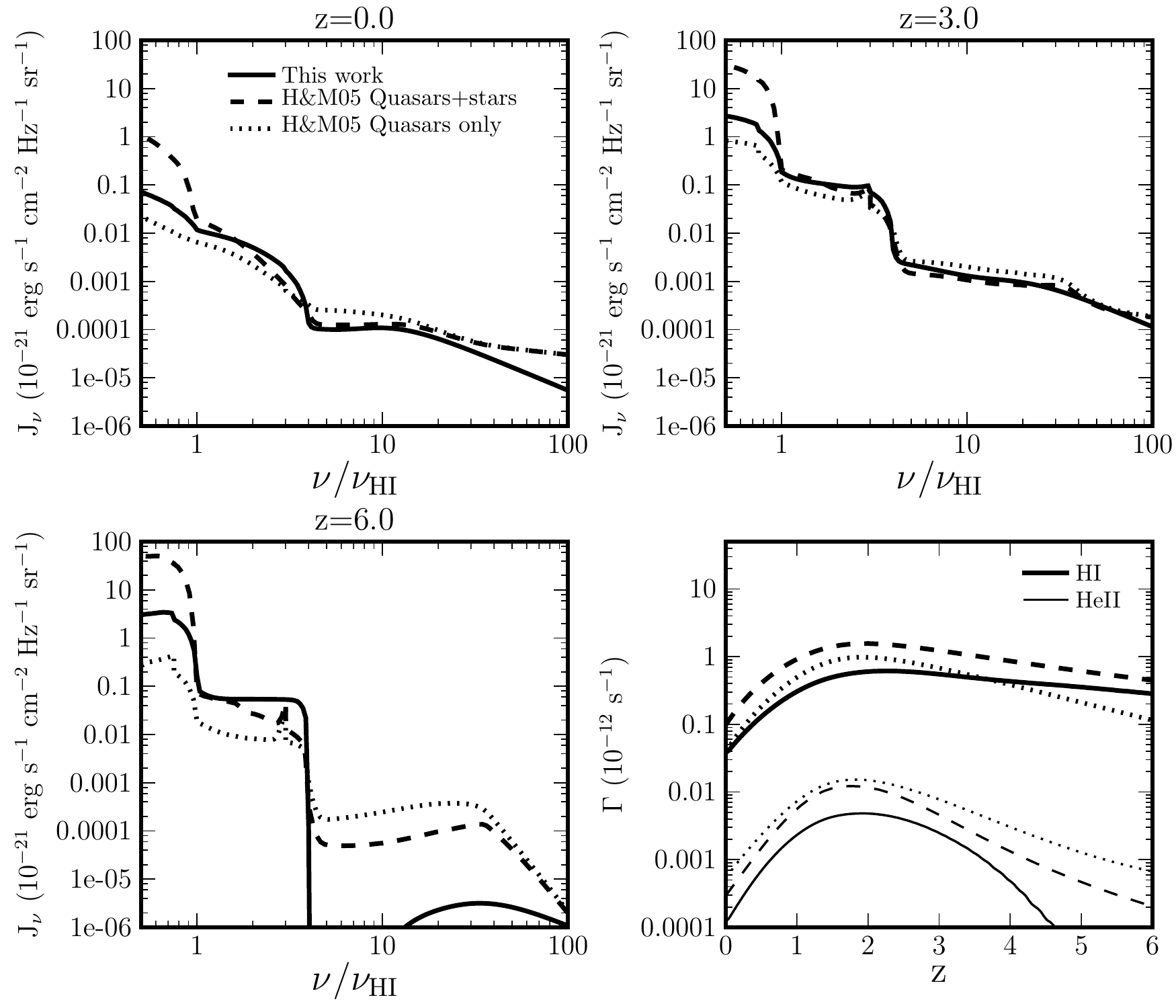}
\end{center} 
\caption{Comparison of our fiducial ionizing background model with models informally released by Haardt \& Madau in 2005 (H\&M05).
Both H\&M05 models include a quasar contribution based on the \cite{2004MNRAS.349.1397C} luminosity function.
One model in addition includes a stellar contribution calculated using a \cite{2003MNRAS.344.1000B} population synthesis code assuming a Salpeter initial mass function, age 0.5 Gyr, constant star formation, and an escape fraction of 10\%.
\emph{First three panels:} Detailed spectra at redshifts $z=0,~3,~{\rm and}~6$.
For this comparison, the H\&M05 models were normalized by a factor of 0.5 to better match our fiducial calculation.
\emph{Lower right:} Photoionization rates of HI and HeII versus redshift.
Here, the H\&M05 models were \emph{not} normalized and so the different amplitudes reflect the difference in the models as provided.
}
\label{HM05 comparison} 
\end{figure}

\section{COMPARISON WITH PREVIOUS WORK}
\label{comparison}
As one of the motivations for our calculation of the ionizing background spectrum was to provide an alternative to the widely-used models of \cite{1996ApJ...461...20H} \citep[see also][]{2001cghr.confE..64H} and their informally-released derivatives, it is useful to directly compare our results with these authors.\\ \\
Before we do so, we wish to emphasize that in this work we attempted to improve on technical aspects of the calculation.
In particular, we performed more self-consistent calculations of the ionization structure of individual absorbers (\S \ref{individual absorbers}) and our treatment of recombination emission (\S \ref{recombinations}) was based on approximating the results of these photoionization calculations rather than on an escape probability formalism.
Moreover, all the numerical calculations presented here were performed using an independently-developed code and our empirical constraints (\S \ref{calculations}) were also obtained independently in our previous work.
The comparison of the final results against those of Haardt \& Madau thus provides an indication of the uncertainty in the resulting spectrum.\\ \\
In Figure \ref{HM05 comparison}, we compare our fiducial model (ignoring HeII reionization) with two models informally released by Haardt \& Madau in 2005 (H\&M05; F. Haardt 2005, private communication).
Both models include a quasar contribution based on the \cite{2004MNRAS.349.1397C} luminosity function.
One model in addition includes a stellar contribution calculated using a \cite{2003MNRAS.344.1000B} population synthesis code assuming a Salpeter initial mass function, age 0.5 Gyr, constant star formation, and an escape fraction of 10\%.
We compare both the detailed spectra at redshifts $z=0,~3,~\rm{and}~6$ and the integrated photoionization rates of HI and HeII versus redshift.
For the detailed spectra, the Haardt \& Madau models were normalized by a factor of 0.5 to better match our fiducial calculations in the ionizing range.
As a non-negligible uncertainty remains in the amplitude of the intergalactic HI photoionization rate \citep[e.g.,][]{2005MNRAS.357.1178B, 2008ApJ...688...85F}, it is fair to renormalize the models before comparing them, a procedure which is also frequently used by simulators to match the observed \Lya~forest mean transmission (see the discussion at the end of \S \ref{spectra results}).
The photoionization rates shown in the last panel have however \emph{not} been renormalized and so reflect the models as provided.\\ \\
It is interesting that in spite of the differences in the technical treatment and the independently obtained empirical constraints, our calculations of the spectral shape in the ionizing range agree quite well with the H\&M05 models between $z=0$ and $z=3$, suggesting that the calculations are relatively robust in this redshift range.
The spectra however diverge increasingly toward higher redshifts as a result of the different source prescriptions this regime.
In our model, the quasar contribution drops more rapidly as $z\to\infty$, while the compensating stellar emissivity increases to maintain the nearly flat total hydrogen photoionization rate measured from the \Lya~forest \citep[][]{2008ApJ...682L...9F, 2008ApJ...688...85F}.
While the \cite{2007ApJ...654..731H} luminosity function we use combines different data sets to constrain the faint-end slope up to $z=4.5$, the \cite{2004MNRAS.349.1397C} luminosity function used in the H\&M05 models is based solely on the 2QZ survey and is only measured up to $z=2.1$.
Our calculations are therefore more reliable at $z\gtrsim3$.
Since we prescribe the HI column density distribution, but self-consistently calculate the HeII distribution from the hardness of the background at each redshift, the decreasing HeII to HI ionizing emissivity ratio results in a reduction of the HeII to HI ionizing photon mean free path, and therefore amplifies the $\Gamma_{\rm HeII}/\Gamma_{\rm HI}$ evolution.
This explains the increasingly strong HeII break in our model.
Although HeII reionization would modify the results above the HeII ionization edge (\S \ref{spectrum during heii reion}), a prediction of our model is a larger ratio of HI to HeII ionizing flux beyond $z\approx3$.
\\ \\
A significant difference between our calculations and the original work of \cite{1996ApJ...461...20H} is with the fraction by which recombination emission boosts the photoionization rates.
For their original model, \cite{1996ApJ...461...20H} found $\Gamma_{\rm HI}^{\rm with~rec}/\Gamma_{\rm HI}^{\rm rec}$ as high as 1.5 and $\Gamma_{\rm HeII}^{\rm with~rec}/\Gamma_{\rm HeII}^{\rm rec}$ peaking at 1.7.
For our fiducial model, we found recombinations to be less important for the photoionization rates, with $\Gamma^{\rm with~rec}/\Gamma^{\rm rec}\lesssim1.1$ at $z\leq4$ for both HI and HeII (Figure \ref{gammas vs z}; the boost factor increases somewhat toward higher redshifts as the mean free paths and leakage due to redshifting below the ionization edges decrease, as discussed in \S \ref{recombination contribution}).
The analytic model developed in Appendix \ref{recombinations analytical model} helps us understand our numerical results and give us confidence in their accuracy. 
The differences with respect to the boost factors found by \cite{1996ApJ...461...20H} must partly be due to the different parameters of the calculations (e.g., the column density distribution and source prescriptions), but may also originate from the different techniques used.
Preliminary investigation suggests that \cite{1996ApJ...461...20H} may have incorrectly used a case B recombination coefficient instead of the case A coefficient in a step of their escape probability calculation, resulting in an overestimate of the recombination boost  which likely explains a large part of the discrepancy (F. Haardt 2009, private communication).
Our results agree better with those of \cite{1998AJ....115.2206F}, who find that recombinations boost the photoionization rates of both HI and HeII by $\sim20$\%.\\ \\
Recently, \cite{2008arXiv0812.0824M} proposed a new effect on the spectrum of the ionizing background which could be important particularly before HeII is reionized.
When a large amount of HeII is present, the opacity arising from HeII Ly$\beta$ and higher Lyman-series resonances produces a sawtooth absorption pattern between 3.56 and 4 Ryd and the HeII \Lya~reemission line at 3 Ryd is boosted by resulting degraded photons.
We do not include this effect in the present work but plan to do so in the future.

\section{DISCUSSION AND CONCLUSION}
\label{discussion}
In this work, we have revisited the calculation of the UV background spectrum.
The three main improvements over previous work are: 
\begin{itemize}
\item The implementation of new empirical constraints on the sources of radiation based on a detailed study of intergalactic absorption and updated luminosity functions \citep[][]{2008ApJ...681..831F, 2008ApJ...682L...9F, 2008ApJ...688...85F}.
In our favored fiducial model, star-forming galaxies play a crucial role and a dominate the HI photoionization rate at $z\gtrsim3$.
\item A reexamination of the radiative transfer within individual absorbers and an exploration of the physical dependences of the calculated background. 
In particular, we perform more self-consistent photoionization calculations including recombination emission, present a new treatment of recombination emission based on them, and clarify how the net enhancement of the photoionization rates is influenced by redshifting  of the recombination photons below the ionization edges and their energy distribution.
\item A treatment of the effects of HeII reionization on background spectrum and on the thermal history of the intergalactic medium.\\
\end{itemize}
The main argument supporting a UV background dominated by stellar emission beyond $z\approx3$ is that while the total HI photoionization rate measured from the \Lya~forest is remarkably constant between at least $z=2$ and $z=4.2$ \citep[e.g.,][]{2008ApJ...682L...9F, 2008ApJ...688...85F}, the quasar luminosity function is strongly peaked near z=2 \citep[e.g.,][]{2007ApJ...654..731H}.
Thanks to large-scale Lyman break and Ly$\alpha$ line surveys \citep[e.g.,][]{1999ApJ...519....1S, 2006ApJ...642..653S, 2006ApJ...653..988Y, 2007ApJ...670..928B, 2008ApJS..175...48R, 2008ApJS..176..301O}, star-forming galaxies are now known to exist numerously at these redshifts and are therefore the leading candidates to account for the remaining ionizing photons.
The quasar luminosity function and measurements of HeII-to-HI column density ratios however indicate that quasars do contributed a large fraction of a HI photoionization rate are their $z\approx2$ peak; in our fiducial model, this fraction is 2/3. \\ \\
The evidence in favor of a large (and dominant at the highest redshifts) contribution of star-forming galaxies to the ionizing background is supported by related and independent studies.
Previous studies of the HI photoionization rate from the \Lya~forest \cite[][]{1997ApJ...489....7R, 2001ApJ...549L.151H, 2005MNRAS.357.1178B} have in fact supported this conclusion, though with somewhat more leeway owing to larger statistical error bars.
The case has also been made independently by combining direct measurements of the UV luminosity function of galaxies and of their escape fraction \citep[e.g.,][]{2001ApJ...546..665S, 2001ApJ...549L..11M, 2006ApJ...651..688S, 2008arXiv0811.1042C}.
Metal line studies provide a further line of evidence, indicating that a mix of stars and quasars best fits measured the abundance ratios of various ions including CIV, SIV, and OVI \citep[][]{2003astro.ph..7557B, 2003ApJ...596..768S, 2004ApJ...602...38A, 2007arXiv0712.1239A}.
In fact, the ionizing spectrum presented herein could be directly confronted against and further constrained by such observations.
Theoretical arguments also suggest that star-forming galaxies should dominate early on \citep[e.g.,][]{2003MNRAS.339..312S, 2003MNRAS.341.1253H, 2008arXiv0811.2222L}.\\ \\
Although the fiducial model detailed in \S \ref{calculations} fits our observational constraints, it is not at present uniquely determined.
As we explore in \S \ref{dependences}, the background depends on the details of both the sources and sinks of radiation.
For instance, we have adopted a hard spectral index $\alpha_{\star}=1$ for star-forming galaxies between 1 and 4 Ryd based on the comparison of theoretical models with observational line diagnostics by \cite{2001ApJ...556..121K}.
However, in spite of a few weak detections, the emergent spectral shape of high-redshift galaxies at these energies has yet to be directly measured owing to the large attenuation by the intervening IGM \citep[e.g.,][]{2001ApJ...546..665S, 2006ApJ...651..688S}.
Moreover, present population synthesis models are at odds with one another in their predictions in this energy range \citep[e.g.,][]{2001ApJ...556..121K, 2003astro.ph..7557B}, making it difficult to rely on them with confidence.
Perhaps the largest uncertainty with respect to the absorbers is the abundance and redshift evolution of Lyman limit systems, which determine the mean free path of ionizing photons.
These are still poorly constrained, especially above $z\approx4$, and introduce a commensurate uncertainty in the calculation of the amplitude of the ionizing background at these redshifts (e.g., Figure \ref{Gamma vs zlow gamma}).
\\ \\
The total ionizing background receives a contribution from recombinations that reemit other ionizing photons.
We reexamined the physics of this recombination contribution.
Interestingly, we find that the enhancement of the photoionization rates from recombinations is not simply a function of the number of recombination photons.
This arises because many of the recombination photons rapidly redshift and leak below their corresponding ionization edges as well as from their distribution in energy.
Focusing on the HI photoionization rate, the main factors determining $\Gamma_{\rm HI}^{\rm with~rec}/\Gamma_{\rm HI}^{\rm rec}$ are the parameters of the column density distribution and the LyC recombination line profile.
Shorter ionizing photon mean free paths, relative to the recombination line width, inhibit leakage.
Wider recombination lines, associated with higher gas temperatures, on the other hand produce higher-energy photons that receive less weight in the photoionization rate owing to the frequency-dependence of the cross section.
The steepness of the column density distribution also plays a role: for steeper distributions, more of the recombinations occur in optically thin absorbers from which essentially all the recombination photons escape into the IGM, in contrast to optically thick absorbers that trap a large fraction.
In general, $\Gamma_{\rm HI}^{\rm with~rec}/\Gamma_{\rm HI}^{\rm rec}$ increases toward higher redshifts as the mean free path diminishes and leakage becomes less significant.
In Appendix \ref{recombinations analytical model}, we give a simple analytical model that quantitatively captures these effects and agrees well with our full numerical calculations.
We conclude that there is not a unique answer for how much recombinations boost the ionization rates, but that it depends both on redshift and on the parameters of the absorbers and their distribution.\\ \\
Our main solutions to the radiative transfer problem assume that the ionizing background is homogeneous.
This approximation will in particular break down during the reionization of HI and HeII.
As HeII reionization may take place under the action of quasars at immediately accessible redshifts $z\sim3-4$ \citep[e.g.,][]{2002MNRAS.332..601S, 2003ApJ...586..693W, 2008ApJ...681....1F, 2008ApJ...688...85F}, we provide a discussion of its effects on both the spectrum of the ionizing background and on the thermal history of the IGM.
In regions that have yet to be reionized, the spectrum is expected to be almost completely suppressed immediately above 54.4 eV by HeII absorption.
However, the universe remains relatively transparent at higher energies owing the frequency-dependence of the photoionization cross section.
As the spectrum recovers, a background of $\gtrsim0.5$ keV photons should thus permeate the entire universe.\\ \\
Another important effect of HeII reionization is to inject heat into the IGM via photoheating.
We provide a simple analytical model to estimate the overall temperature increase owing to HeII reionization based on energy conservation.
In this model, which agrees well with the 3-D radiative transfer simulations of \cite{2008arXiv0807.2799M}, the total temperature increase depends most sensitively on the far UV quasar spectral index.
For a value $\alpha_{\rm UV}=1.5$, the temperature increase could be as much as $15,000$ K, though the effect is mitigated by simultaneous adiabatic cooling.
Harder spectral energy distributions lead to more energy injection.
The model is extended to understand the effects on the temperature-density relation under the ansatz that the local heat input scales with the effective exposure to the high-energy background.
The main effects are to flatten the mean temperature-density relation with respect to the early HI reionization limit \citep[][]{1997MNRAS.292...27H} and introduce a large scatter.
The flattening arises because the high-energy background heat deposition per atom is independent of the local density: the temperature of initially cooler lower-density regions is increased by the same amount as that of initially hotter ones of higher density.
The temperature-density relation does not become fully isothermal as the HeII reionization heat input is comparable to the initial thermal energy of the gas and so the memory of the latter is not erased.
In our model, the scatter of the temperature-density relation originates from the scatter in the reionization times of different gas elements and can be calculated from the quasar luminosity function.
\\ \\
What remains to be done?
As outlined above, the ionizing background spectrum is not yet uniquely determined and further direct constraints (for example, using metal line ratios) as well as constraints on its sources and sinks (which enter the theoretical calculation) are sure to continue to refine it, especially at the highest redshifts.
The theoretical framework itself needs to be improved to take into account the fluctuations in the ionizing background that necessarily exist at some level and are certain to be important at least during reionization events \citep[for work in this direction, see][]{2006MNRAS.366.1378B, 2008arXiv0812.3411F, 2009arXiv0901.2584F}.
In this vein, while our idealized discussion of the effects of HeII reionization provides some basic physical understanding, it leaves ample opportunity for improvement.
This a particularly exciting area for future progress as studies of HeII reionization are currently blossoming, with much HI~\Lya~forest data already available and new HeII~\Lya~forest lines of sight that the \emph{Cosmic Origins Spectrograph} \cite[][]{2000SPIE.4013..352G} to be installed aboard the \emph{Hubble Space Telescope} is poised to deliver soon \cite[][]{2008arXiv0809.0765S}, and the accompanying surge of theoretical interest.
In particular, several groups are now beginning to numerically tackle the full problem of 3-D radiative transfer and its thermal effects \citep[][]{2002MNRAS.332..601S, 2007arXiv0711.1904P, 2008arXiv0807.2799M, 2008arXiv0811.0315M}.
Another interesting observational opportunity for probing the extragalactic radiation background is provided by the newly-launched \emph{Fermi Gamma-Ray Space Telescope}.
Indeed, the intervening extragalactic background light should attenuate the $\gamma-$rays from distant sources through electron-positron pair production and thus give us an additional handle on it, particularly below the HI ionization edge \citep[e.g.,][]{1996ApJ...456..124M, 2001ApJ...553...25O, 2008arXiv0807.4294R}. \\ \\
To facilitate use and extension of the results presented in this work and their comparison with observations, numerical tables are available in electronic form on the web at http://www.cfa.harvard.edu/$\sim$cgiguere/uvbkg.html.
 
\acknowledgements
We thank Hy Trac for useful discussions, Francesco Haardt for comments that helped us clarify the manuscript, and the referee for a thorough and careful review.
We are grateful to Matt McQuinn for physical insights in the effects of HeII reionization on the thermal history of the IGM.
CAFG is supported by a NSERC Postgraduate Fellowship and the Canadian Space Agency. 
This work was supported in part by NSF grants ACI 96-19019, AST 00-71019, AST 02-06299, AST 03-07690, and AST 05-06556, and NASA ATP grants NAG5-12140, NAG5-13292, NAG5-13381, and NNG-05GJ40G. 
Further support was provided by the David and Lucile Packard, the Alfred P. Sloan, and the John D. and Catherine T. MacArthur Foundations.

\appendix
\section{A. PHOTOIONIZATION CALCULATIONS}
\label{photoionization appendix}
\begin{figure}[ht] 
\begin{center} 
\includegraphics[width=0.45\textwidth]{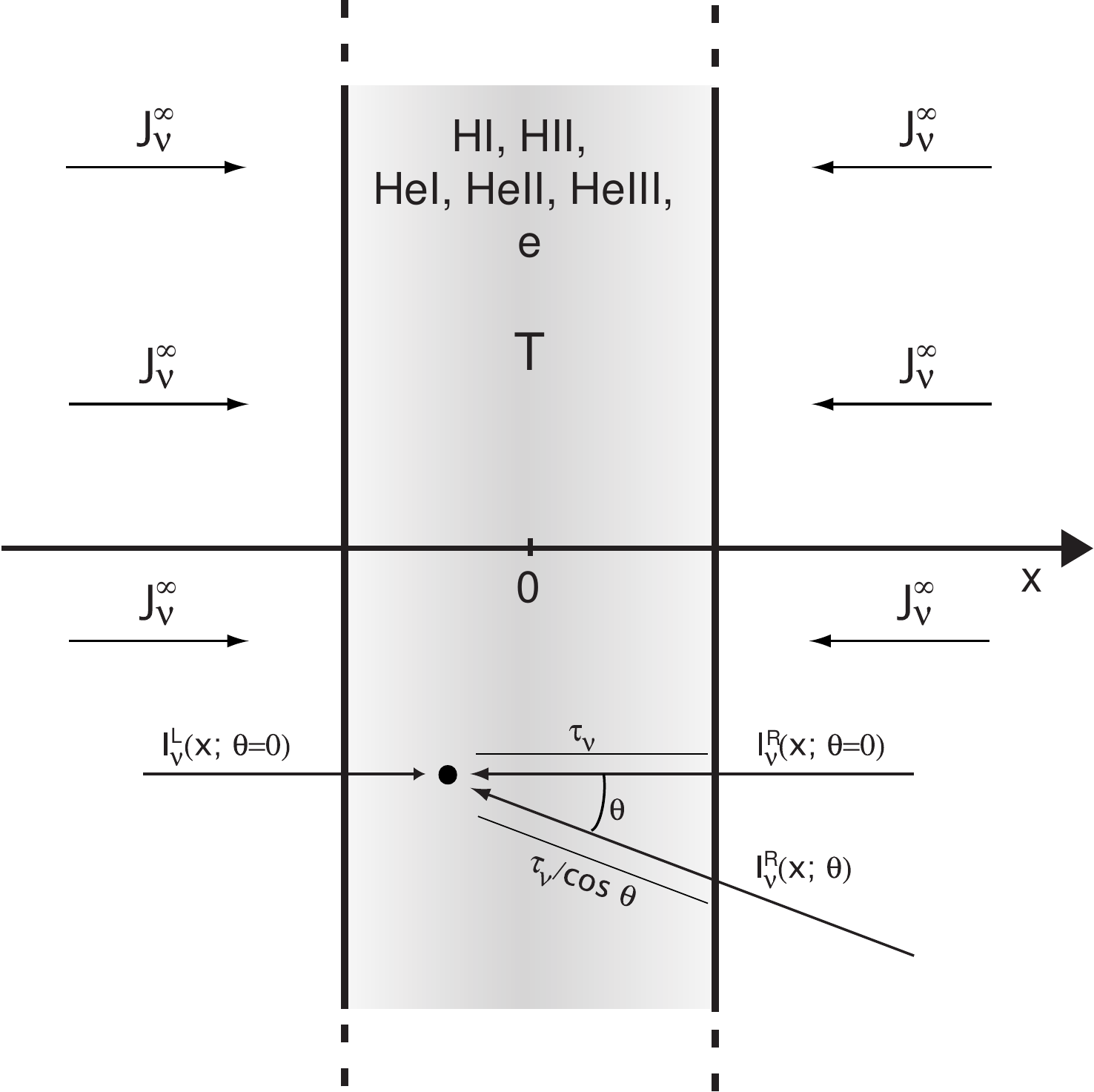} 
\end{center} 
\caption{Definition of the geometry for our photoionization calculations.
Absorbers are modeled as sheets infinite in extent, but finite in thickness.
The ionization state of the ions of hydrogen and helium are calculated for a prescribed gas temperature $T$ and isotropic external radiation field of specific intensity $J_{\nu}^{\infty}$.} 
\label{slab geometry} 
\end{figure}

In this Appendix, we detail the code used to calculate the photoionization equilibrium structure of individual cosmic absorbers outlined in \S \ref{individual absorbers}.\\ \\
We approximate the absorbers as sheets infinite in extent but finite in thickness, with geometry defined in Figure \ref{slab geometry}.
Although this geometry is restrictive, the calculation is otherwise three-dimensional in the sense that it takes into account that rays incident at different angles encounter different optical depths.
This geometry is clearly adequate for the sheets of the cosmic web, but somewhat inexact for the filamentary and clumpy structures, and the results may therefore be off by a corresponding geometrical factor.
Nevertheless, the calculations retain the essence of the problem and significantly improve over previous work that either assumed a semi-infinite geometry and an escape probability formalism \citep[][]{1996ApJ...461...20H} or gray cross sections \citep[][]{1998AJ....115.2206F}.
We also for the first time self-consistently treat the coupling between hydrogen and helium arising from their recombination emission, as explained below.
A more accurate approach would consider a distribution of three-dimensional absorber geometries (e.g., obtained from a cosmological simulation) for each line-of-sight optical depth considered, but would be much more involved and is beyond the scope of this work.\\ \\
We assume that the slab is composed of hydrogen and helium, with cosmic mass fractions $X=0.75$ and $Y=0.25$ \citep[e.g.,][]{2001ApJ...552L...1B}.
The temperature of the gas is set to $T=2\times10^{4}$ K, as estimated for optically thin \Lya~forest absorbers \citep[][see also Schaye et al. 2000\nocite{2000MNRAS.318..817S} and Ricotti et al. 2000\nocite{2000ApJ...534...41R}]{2001ApJ...562...52M, 2001ApJ...557..519Z}.
While the gas temperature may differ and be more complex in structure in damped Ly$\alpha$ absorbers (DLA) with $N_{\rm HI}\geq2\times10^{20}$ cm$^{-2}$ that are able to cool and form stars \citep[e.g.,][]{2005ARA&A..43..861W}, the detailed properties of these absorbers are not crucial since absorbers with $N_{\rm HI}\gg10^{17.2}$ cm$^{-2}$ are completely opaque to ionizing photons regardless of their exact column density.
The thickness of the slab is assumed to be equal to its Jeans length $L_{J}=\sqrt{\pi \gamma k T/G \rho \mu m_{p}}$,\footnote{Here, $\gamma$ is the adiabatic index and $\mu$ the mean molecular weight of the gas. These are taken to be 5/3 and 0.59, respectively, corresponding to a monatomic and fully ionized gas of cosmic composition.} which is both theoretically motivated and provides a good match to observations \citep[][]{2001ApJ...559..507S}.
In practice, we prescribe the physical mass density $\rho$ and derive the ionic column densities in photoionization equilibrium so that we do not need to explicitly assume a relation between $N_{\rm HI}$ and $n_{\rm H}$.
Because pressure also smooths the gas on this scale, the absorbers are assumed to be uniform in density, but we have explored other density profiles and found that our numerical results are only marginally affected, and our broad conclusions unaltered.
\\ \\
Specifically, we solve the following set of equilibrium equations under external illumination from both sides by an isotropic radiation field of specific intensity $J_{\nu}^{\infty}$ equal to the cosmological $J_{\nu}$ at each point in the absorber:
\begin{equation}
\label{HI photoeq}
\alpha_{\rm HI}(T)n_{\rm HII}n_{\rm e}=\Gamma_{\rm HI}n_{\rm HI},
\end{equation}
\begin{equation}
\label{HeI photoeq}
\alpha_{\rm HeI}(T)n_{\rm HeII}n_{\rm e}=\Gamma_{\rm HeI}n_{\rm HeI},{\rm~and}
\end{equation}
\begin{equation}
\label{HeII photoeq}
\alpha_{\rm HeII}(T)n_{\rm HeIII}n_{\rm e}=\Gamma_{\rm HeII}n_{\rm HeII}
\end{equation}
subject to the constraints $n_{\rm H}=n_{\rm HI}+n_{\rm HII}$, $n_{\rm He}=n_{\rm HeI}+n_{\rm HeII}+n_{\rm HeIII}$, and $n_{\rm e}=n_{\rm HII}+n_{\rm HeII}+2n_{\rm HeIII}$.
The photoionization equilibrium assumption is generally accurate as the ionization timescale is much smaller than the Hubble time and collisional processes are negligible at the densities and temperatures considered \citep[][]{1996ApJ...461...20H}.\\ \\
To close the system of equations \ref{HI photoeq}$-$\ref{HeII photoeq}, we must specify how to calculate the photoionization rates.
Let $J_{\nu}=(1/4\pi)\int d\Omega I_{\nu}(\theta)$ be the angle average of the total specific intensity at any given point.
Then the local photoionization rate for species $i\in \{\rm{HI, HeI, HeII}\}$ is given by
\begin{equation}
\Gamma_{i}(x)= 4\pi
\int_{\nu_{i}}^{\infty}
\frac{d\nu}{h\nu}
\sigma_{i}(\nu)J_{\nu}(x).
\end{equation}
The specific intensity along a particular ray will in general depend on the angle $\theta$ of incidence 
inside the slab because different rays encounter different optical depths.
We take the total specific intensity at any given point to be the sum of the radiation originating from the external background and of the radiation from the recombination processes to HI and HeII (HI LyC, HeII BalC, HeII \Lya, and HeII LyC) within the absorber calculated as in \S \ref{recombinations} and Appendix \ref{recombination emission appendix}.
For this purpose, the recombination line photons are treated as $\delta-$functions, which is a good approximation within individual absorbers.
Each component is attenuated in magnitude depending on the optical depth from its source following the usual transfer equation.\\ \\
In particular, our approach explicitly takes into account the coupling between different species arising from the reabsorption of recombination photons \emph{by a different species}.
For example, ionizing HeII LyC recombination photons will in general not only be reabsorbed by HeII, but also by HI and HeI.
Because of this explicit treatment of recombinations, case A recombination coefficients are appropriate in equations \ref{HI photoeq}$-$\ref{HeII photoeq}.
As HeI is found to play a negligible role in our calculations, we do not treat its recombination processes in detail, but we do approximate their effect by using the case B recombination coefficient in the optically thick regime.\\ \\
Our calculations use 100 spatial bins, 100 logarithmically-spaced frequency bins, and 20 angular bins covering $0\leq \theta<\pi \leq2$. 
The equations are solved iteratively until convergence to better than one part in $10^{3}$ is attained at each point.

\section{B. RECOMBINATION EMISSION}
\label{recombination emission appendix}
Here we provide the technical details of our treatment of recombination emission from individual absorbers outlined in \S \ref{recombinations}, including the analytic approximations to the self-consistent numerical photoionization results.

\subsection{General Formalism}
\label{recombinations general formalism}
Consider a generic line recombination process with source function $S_{\nu}^{\rm rec}\equiv{j_{\nu}^{\rm rec}}/{\alpha_{\nu}}$, where 
\begin{equation}
j_{\nu}^{\rm rec}
=
\frac{h \nu_{\rm rec}}{4\pi}
\alpha_{\rm rec}(T) n_{i+1} n_{e}
\phi_{\rm rec}(\nu)
\end{equation}
is the relevant emission coefficient and $\alpha_{\nu}=\sum_{i} n_{i} \sigma_{i}(\nu)$ is the absorption coefficient accounting for absorption by all species.
Here, $\nu_{\rm rec}$ is the frequency of the recombination line of interest; we make the approximation that the line is sufficiently narrow that it can be represented by a single frequency for energetic purposes and the frequency dependence of the line profile is captured by the function $\phi_{\rm rec}(\nu)$, discussed in \S \ref{recombination processes}.
The effective recombination coefficient $\alpha_{\rm rec}(T)$ accounts for all the channels leading to the transition of interest.
We use the subscript ``$i+1$'' to refer to singly ionized species $i$.
For instance, if $i$ is HI, then $i+1$ is HII, and if $i$ is HeII, then $i+1$ is HeIII.\\ \\
A given ray crossing a slab absorber with incidence angle $\theta$ will emerge with extra recombination photons following the general solution to the radiative transfer equation \citep[e.g.,][]{1979rpa..book.....R}:
\begin{equation}
\label{recombination transfer solution}
I^{\rm rec}_{\nu}(N_{\rm HI};~x=\infty,~\mu)=
\int_{0}^{\tau_{\nu}(x=\infty)}
d(\tau_{\nu}'/\mu)
e^{-(\tau_{\nu}-\tau_{\nu}')/\mu}
S_{\nu}^{\rm rec}(\tau_{\nu}'),
\end{equation}
where $\tau_{\nu}$ is the optical depth normal to the slab and $\mu=\cos{\theta}$.
The recombination intensity can also be calculated in this way at any point $x$ within the slab to model the effects of recombinations on the ionization structure of the slab itself (\S \ref{individual absorbers} and Appendix \ref{photoionization appendix}). 
Since $\partial^{2}N/\partial z \partial N_{\rm HI}$ is the observed column density distribution, the appropriate value of $\mu$ to use in calculating the cosmological emissivity (eq. \ref{epsilon nu rec}) is one and we define $I_{\nu}^{\rm rec}(N_{i})\equiv I_{\nu}^{\rm rec}(N_{i};~x=\infty,~\mu=1)$.\\ \\
In the optically thin limit $\tau_{\nu}(x=\infty) \ll 1$,
\begin{equation}
\label{irec optically thin}
I^{\rm rec}_{\nu}(N_{\rm HI})
\stackrel{\tau_{\nu} \ll 1}{\to}
\frac{h \nu_{\rm rec}}{4\pi}
\frac{\alpha_{\rm rec}(T)}{\alpha^{A}_{i}(T)}
N_{i}
\Gamma_{i}
\phi_{\rm rec}(\nu),
\end{equation}
so that the recombination emission is proportional to the column density of the absorber ($N_{i}$), times the number of incident ionizing photons ($\Gamma_{i}$), times the fraction of recombinations that lead to the recombination process of interest ($\alpha_{\rm rec}(T)/\alpha^{A}_{i}(T)$).
As an absorber becomes optically thick, the number of incident photons it absorbs saturates as $1-e^{-\tau_{\nu}}$, so that we expect that the recombination emission will also saturate accordingly.
As the numerical calculations in Figures \ref{gammarec vs nhi} and \ref{HI Lya reem} show, this is the case and will be the basis for the analytical approximations to the reemission that we develop next.

\subsection{Analytic Approximations}
\label{recombinations analytic approximations}
A good analytic approximation to the emergent reemission from an absorber, capturing the optically thin limit and the optically thick saturation, is given by
\begin{equation}
\label{Inu rec}
I_{\nu}^{\rm rec}(N_{i})=\frac{h\nu_{\rm rec}}{4\pi}
\frac{\alpha_{\rm rec}(T)}{\alpha^{A}_{i}(T)}
\min(N_{i},~N_{\rm i,~thresh})
\Gamma_{i}
\phi_{\rm rec}(\nu),
\end{equation}
where and $N_{\rm thresh}$ is a threshold column density, near the optically thick transition, at which the recombination intensity saturates.
To make the approximation smoothly varying with column density, it is convenient to make the replacement min$(N_{i},~N_{i,\rm thresh}) \to N_{i,\rm thresh}(1-e^{-N_{i}/N_{i,\rm thresh}})$, which we do throughout in our numerical evaluations.
By inspection, we find that $N_{\rm HI,~thresh}=10^{16.75}$ cm$^{-2}$ and $N_{\rm HeII,~thresh}=10^{17.3}$ cm$^{-2}$ give good approximations to the full numerical results for the HI LyC and HeII LyC processes, respectively (Fig. \ref{gammarec vs nhi}).\\ \\
For HeII BalC reemission, a more robust approximation is obtained by noting that in the optically thick regime, the absorbers have a ``skin'' (analogous to a region behind an ionizing front) that is nearly uniform in HI and HeII.
While the recombinations are to HeII, the opacity to the 1 Ryd recombination photons owes to HI.
Because this skin is optically thick at the recombination energy, the emergent intensity simply approaches the source function.
The value of the source function near the outer edge of the absorber can be approximated using optically thin photoionization abundance ratios:
\begin{equation}
I_{\nu}^{\rm HeII~BalC}(N_{\rm HI}) \to 
S_{\nu}^{\rm HeII~BalC}({\rm skin})
\approx
\frac{h \nu_{\rm HeII~BalC}}{4\pi}
\frac{\alpha_{{\rm HeII},n=2}(T)}{\alpha^{A}_{\rm HeII}(T)}
\frac{\eta_{\rm thin} \Gamma_{\rm HeII}^{\rm ext}}
{\sigma_{\rm HI}(\nu_{\rm HI})}
\phi_{\rm rec}(\nu).
\end{equation}
For this process, we approximate the reemission as the minimum of the optically thin limit and this optically thick result.\\ \\
As Ly$\alpha$ is a resonant transition, its recombination photons will scatter until they diffuse out of the absorber unless they are destroyed by dust, metals, or HI continuum opacity in the case of HeII Ly$\alpha$ which can ionize it.
Of particular interest for the destruction by metals is the coincidence between the OIII $\lambda$303.799 line and the HeII \Lya~line at 303.783~\AA~\citep{1934PASP...46..146B}.
Studies of this process (including HI opacity) in the context of dense and enriched planetary nebulae and galactic nuclei \citep[][]{1969ApJ...157.1201W, 1980ApJ...242..615K, 1985ApJ...299..785E} suggest that the vast majority of the HeII \Lya~emission created by recombinations diffuse unimpeded into the IGM for \Lya~forest systems.
Only the most metal-rich Lyman limit systems and damped \Lya~absorbers potentially lose a significant fraction of their HeII \Lya~recombination photons to OIII, which should have little global impact.
Similarly, only the highest column density and most chemically evolved systems are likely to contain enough dust to efficiently destroy HI \Lya~photons.
We therefore adopt a simplified treatment of HI and HeII \Lya~recombination emission in which we assume that all recombination photons escape into the IGM.
Instead of integrating the source function over the absorber, we thus simply integrate the emission coefficient $j_{\nu}^{\rm HI/HeII~\Lya}$: 
\begin{equation}
I^{\rm HI/HeII~\Lya}_{\nu}(N_{\rm HI})=
\int_{-\infty}^{\infty}
dx
j_{\nu}^{\rm HI/HeII~\Lya}(x).
\end{equation}
By the preceding arguments, this is likely a reasonable approximation, but certainly an upper bound.
We assume case B conditions in which higher Lyman-series photons are ultimately degraded into lower-energy photons, with the ultimate production of extra \Lya~photons, which we implement by taking the appropriate case B emission coefficient.
\\ \\
For the Ly$\alpha$ processes, particularly good approximations to the numerical results can be developed as the optically thick limit can be accurately estimated.
The key observation is that the fraction of recombinations ultimately leading to the reemission of a \Lya~photon is a fixed number (at a given temperature) equal to the sum of all the production channels allowed by the selection rules.
Recombinations directly to $n=1$ produce LyC photons, but unless the recombination occurs near the skin of the absorber this photon will be reabsorbed before escaping and one must account for the probability that it will ultimately result in a \Lya~photon.
Averaging over angles, the number of ionizing photons incident from one side of the absorber that are absorbed per unit time per unit area per unit solid angle is given by
\begin{equation}
\label{Nabs eq}
\dot{N}_{\rm abs}^{i}(N_{\rm HI}) = \int_{0}^{1}d\mu
\int_{\nu_{i}}^{\infty}
\frac{d\nu}{h \nu}
J_{\nu}^{\infty} [1 - e^{\tau_{\nu}/\mu}]=
\int_{\nu_{i}}^{\infty}\frac{d\nu}{h \nu}J_{\nu}^{\infty} [1 - \tau_{\nu}\Gamma(-1,~\tau_{\nu})],
\end{equation}
where here $\Gamma(.,.)$ is the upper incomplete gamma-function and is unrelated to the photoionization rate, and $i\in \{{\rm HI,~HeII}\}$ indicates the species of interest.
Since in equilibrium the total number of recombinations equals the number of ionizations, the \Lya~reemission in the very optically thick regime can then be written as
\begin{equation}
\label{Lya reemission approximation}
I_{\nu}^{\rm HI/HeII~\Lya}(N_{\rm HI}) = f_{\alpha}^{\rm thick} h \nu_{\rm HI/HeII~\Lya} \dot{N}_{\rm abs}^{\rm HI/HeII~\Lya}(N_{\rm HI}) \delta(\nu-\nu_{\rm HI/HeII~\Lya}),
\end{equation}
where $f_{\alpha}^{\rm thick}=0.4$, set by matching the numerical calculations (Fig. \ref{HI Lya reem}), accounts for the efficiency of \Lya~photon production and geometrical effects.
In general, we again approximate the reemission as the minimum between the optically thin and optically thick results.\\ \\
Note that an alternative approach for approximating the recombination emission can be derived from the fact that the cosmological emissivity (eq. \ref{epsilon nu rec}) ultimately does not depend on the detailed function $I_{\nu}^{\rm rec}(N_{i})$ but on its integral over the column density distribution.
The part of the integral involving $N_{i}$ can thus be factored out and pre-computed for a given column density distribution.
\cite{1998AJ....115.2206F} used an approach along these lines and we also make use of this fact in the next section (eqs \ref{chi definition} and \ref{chi expression}) in developing an analytic model for the integrated recombination contribution to the ionizing background.

\section{C. ANALYTIC MODEL FOR THE RECOMBINATION CONTRIBUTION TO THE IONIZING BACKGROUND}
\label{recombinations analytical model}
Here we provide a quantitative analytic basis for understanding for the full numerical results on the contribution of recombinations to the ionizing background obtained in \S \ref{recombination contribution}.
We focus on the HI photoionization rate and assume that the mean free path is sufficiently short that the local source approximation of equation \ref{local source approximation} applies, which is valid at $z\gtrsim2$.
Based on \S \ref{recombinations} and in particular Figure \ref{gammarec vs nhi}, we assume that HI LyC reemission is the dominant process.
Following the notation of the main text, let us define $\Gamma_{\rm HI}^{\rm with~rec}\equiv\Gamma_{\rm HI}^{\rm no~rec}+\Gamma_{\rm HI}^{\rm rec}$ so that $\Gamma_{\rm HI}^{\rm rec}$ is the portion of the photoionization rate that is contributed by recombinations.\\ \\
As outlined in \S \ref{recombination contribution}, the recombination contribution depends on several factors.
Quantitatively, we write
\begin{equation}
\frac{\Gamma_{\rm HI}^{\rm rec}}{\Gamma_{\rm HI}^{\rm with~rec}}
=
\left.
\frac{\Gamma_{\rm HI}^{\rm rec}}{\Gamma_{\rm HI}^{\rm with~rec}}
\right|_{\rm max}
\times
f_{\rm leak}
\times
f_{\rm \sigma_{i}},
\end{equation}
where the first term is the maximum that would be attained if all the reemitted photons contributed to the ionizing background and all had frequency $\nu_{\rm HI}$, $f_{\rm leak}$ accounts for leakage of the photons below the ionization edge, and $f_{\rm \sigma_{i}}$ accounts for the frequency-dependence of the photoionization cross section and recombination line profile.\\ \\
We first consider the maximum contribution by assuming that a negligible fraction of the recombination photons leak below the ionization edge before being reabsorbed and taking the recombination line profile to be  a pure $\delta-$function, $\epsilon_{\nu}^{\rm rec} \equiv \tilde{\epsilon}_{\nu_{\rm HI}}^{\rm rec} \delta(\nu-\nu_{\rm HI})$.
Then,
\begin{equation}
\Gamma_{\rm HI}^{\rm rec} = 4 \pi
\int_{\nu_{\rm HI}}^{\infty}
\frac{d\nu}{h\nu}
J_{\nu}^{\rm rec}
\sigma_{\rm HI}(\nu)
=
\frac{
\Delta l_{\rm mfp}(\nu_{\rm HI})
\sigma_{\rm HI}
\tilde{\epsilon}_{\nu_{\rm HI}}^{\rm rec}
}{h \nu_{\rm HI}}.
\end{equation}
Combining equations \ref{epsilon nu rec} and \ref{Inu rec} for the cosmological recombination emissivity and the special case of HI LyC gives
\begin{equation}
\tilde{\epsilon}_{\nu_{\rm HI}}^{\rm rec}
=
h\nu_{\rm HI}
\frac{\alpha_{\rm HI,n=1}(T)}{\alpha_{\rm HI}^{\rm A}(T)}
\Gamma_{\rm HI}^{\rm with~rec}
\frac{dz}{dl}
\int_{0}^{\infty}
dN_{\rm HI}
\frac{\partial^{2}N}{\partial z \partial N_{\rm HI}}
N_{\rm HI, thresh}(1-e^{-N_{\rm HI}/N_{\rm HI,thresh}})
\end{equation}
and therefore
\begin{equation}
\left.
\frac{\Gamma^{\rm rec}_{\rm HI}}{\Gamma^{\rm with~rec}_{\rm HI}} 
\right|_{\rm max}
= 
\frac{\alpha_{\rm HI,n=1}(T)}{\alpha_{\rm HI}^{\rm A}(T)}
f_{\rm eff},
\end{equation}
where
\begin{equation}
\label{chi definition}
f_{\rm eff} \equiv 
\Delta l_{\rm mfp}(\nu_{\rm HI})
\sigma_{\rm HI}
\frac{dz}{dl}
\int_{0}^{\infty}
dN_{\rm HI}
\frac{\partial^{2}N}{\partial z \partial N_{\rm HI}}
N_{\rm HI, thresh}(1-e^{-N_{\rm HI}/N_{\rm HI,thresh}})
\end{equation}
is a dimensionless efficiency factor whose value depends on how recombination emission saturates with column density relative to absorption (the mean free path term).
Using equations \ref{taueff poisson expression} and \ref{mfp general}, we can express the mean free path at the ionization edge as
\begin{equation}
\Delta l_{\rm mfp}(\nu_{\rm HI}) = 
\frac{dl}{dz}
\left[
\int_{0}^{\infty}
dN_{\rm HI}
\frac{\partial^{2}N}{\partial z \partial N_{\rm HI}}
(1- e^{-\sigma_{\rm HI}N_{\rm HI}})
\right]^{-1}
\end{equation}
and thus
\begin{equation}
\label{chi expression}
f_{\rm eff} = 
\frac{
\int_{0}^{\infty}
dN_{\rm HI}
\partial^{2}N/\partial z \partial N_{\rm HI}
(\sigma_{\rm HI} N_{\rm HI, thresh})(1-e^{-N_{\rm HI}/N_{\rm HI,thresh}})
}
{
\int_{0}^{\infty}
dN_{\rm HI}
\partial^{2}N/\partial z \partial N_{\rm HI}
(1- e^{-\sigma_{\rm HI}N_{\rm HI}})
}.
\end{equation}
For a power-law column density distribution as in equation \ref{column density distribution eq}, the redshift dependence cancels out in the ratio and the integrals over $N_{\rm HI}$ can be done analytically in terms of gamma-functions.
The latter also cancel out, leaving a very simple result:
\begin{equation}
\label{chi simple}
f_{\rm eff} = (\sigma_{\rm HI} N_{\rm HI, thresh})^{2-\beta}.
\end{equation}
In \S \ref{recombinations analytic approximations}, we found that $N_{\rm HI, thresh}\approx10^{16.75}$ cm$^{-2}$ provides a good approximation for HI LyC.
Since $\sigma_{\rm HI} N_{\rm HI, thresh}<1$, equation \ref{chi simple} shows quantitatively how the recombination contribution increases with the steepness of the column density distribution (large $\beta$).
As noted in \S \ref{recombination contribution}, this physically arises because a larger fraction of recombinations occur in optically thin absorbers which return more of their recombination photons into the IGM.\\ \\
For a hydrogenic atom of integer charge $Z$, the fraction of recombinations that are to level $n$ is
\begin{equation}
\frac{\alpha_{\rm HI,n}(T)}{\alpha_{\rm HI}^{\rm A}(T)} = 
\frac{n^{-3}e^{\phi_{n}/kT} Ei(\phi_{n}/kT)}
{\sum_{i=1}^{\infty} i^{-3}e^{\phi_{i}/kT} Ei(\phi_{i}/kT)},
\end{equation}
where $\phi_{i}=Z^{2} h \nu_{\rm HI}/i^{2}$ is the ionization energy of level $i$ and $Ei(x)\equiv \int_{x}^{\infty} dx' e^{-x'}/x'$ \citep{1932MNRAS..92..820C}.\\ \\

\begin{figure}[ht]
\begin{center} 
\includegraphics[width=1.0\textwidth]{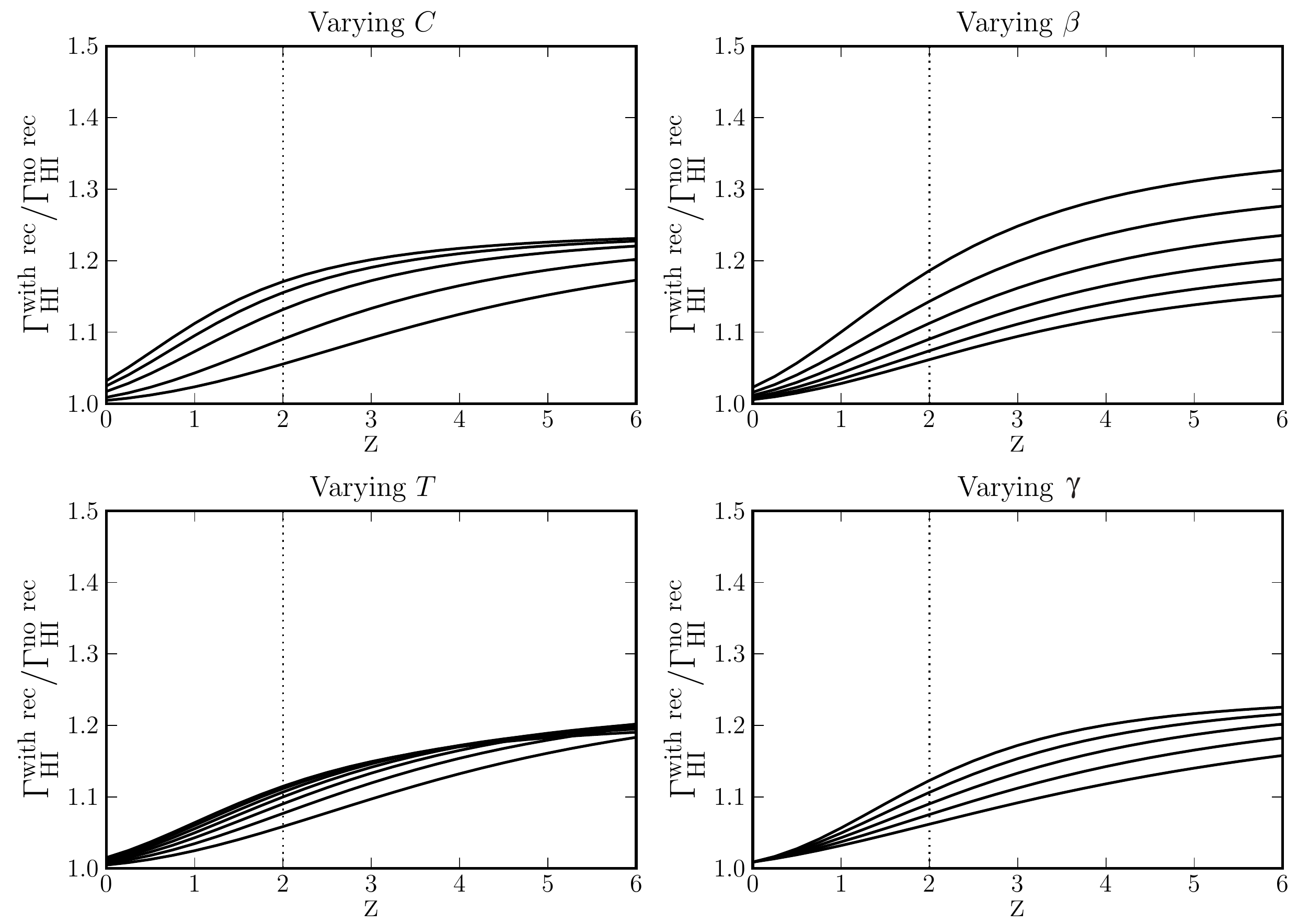} 
\end{center} 
\caption{
Dependences of the recombination contribution to HI photoionization rate calculated with the analytic model of Appendix \ref{recombinations analytical model}.
Shown are ratios of the total HI photoionization rate, including recombination emission, to the same calculation without recombination emission.
We vary the parameters of the column density distribution and temperature of the absorbers.
\emph{Top left:} $C=0.125,~0.25,~0.5,~0.75,~\textrm{and}~1$ from bottom up.
\emph{Top right:} $\beta=1.2,~1.3,~1.4,~1.5,~1.6~\textrm{and}~1.7$ from bottom up.
\emph{Bottom left:} $T=1.0,~1.5,~2.0,~2.5,~3.0,~3.5,~\textrm{and}~4.0\times10^{4}$ K from bottom up.
\emph{Bottom right:} $\gamma=1.0,~1.25,~1.5,~1.75,~\textrm{and}~2.0$ from bottom up.
The first three panels can be directly compared with the corresponding full numerical calculations shown in Figure \ref{Gammarec dependences}.
The main features and dependences are well reproduced by this simple model.
The differences at $z\lesssim2$, where the analytic model is inapplicable, arise because the local source approximation used becomes invalid as the radiation field begins to be limited by the cosmological horizon.
}
\label{Gammarec analytic} 
\end{figure}

The actual contribution of recombinations to the ionizing background is smaller because many recombination photons rapidly redshift below the ionization threshold.
What fraction of photons are lost through this leakage?
Focusing again on LyC recombinations, a photon is reemitted just above the ionization edge with energy $\nu'=\nu_{\rm HI}+\Delta \nu$, where $\Delta \nu/\nu_{\rm HI}\ll1$.
Supposing this photon is reemitted at redshift $z_{\rm rec}$, it will redshift below $\nu_{\rm HI}$ after traveling a proper distance
\begin{equation}
\Delta l_{\rm leak}(z_{\rm rec};~\Delta \nu) \approx
\frac{dl}{dz}(z_{\rm rec})
\frac{\Delta \nu}{\nu_{\rm HI}} (1+z_{\rm rec}).
\end{equation}
The local source approximation of equation \ref{local source approximation} is valid because in this regime we can write $J_{\nu}=(4\pi)^{-1} \int_{0}^{\infty}dl \epsilon_{\nu} e^{-l/\Delta l_{\rm mfp}}$, with the emissivity treated as a constant.
The leakage of recombination photons owing to redshifting implies that the emissivity should really integrated over a maximum distance $\Delta l_{\rm leak}$:
\begin{equation}
J_{\nu}^{\rm rec} = 
\frac{1}{4\pi}
\int_{0}^{\Delta l_{\rm leak}}
dl \epsilon_{\nu}^{\rm rec} e^{-l/\Delta l_{\rm mfp}}
=
\frac{1}{4\pi}
\Delta l_{\rm mfp}
\epsilon_{\nu}^{\rm rec}
[1 - f_{\rm lost}],
\end{equation}
where $f_{\rm lost}(z_{\rm rec};~\Delta \nu) \equiv e^{-\Delta l_{\rm leak}/\Delta l_{\rm mfp}}$ is the fraction of recombination photons emitted with frequency above the ionization edge that are lost to redshifting.\\ \\
The overall fraction of photons lost is then an average over the recombination line profile:
\begin{equation}
f_{\rm lost}(z_{\rm rec}) = \int_{\nu_{\rm  HI}}^{\infty} d\nu' \phi_{\rm rec}(\nu') f_{\rm lost}(z_{\rm rec};~\Delta \nu=\nu'-\nu_{\rm HI}).
\end{equation}
The result is simplified if the line profile is taken to be purely exponential, $\phi_{\rm rec}(\nu')=(h/kT)e^{-h\Delta \nu/kT}$ for $\Delta \nu \geq 0$, in which case
\begin{equation}
f_{\rm lost}(z_{\rm rec}) = 
\frac{1}{1+y(z_{\rm rec};~\Delta l_{\rm mfp},~T)},
\end{equation}
where
\begin{equation}
y(z_{\rm rec};~\Delta l_{\rm mfp},~T) \equiv
\frac{kT/h\nu_{\rm HI}}{\Delta l_{\rm mfp} (dz/dl)/(1+z_{\rm rec})}
\end{equation}
The mean free path depends on redshift and on the parameters of the column density distribution.
Using the analytical expression \ref{mean free path analytical equation} for $\Delta l_{\rm mfp}$, we can express $f_{\rm lost}$ directly in terms of these basic parameters:

\begin{equation}
f_{\rm lost}(z_{\rm rec};~C,~\gamma,~\beta,~T) = 
\frac{1}
{
1 + C\Gamma(2-\beta)(1+z_{\rm rec})^{1+\gamma} kT/h\nu_{\rm HI}
}
\end{equation}
and finally define $f_{\rm leak}\equiv 1 - f_{\rm lost}$.\\ \\
The last term to consider is the suppression factor that arises because the photoionization cross section entering in the photoionization rate from recombinations is frequency-dependent and photons are really reemitted with finite energy above the ionization threshold.
This is simply a frequency average of the cross section, over its maximum at $\nu_{\rm HI}$:
\begin{equation}
f_{\sigma_{i}} = 
\int_{\nu_{\rm HI}}^{\infty}
d\nu'
\phi_{\rm rec}(\nu')(\nu/\nu_{\rm HI})^{-3}
=
\left( \frac{h \nu_{\rm HI}}{k T} \right)^{3}
e^{-h\nu_{\rm HI}/kT}
\Gamma(-2,~h\nu_{\rm HI}/kT),
\end{equation}
where we have again approximated the recombination radiation to have a purely exponential profile and that the cross section scales as $\nu^{-3}$ just above the ionization edge.\\ \\
In Figure \ref{Gammarec analytic}, we combine these analytic results and show how $\Gamma_{\rm HI}^{\rm with~rec}/\Gamma_{\rm HI}^{\rm no~rec}=1/(1-\Gamma_{\rm HI}^{\rm rec}/\Gamma_{\rm HI}^{\rm with~rec})$ varies as a function of the parameters of the column density distribution and the temperature of the absorbers.
Note that the main features and dependences of the corresponding full numerical calculations shown in Figure \ref{Gammarec dependences} are well reproduced by this simple model.
The differences at $z\lesssim2$ arise because the local source approximation used in the analytic calculations becomes invalid as the radiation field begins to be limited by the cosmological horizon; the ``effective mean free path'' is then shorter than the mean absorption distance and more recombination photons are thus in reality retained in the ionizing range.

\section{D. SPECTRAL FILTERING}
\label{spectral filtering}
The spectral index of a radiation background in general differs from the spectral index of its sources owing to filtering along the line of sight.
In this section, we explore different filtering cases relevant to the ionizing background in order to provide physical understanding of our numerical solutions of the radiative transfer equation and provide analytical results referred to in the main text.\\ \\
An important characteristic of the ionizing background at high redshifts is that it is local in the sense that the specific intensity $J_{\nu}$ depends only on the local value of the specific emissivity $\epsilon_{\nu}$:
\begin{equation}
\label{local source approximation}
J_{\nu}(z)\approx
\frac{1}{4\pi}
\Delta l_{\rm mfp}(\nu,~z) \epsilon_{\nu}(z). 
\end{equation}
Here, $\Delta l_{\rm mfp}(\nu,~z)$ is the mean free path of photons of frequency $\nu$ and redshift $z$ and is given by
\begin{equation}
\label{mfp general}
\Delta l_{\rm mfp}(\nu_{0},~z_{0})=
\frac{dl}{dz}(z_{0})
\left(\frac{d\bar{\tau}(\nu_{0},~z_{0},~z)}{dz}\right)^{-1}(z_{0}).
\end{equation}
This limit of equation \ref{transfer equation solution} is valid whenever photons are absorbed so close to their point of emission that they redshift only negligibly.
For HI ionizing photons of wavelength 912~\AA, the ``breakthrough'' point above which this approximation holds is approximately $z=2$ \citep[][]{1999ApJ...514..648M}.
In this regime, cosmological effects are unimportant and calculations can be performed in ordinary Euclidean geometry.
Then the effective optical depth at frequency $\nu$ over a proper length $l$ at redshift $z$ can be written as $\bar{\tau}(\nu,~z,~l)$ and $\Delta l_{\rm mfp}(\nu,~z)=(d\bar{\tau}/dl)^{-1}$.
A useful analytical expression for the mean free path can be obtained at frequencies where the photoionization cross section $\sigma_{i}\propto \nu^{-3}$ and the column density distribution is described by single power laws $\beta$ and $\gamma$ (see eq. \ref{column density distribution eq}).
In the case of HI,
\begin{equation}
\label{mean free path analytical equation}
\Delta l_{\rm mfp}(\nu_{0},~z_{0}) \approx
\frac{(\beta-1)c}{\Gamma(2-\beta)N_{0}\sigma_{\rm HI}^{\beta-1}} 
\left(\frac{\nu_{0}}{\nu_{\rm HI}}\right)^{3(\beta-1)}
\frac{1}
{(1+z_{0})^{\gamma+1}H(z_{0})},
\end{equation}
and it is straightforwardly generalized to any other single ion.
\\ \\
An implicit assumption in equation \ref{local source approximation} (as well as throughout much of this paper, such as as in eqs. \ref{transfer equation} and \ref{transfer equation solution}) is that a sphere of radius one mean free path contains sufficiently many sources that their effect is well-captured by the use of a volume-averaged uniform emissivity.
This is generally valid when the sources are star-forming galaxies, which are very numerous.
Quasars, however, are much rarer and this assumption in general fails. 
(See Figure \ref{scales} and the discussion in \S \ref{spectrum during heii reion}.)
We will therefore also consider the case of filtering of radiation from an isolated source.
For each case, we will consider both the cases in which the intervening absorbers are uniformly distributed and the one in which discrete absorbers are Poisson-distributed following a column density distribution.

\subsection{Uniform Emissivity and Absorbing Material}
In the case in which both the emissivity and the absorbing material are spatially uniform, 
\begin{equation}
J_{\nu}(z)\approx \frac{1}{4\pi} \Delta l_{\rm mfp}(\nu,~z) \epsilon_{\nu}(z) \textrm{~with~} \bar{\tau}(\nu,~z,~l)=n_{\rm abs} \sigma_{\rm abs}(\nu) l,
\textrm{~so that~}
J_{\nu}(z)\approx
\frac{1}{4\pi}
\frac{\epsilon_{\nu}(z)}{n_{i}\sigma_{i}(\nu)},
\end{equation}
where $n_{i}$ is the number density of the absorbing material and $\sigma_{i}(\nu)$ is its cross section.
For photons with $\nu_{\rm HI} \leq \nu < \nu_{\rm HeII}$, the dominant source of continuum opacity owes to HI photoionization, so $n_{i}=n_{\rm HI}$ and $\sigma_{i}(\nu)=\sigma_{\rm HI}(\nu)$.
For $\nu \geq \nu_{\rm HeII}$, HI continuum opacity is fractionally small in the cosmological context 
and we can take $n_{i}=n_{\rm HeII}$ and $\sigma_{i}(\nu)=\sigma_{\rm HeII}(\nu)$.
Since both absorbing ions are hydrogenic, they similarly harden spectra following $J_{\nu} \propto \epsilon_{\nu}/\sigma_{i}(\nu) \approx \epsilon_{\nu} \nu^{3}$.
The last equality is approximately valid near the ionization edge.
At high energies, the photoionization cross section decreases more slowly and the hardening becomes negligible.

\subsection{Uniform Emissivity and Discrete Absorbers}
If the emissivity is spatially uniform, but the absorbers are discrete and Poisson-distributed, the result is similar, but with the effectively optical depth calculated from the column density distribution as in equation \ref{taueff poisson expression}. 
If the column density distribution of the absorbers follow a power-law $dN/dN_{i}\propto N_{i}^{-\beta}$, then just above the ionization edge where the cross section $\sigma_{i}(\nu)\propto \nu^{-3}$, the mean free path $\Delta l_{\rm mfp}(\nu,~z)\propto \nu^{3(\beta-1)}$ \citep[][]{1993ApJ...418...28Z, 2008ApJ...688...85F} and therefore $J_{\nu} \propto \epsilon_{\nu}\nu^{3(\beta-1)}$.
For HI, we may take $\beta=1.4$, as measured by \cite{2007AJ....134.1634M}, in which case the hardening $\alpha \to \alpha-1.2$ is significantly weaker than the $\alpha \to \alpha-3$ in the uniform absorbing material case above the HI ionization edge.
If the proportionality factor $\eta$ between $N_{\rm HI}$ and $N_{\rm HeII}$ were a constant throughout, the radiation above the HeII ionization edge would be identically hardened.
However, the complex behavior of $\eta$ in the optically thick regime (\S \ref{column density relations}) may alter this result; this behavior is taken into account in our numerical calculations.

\subsection{Point Source and Uniform Absorbing Material}
The case of an isolated point source is quite different.
In this case, the specific intensity (which can no longer be assumed to be isotropic, so we denote it by $I_{\nu}$ instead of $J_{\nu}$), is exponentially suppressed with increasing optical depth from the source:
\begin{equation}
I_{\nu}= I_{\nu}(l=0) e^{-\tau_{\nu}(l)}.
\end{equation}
We again assume that HI continuum opacity dominates for $\nu_{\rm HI} \leq \nu < \nu_{\rm HeII}$ and that HeII continuum opacity dominates for $\nu>\nu_{\rm HeII}$.
In each regime, $\tau_{\nu}(l)=n_{i}\sigma_{i}(\nu)l$.
The spectrum is attenuated relative to its value at $\nu_{\rm i}$ by a factor
\begin{equation}
\frac{I_{\nu}}{I_{\nu_{i}}} =
\frac{e^{-\tau_{\nu}}}
{e^{-\tau_{\nu_{i}}}}
= e^{\tau_{\nu_{i}}}
e^{-\tau_{\nu_{i}}[\sigma_{i}(\nu)/\sigma_{i}(\nu_{i})]}
\propto
\left( e^{-\tau_{\nu_{i}}} \right)
^{\sigma_{i}(\nu)/\sigma_{i}(\nu_{i})}
\approx
\left( e^{-\tau_{\nu_{i}}} \right)
^{(\nu/\nu_{i})^{-3}},
\end{equation}
where the last equality again holds approximately just above the ionization edge.

\subsection{Point Source and Discrete Absorbers}
For a point source attenuated by Poisson-distributed discrete absorbers, we simply replace the ordinary optical depth by the effective optical depth to obtain the average spectrum:
\begin{equation}
I_{\nu}(l)=I_{\nu}(l=0) 
e^{-\bar{\tau}_{\nu}(l)}.
\end{equation}
Drawing on our previous results,
\begin{equation}
\frac{I_{\nu}}{I_{\nu_{i}}} =
\frac{e^{-\bar{\tau}_{\nu}}}
{e^{-\bar{\tau}_{\nu_{i}}}}
= e^{\bar{\tau}_{\nu_{i}}}
e^{-\bar{\tau}_{\nu}}
\propto e^{-\bar{\tau}_{\nu}}
\approx
\left( e^{-\bar{\tau}_{\nu_{i}}} \right)
^{(\nu/\nu_{\rm i})^{-3(\beta-1)}}.
\end{equation}\\ \\
It is worth noting, as explained in \S \ref{quasar within ionized bubble}, that at fixed optical depth from a source the hardening is the same regardless of how (smoothly or discretely) the intervening material is distributed.
The different average hardening differs in the discrete case really because of the added \emph{stochastic} nature of the intervening optical depth at fixed distance.

\section{E. ATOMIC PHYSICS}
\label{atomic physics values}
The radiative transfer calculations in this paper are ultimately rooted in atomic physics.
The recombination rates and HeI photoionization cross sections are taken from the appendix of \cite{1997MNRAS.292...27H}.
For the \Lya~emission coefficients and the HI and HeII photoionization cross sections we take the expressions given by \cite{2006agna.book.....O}.
In particular, given the prominent role it plays in our calculations, the photoionization cross section of a hydrogenic atom of atomic number $Z$ ($Z=1$ and 2 for HI and HeII) is given by
\begin{equation}
\sigma_{i}(\nu) = 
\frac{A_{0}}{Z^{2}}
\left( \frac{\nu_{1}}{\nu} \right)^{4}
\frac{\exp\{ 4 - [4\tan^{-1}{\epsilon}]/\epsilon \}}
{1-\exp(-2\pi/\epsilon)}
\end{equation}
for $\nu\geq \nu_{1}$ and 0 otherwise.
Here,
\begin{equation}
A_{0} = \frac{2^{9}\pi}{3 e^{4}} \alpha \pi a_{0}^{2} = 6.30 \times 10^{-18} {\rm~cm}^{2},
\end{equation}
where $\alpha$ is the fine structure constant and $a_{0}$ is the Bohr radius, $\epsilon=\sqrt{\nu/\nu_{1}-1}$, and $h \nu_{1} = Z^{2} h \nu_{\rm HI} = 13.6Z^{2}{\rm~eV}$.
Just above the photoionization edge $\nu_{1}$, i.e. for $\epsilon \ll 1$, $\sigma_{i}(\nu)\propto \nu^{-3}$ but the cross section drops more slowly as $\nu\to \infty$, explaining the lack of spectral hardening in this limit.\\ \\
Also of interest is the recombination line profile for free-bound transitions, which is important in determining the contribution of recombination emission to the ionizing background (\S \ref{recombinations}).
The probability that a recombination yields a continuum photon of frequency between $\nu$ and $\nu+d\nu$, by definition, is $\phi_{\rm rec}(\nu)d\nu$.
The velocity $u$ of the recombining electron relative to the nucleus\footnote{Note that since $m_{p}/m_{e}\approx2000$, the nuclei can be assumed to be at rest in the frame of the gas.} is related to the frequency by $h\nu=m_{e}u^{2}/2+h\nu_{\rm rec}$, where $\nu_{\rm rec}$ is the ionization edge frequency.
Now, the probability that the recombining electron has velocity between $u$ and $u+du$ scales as the probability that an electron in this velocity range recombines when it encounters a nucleus, times the number of electrons with velocity in this range.
If $\sigma_{\rm rec}(u)$ is the velocity-dependent recombination cross section, then the first term is $\propto \sigma_{\rm rec}(u) u$.
In thermal equilibrium, the second term is given by the Maxwell-Boltzmann speed distribution, $f_{\rm M-B}(u)\propto u^{2} \exp(-m_{e}u^{2}/2kT)$.
Thus,
\begin{equation}
\phi_{\rm rec}(\nu) d\nu \propto \sigma_{\rm rec}(u) u f_{\rm M-B}(u) du.
\end{equation}
The Milne detailed balance relation relates the recombination cross section to the corresponding photoionization cross section: 
\begin{equation}
\sigma_{\rm rec}(u) = 2 
\left(
\frac{h \nu}{m_{e} u c}
\right)^{2}
\sigma_{i}(\nu)
\end{equation}
\citep[e.g.,][]{2006agna.book.....O}.
Using the approximation $\sigma_{i}(\nu)\propto \nu^{-3}$, we can solve and find, after normalizing,
\begin{equation}
\phi_{\rm rec}(\nu) = \frac{(\nu/\nu_{\rm rec})^{-1}\exp{(-h\nu/kT})}{\Gamma(0,~h\nu_{\rm rec}/kT)} \frac{\theta(\nu-\nu_{\rm rec})}{\nu_{\rm rec}}.
\end{equation}
This profile will also be thermal broadened and shifted owing to the peculiar velocity of the emitting gas; these corrections are however negligible in comparison to the width of the recombination line profile.
For instance, the recombination line width $\Delta \nu_{\rm rec}/\nu_{\rm rec}\approx kT/h\nu_{\rm rec}\approx0.13$ for HI LyC at $T=2\times10^{4}$ K, while the Doppler broadening at the same temperature $\Delta \nu_{D}/\nu_{\rm rec}\approx0.002$ and the peculiar velocity shift $\Delta \nu_{\rm pec}/\nu_{\rm rec}\approx0.0003$ for $v_{\rm pec}=100$ km s$^{-1}$.

\bibliography{references} 
 
\end{document}